\documentclass[twocolumn]{raa} 

\usepackage{graphicx,times}
\usepackage{natbib}
\usepackage{amssymb,amsmath}
\bibpunct{(}{)}{;}{a}{}{,}
\usepackage[pagebackref=true]{hyperref}

\usepackage{float}
\usepackage{mathrsfs}	
\usepackage{color}
\usepackage{multirow}
\usepackage{blindtext}
\usepackage{longtable,booktabs,supertabular}

\usepackage{setspace}

\def\gsim{~\rlap{$>$}{\lower 1.0ex\hbox{$\sim$}}}
\def\lsim{\mathrel{\rlap{\lower3.5pt\hbox{\hskip0.5pt$\sim$}}
            \raise0.5pt\hbox{$<$}}}
            
\usepackage{color}
\newcommand{\Zhong}[1]{{\color{black}{#1}}}

\def\Gone{GaSNet-L1}
\def\Gtwo{GaSNet-L2}
\def\Gthr{GaSNet-L3}
\def\Gs{GaSNets}
\def\eboss{eBOSS}
\usepackage{comment}

\begin{document}

\title{Galaxy Spectra neural Networks (GaSNets). I. Searching for strong lens candidates in \eboss\ spectra using Deep Learning}

\volnopage{ {\bf 20XX} Vol.\ {\bf X} No. {\bf XX}, 000--000}
   \setcounter{page}{1}

\author{
        Fucheng Zhong \inst{1,2},
        Rui Li \inst{3,4}
        \thanks{E-mail: lrui@bao.ac.cn},
        Nicola R. Napolitano \inst{1,2} 
        \thanks{E-mail: napolitano@mail.sysu.edu.cn}
        }
        
\institute{
School of Physics and Astronomy, Sun Yat-sen University, Zhuhai Campus, 2 Daxue Road, Xiangzhou District, Zhuhai, P. R. China; \\
        \and
                CSST Science Center for Guangdong-Hong Kong-Macau Great Bay Area, Zhuhai, China, 519082\\
            \and
                    School of Astronomy and Space Science, University of Chinese Academy of Sciences, Beijing 100049, China; \\
        \and 
                National Astronomical Observatories, Chinese Academy of Sciences, 20A Datun Road, Chaoyang District, Beijing 100012, China\\
        \vs \no
   {\small Received 20XX Month Day; accepted 20XX Month Day}
}

\abstract{With the advent of new spectroscopic surveys from ground and space, observing up to hundreds of millions of galaxies, spectra classification will become overwhelming for standard analysis techniques. To prepare for this challenge, we introduce a family of deep learning tools to classify features in one-dimensional spectra. As the first application of these Galaxy Spectra neural Networks (GaSNets), we focus on tools specialized \Zhong{in} identifying emission lines from strongly lensed star-forming galaxies in the eBOSS spectra. We first discuss the training and testing of these networks and define a threshold probability, $P_L$, of 95\% for the \Zhong{high-quality} event detection. Then, using a previous set of spectroscopically selected strong lenses from eBOSS, confirmed with HST, we estimate \Zhong{a completeness} of 
$\sim$80\% as the fraction of lenses recovered above the adopted $P_L$. 
We finally apply the GaSNets to $\sim1.3$M eBOSS spectra
to collect \Zhong{the} first list of $\sim$430 new \Zhong{high-quality} candidates identified with deep learning from spectroscopy
and visually graded as highly probable real events. A preliminary check against ground-based observations tentatively shows that {this sample 
has a confirmation rate of 38\%, 
in line with previous samples selected with standard (no deep learning) classification tools and confirmed by the Hubble Space Telescope.}  
This first test shows that machine learning can be efficiently extended to feature recognition in the wavelength space, which will be crucial for 
future surveys like 4MOST, DESI, Euclid, and the China Space Station Telescope (CSST).
}

\authorrunning{F. Zhong et al.}
\titlerunning{GaSNets: searching strong lenses in eBOSS}%

\maketitle


\section{Introduction}
\label{sec:intro}
Strong gravitational lensing (SGL) is a powerful tool to investigate a large variety of open questions in cosmology. The formation of the distorted images of background galaxies, the ``sources'', \Zhong{depends} on the total mass of the foreground gravitational systems acting as ``deflectors'' or ``lenses''. In case this latter are galaxies, SGL provides us 
accurate constraints on different properties 
correlated to their total mass, like the mass-to-light ratio \citep{2021MNRAS.506.6144G}, the dark matter fraction \citep{2009ApJ...705.1099A,2010ApJ...721L...1T}, the slope of the total density profile (\citealt{2006ApJ...649..599K,2009ApJ...703L..51K}, \citealt{2009ApJ...705.1099A}) and its relation with other parameters (\citealt{2012ApJ...757...82B},  \citealt{2015ApJ...803...71S}; \citealt{2018MNRAS.480..431L}). \Zhong{SGL is also used to constrain the evolution} of galaxies via merging (\citealt{2012ApJ...757...82B}, \citealt{2013ApJ...777...98S,2014ApJ...786...89S,2015AAS...22530902S}), 
the initial mass function in massive ellipticals (\citealt{2011MNRAS.417.3000S, 2012MNRAS.423.1073B}), and study the dark substructures around large galaxies (\citealt{ 2018MNRAS.481..819G,2019A&A...631A..40S}). 

Moving to more cosmological constraints, SGL is used to measure the Hubble constant($H_0$), and other cosmological parameters (\citealt{2013ApJ...766...70S, 2017MNRAS.468.2590S,2019MNRAS.490..613S,2020MNRAS.498.1440R,2020MNRAS.498.1420W}). 

Strong lenses are generally searched in imaging data, where one can clearly distinguish the lensing features in the form of arcs, or multiple images of compact sources, like galaxies or quasars
(\citealt{2008ApJ...682..964B}; \citealt{2012ApJ...744...41B}; \citealt{2013ApJ...777...98S}; \citealt{2013ApJ...766...70S}). Here, a great impulse to lens hunting has been recently provided by automatized tools for lens finding \citep{2014ApJ...785..144G}. In particular, machine learning (ML) techniques have been lately found to be very powerful in collecting hundreds of high quality (HQ) candidates (Dark Energy Survey -- DES: \citealt{2019yCat..22430017J}, Kilo Degree Survey -- KiDS: \citealt{2019MNRAS.482..807P}; \citealt{2020ApJ...899...30L, 2021ApJ...923...16L}, Hyper Supreme-Cam -- HSC: \citealt{2018PASJ...70S..29S}).

After the identification of HQ candidates, a spectroscopical follow-up is needed to confirm their gravitational lensing nature (\citealt{2019A&A...625A.119M}; \citealt{2021MNRAS.508.1686W}). In practice, one needs to collect the spectra of the lens and the source and measure their relative redshift, confirming that the lens is located in front of the source as expected from 
ray-tracing lensing models 
(see \citealt{2018ApJ...853..148C}, \citealt{2020ApJ...904L..31N}).
This is a severe \Zhong{bottleneck} in the SGL studies and, so far, there have been only sparse programs dedicated to \Zhong{these} follow-up observations (\citealt{2019MNRAS.483.3888S,2019MNRAS.485.5086S}; \citealt{2020MNRAS.494.3491L}; \citealt{2020MNRAS.494.1308N}). However, future large sky spectroscopic surveys (e.g. Taipan: \citealt{2017PASA...34...47D} 4MOST: \citealt{2012SPIE.8446E..0TD}, DESI: \citealt{2016arXiv161100036D}) will provide an unprecedented opportunity for massive 
follow-ups of lensing candidates, e.g. by reserving them dedicated observing nieces in wide programs \Zhong{or} by accommodating them as filler targets in large-sky, multi-purpose surveys.

More interestingly, these large spectroscopic sky surveys will offer a unique chance to be used
as a playground for lens finding, e.g. by looking for blended emission lines of background \Zhong{``lensed''} galaxies, e.g. star-forming systems, in the spectrum of a forward massive \Zhong{systems}.
This method has been extensively used in the last years to produce tens of discoveries of new unknown lens candidates. 

The first example of a search of this kind was presented by
\citet{2004AJ....127.1860B}, within The Sloan Lens ACS (SLACS). They found 49 SGL candidates in 50\,996 Sloan Sky Digital Survey (SDSS) spectra of luminous red galaxies (LRG).
They used 
the principal-component analysis to subtract the main components of the foreground LRG spectrum and a Gaussian kernel to find the best emission lines in the residual flux. They mainly focused on [OII] (3728\AA), [OIII] (4960\AA, 5007\AA)\Zhong{, and} $H_\beta$ (4863\AA) lines, hence exploring a redshift range of $z=(0.16-0.49)$ for the lenses and $z=(0.25-0.81)$ for the sources.
Later analyses increased the number of SLACS candidates to 131 (\citealt[][SLACS hereafter]{2008ApJ...682..964B}). 
Within the BOSS Emission-Line Lens Survey (BELLS, hereafter), \citet[][]{2012ApJ...744...41B} 
extended the spectroscopic search, \Zhong{previously performed} in SLACS, 
to higher redshift, by looking for lenses up to $z=0.7$ and the background sources
up to $z=1.4$, with no color pre-selection. This allowed them to finally find 
45 SGL candidates in 133\,852 SDSS galaxy spectra.
\Zhong{Along} the same line of approaches, in the SLACS Survey for the Masses (S4TM) project, \citet[][S4TM hereafter]{2015ApJ...803...71S,2017ApJ...851...48S} have extended the \Zhong{search} for SGL candidates to lower masses 
and found 118 new lens candidates.
On the other hand, \citet[][BELLS GALLERY]{ 2016ApJ...824...86S,2016ApJ...833..264S} and \cite{2020MNRAS.499.3610C} looked for high-redshift Ly$\alpha$ emitters as background sources \Zhong{and} found 361 candidates.

The main disadvantage of these \Zhong{spectroscopy-selected} samples is the missing information from images. Indeed, 
even if spectra can provide the evidence of two different emitting sources along the line of sight located at different redshifts, they cannot guarantee that they represent an SGL event.
Hence, high-resolution imaging
from space telescopes or adaptive optics \Zhong{is} needed to have a visual confirmation of the lenses. 
Currently, there are 135 confirmed lenses with Hubble Space Telescope (HST) observations of the 294 selected using optical lines (70/131 from SLACS, 25/45 from BELLS, 40/118 from S4TM), and 17/21 Ly$\alpha$ candidates from BELLS GALLERY.
With the lesson learned from SDSS/BOSS, other experiments have combined the spectroscopic selection and imaging: \citet{2016ApJ...832..135C} matched 45 spectra from the Galaxy And Mass Assembly (GAMA) survey and confirmed 10 of them with Hyper Suprime-Cam (HSC) imaging; \citet{2022MNRAS.510.2305H} selected lens candidates in AAOmega spectra and followed up 56 of them with HST to find 9 confirmations. 

The discovery power of this approach will be pushed to unprecedented limits by future surveys combining spectroscopy and imaging from space (e.g. Euclid mission and CSST)
and produce a revolution in the lensing searches.
However, this revolution will stand on the ability to effectively analyze 
gigantic spectroscopic data loads,
which will imply the inspection of millions of spectra and the identification of (\Zhong{sometimes} very faint) emission lines from background lensed systems. This is a prohibitive task for standard human-driven analyses, \Zhong{unless} one adopts severe selections \Zhong{to maximise the number of detection but reduce} the spectra to visually inspect.


Machine learning techniques can provide, instead, fast and efficient methods to overcome these difficulties and systematically search for lensing features in spectra. For instance, Convolutional Neural Networks (CNNs) have been previously applied
for lens searches 
\cite[see][Li+19, hereafter]{2019MNRAS.482..313L}. In particular, they have 
focused on the identification of Ly$\alpha$ emitters 
at higher redshift ($2<z<3$) in the spectra of lower redshift early-type galaxies ($z<0.6$), and showed that
these techniques can be efficiently used as a classifier for galaxy spectra.

In this paper, we expand this approach and develop a new CNN tool to look for SGL 
in the Baryon Oscillation Spectroscopic Survey (BOSS) spectroscopic database (\citealt{2016AJ....151...44D}). Since these sources are usually \Zhong{star-forming} galaxies,
we plan to use machine learning techniques to search for higher redshift emission lines such as [OII], [OIII], $H_\alpha$, $H_\beta$ and $H_\delta$ 
mixed in the foreground galaxy spectra. 
To do that, we build 3 CNN models: 
a classifier, to search for reliable emission lines in spectra, and two regression models, to measure the foreground galaxy and the background source redshifts, respectively.
Then, we 
combine the predictions of the 3 CNNs to provide a list of high probability events that we visually inspect to select HQ candidates.

Finally, we compare this {\it first deep learning spectroscopic-selected sample} with the most complete spectroscopic sample of SGL candidates in BOSS observations from \citet[][T+21 hereafter]{2021MNRAS.502.4617T}, obtained with standard cross-correlation techniques. This catalog consist of 838 likely, 448 probable, and 265 possible strong lens candidates, for a total of 1551 objects. They have also obtained a preliminary confirmation of 477 of them with low-resolution imaging.






The paper is organized as follows. In \S2, we will introduce the whole idea of this project and the details of the new CNN models. In \S3, we introduce the \Zhong{modeled} emission lines and the construction of training data. In \S4, we show the training and testing results of the new CNNs. In \S5, we will apply the new CNNs to the BOSS spectra and derive a list of candidates
that we qualify via visual inspection of their spectra, finally providing a catalog of HQ candidates. In \S6, we will discuss the results and estimate a tentative confirmation rate based on the match with ground-based imaging. We also discuss some avenues for improvement of future CNNs. In the final \S7, we draw some conclusions.

\section{ \Zhong{Methodology} }
\Zhong{
In this work we want to apply ML techniques to spectroscopic data. In particular, we want to use the 
ability of these techniques to perform feature recognition and classification. ML has been widely used in astronomical data analysis: 1) to find or classify different astronomical target candidates, like AGN \citep{2018MNRAS.478.3177T,Zhu:2020hwk,2021ApJ...920...68C}, quasar \citep{2019A&A...632A..56K,2021RAA....21...99Y}, star clusters \citep{2021MNRAS.503..236J,2021RAA....21...93H} or 2) to measure or predict physical parameters of astronomical targets, like redshifts \citep{2007ApJ...663..774B,2021GeneticKNN}, masses \citep{2016ApJ...831..135N,2019A&A...622A.137B}, or velocity profiles \citep{2021RSPTA.37900171M}. 
Here we want to test the possibility to 
use ML techniques to 
efficiently search for strong lenses on vast amounts of spectra and predict the redshift of their lenses and sources.
}

\subsection{The challenge of searching for strong lenses in spectra}
\label{sec:challange}
Next generation spectroscopic surveys will target tens of millions of galaxies (\citealt{2019BAAS...51c.363M}). These huge samples will allow us to 
systematically search for high-probability candidates from integrated spectra, as the number of expected events is noticeable, given the large number of background galaxies potentially giving rise to lensing events.

Using the set of predictions from Collett (2015)\footnote{https://github.com/tcollett/LensPop}, we have estimated that the number of lenses with a $1''$ Einstein radius, $R_{\rm E}$, producing lensed images of the source, observable with a spectroscopic survey with a 2$''$ diameter fiber, over 15\,000 deg$^2$ of the sky, is of the order of 7\,000. This is obtained assuming that the source is bright enough in some visual band (e.g. Euclid visual mag$=24.5$) to make also the signal-to-noise ratio of the emission lines high enough to be detected from the ground for typical spectroscopic surveys (e.g. 4MOST or DESI).
This estimate is subject to different factors, including some flux loss, but it also excludes the contribution from sources with slightly larger $R_{\rm E}$s that might eventually scatter part of their light into a 2$''$ fiber. Hence, combining all these effects, this forecast is possibly not far from realistic. This is a wealth of data extremely valuable \Zhong{because} it provides, for free, the information on the lens and the source redshifts, which are crucial for the lensing modeling. Standard techniques based on sophisticated selection criteria (T+21) still require rather \Zhong{time-consuming} visual inspection. Hence, \Zhong{a} more practical solution to perform a systematic search of lens candidates in these datasets is mandatory.

This is possibly true also for current spectroscopic surveys.
For instance, using the same set of predictions
for the BOSS area ($\sim 10000$ deg$^2$), and assuming a fainter limiting magnitude for the sources, $r=$23.5, we get $\sim 920$ lenses within a 2$''$ fiber, which become $\sim 1470$ within a 3$''$ fiber (e.g. the one available for SDSS releases earlier than 12). Currently, the largest collection of candidate lenses with BOSS spectra consists of 477 objects with lensing evidence from low-resolution images (T+21).
Taking this sample as a {\it bona fide} high-completeness sample, this is rather far  from the expected number of discoverable lenses, meaning that there might \Zhong{be} more lenses to find in the full BOSS dataset. Given the full set of BOSS spectra available, i.e. $\sim2.6$M items (\citealt{2020ApJS..249....3A}), this means that we should expect one real blended emission line object every $\sim$3500 spectra. 

In this work, we want to tackle the problem of
systematic searches of lens candidates in spectra with
deep learning \Zhong{and} use the BOSS dataset to test the efficiency of this approach. 

\begin{figure}
    \centering 
    \includegraphics[width=0.8\linewidth]{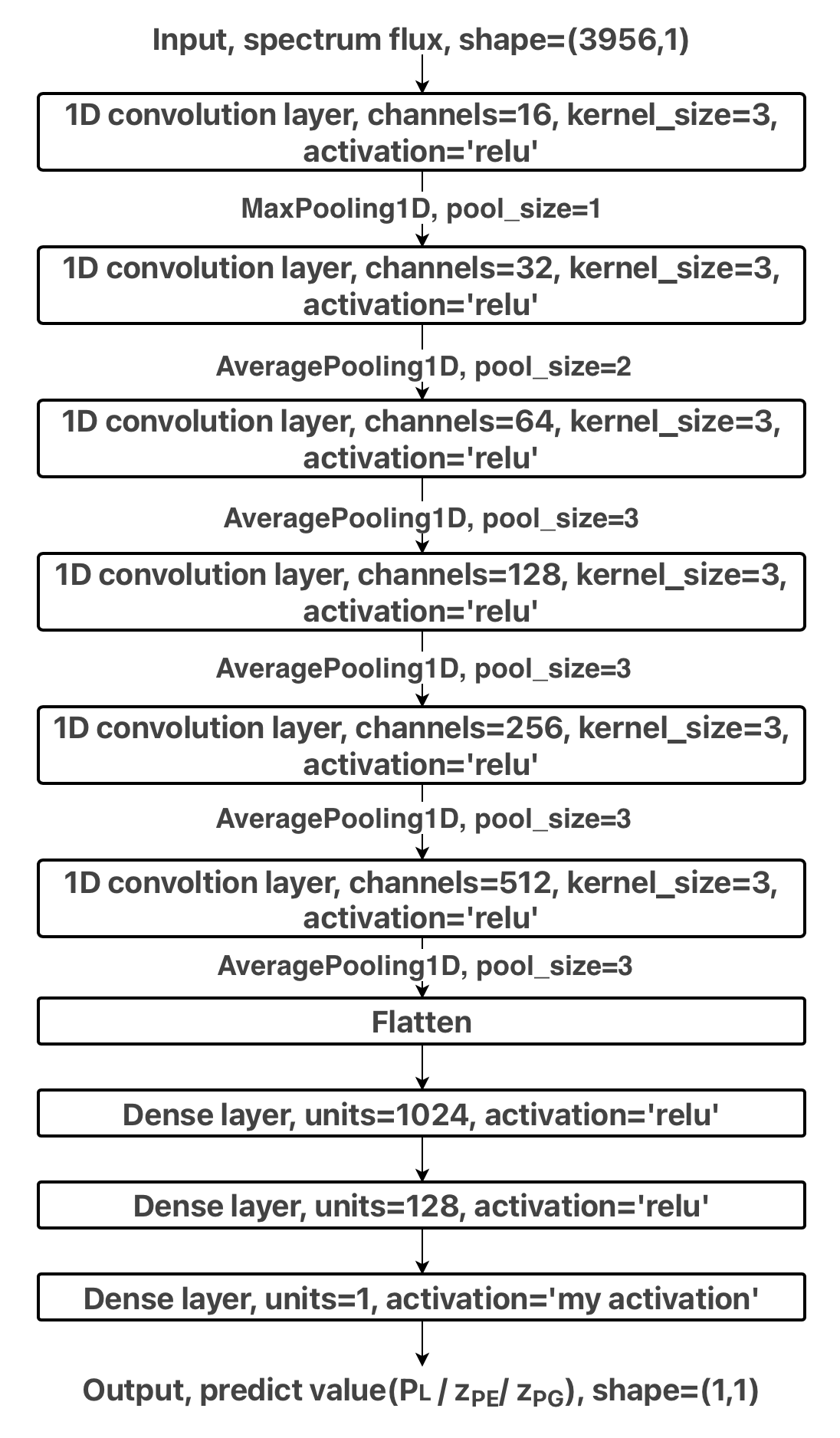}
    \caption{The CNN model adopted for the GasNet-L1, L2, and L3. The network structure is the same for the three \Gs, except \Zhong{for the} activation and loss functions as reported in Table \ref{table:my activation and loss}. 
    }
    \label{CNN_model_detail}
\end{figure}

\subsection{Convolutional Neural Networks as lens classifiers in 1D spectra}
When searching for strong lens candidates in 1-dimension (1D) spectra, one needs to 
identify two main features: 1) the potential emission lines from the background sources, to determine 
the redshift of the source, and
2) the absorption or emission lines of the foreground galaxies, to determine the redshift of the lens and compare this with \Zhong{one} of the putative \Zhong{sources} to possibly qualify the whole system as a lensing candidate. 
In most of the current and planned surveys, the redshift of the main galaxy (the lens) is a standard data product, hence this can be assumed to be a label of the\Zhong{ spectroscopic catalog}. 
This can be either used as a first guess for the lens classifier, to estimate the lens redshift itself \Zhong{or}
 kept fixed, asking the CNN to identify tentative background lines (see below).

\subsection{Galaxy Spectra convolutional neural Networks (GaSNets)}
\label{sec:galnets}
In this work, we present the first set of Galaxy Spectra convolutional neural Networks (GaSNets) for Lensing (-L). These are CNNs trained to identify strong lensing event candidates in 1D galaxy spectra. To perform this task, we have built 
three different GaSNet models: 
\begin{enumerate}
\item \Gone. This CNN is a classifier, trained to look for the presence of emission lines blended in the features of the foreground galaxy and give the probability to be a lens ($P_L$). In doing this, we do not assume any specific morphology for the lens, which can be either a standard early-type galaxy (ETG), dominated by absorption line features, or a late-type galaxy (LTG), with ongoing star-formation. The \Gone\ will learn whether, in the spectra of either \Zhong{kind}, there are higher-redshift emission lines, to finally give the $P_L$. 
\item \Gtwo. This CNN is a regression algorithm, trained to identify potential emission lines, among a list of standard features from star-forming galaxies, overlapping a foreground galaxy spectrum and predict their redshift ($z_{PE}$).
\item \Gthr. This CNN is also \Zhong{a} regression algorithm, trained to predict the redshift of the foreground galaxy ($z_{PG}$) from the combination of continuum plus a) classical absorption features from ETG spectra or b) emission lines of LTGs.
Having such an output \Zhong{will} make the overall Network general enough to be applied to spectroscopic databases, regardless these have gone \Zhong{through} a pipeline to estimate galaxy redshifts. In our analysis below, even though 
we can assume that the redshift of the lenses \Zhong{is} given (as they are provided with the BOSS spectra, see Sect. \ref{sec:data}), we opt to use the redshift predictions of our \Gthr\ for the candidate selection and use the BOSS redshifts as ground truth to assess the accuracy of the deep learning estimates.   
\end{enumerate}

The three CNN models have the same structure. They are built by 6 convolution layers and 3 total connected layers (see Fig. \ref{CNN_model_detail}), assembled by Python modules TensorFlow and Keras. 
{
In the last layer in Fig. \ref{CNN_model_detail}, due to the different \Zhong{tasks} to perform (classification vs. regression), for \Gone\ we use a ``sigmoid" activation  function (labeled as ``my activation''), while for \Gtwo\ and L3 we need no activation.}
For the same reason, we also use different loss functions.
For \Gone\ we adopt a ``binary cross-entropy" loss, which is commonly used for a binary classifier. For \Gtwo\ and \Gthr, which are two regression models, instead of the commonly used MAE and MSE loss functions, we apply the ``Huber" loss. This is defined as 
\begin{equation}
	L_{\delta}(a) =
	\begin{cases}
	    \dfrac{1}{2}(a)^2,  \ \ \ |a|\leq\delta\\
	    \delta\cdot(|a|-\dfrac{1}{2} \delta),  \ \ \ {\rm otherwise}.
    \end{cases}
\end{equation}
where $a=y_{\rm true}-y_{\rm pred}$, $y_{\rm true}$ is the real redshift, $y_{\rm pred}$ is the predicted redshift by the CNNs \Zhong{, and} $\delta$ is a parameter that can be preset ($0.001$ in this work). The choice of the ``Huber" Loss has been made because, as shown in CNN regression models for galaxy light profiles (i.e. the GaLNets, \citealt{2021arXiv211105434L}), it can achieve higher accuracy than MAE and MSE and better convergence.
Both ``activation'' and ``loss functions'' 
are summarized in Table \ref{table:my activation and loss}. 
\begin{table}
\footnotesize
\caption{CNN ``my activation'' and ``loss function''
}
    \begin{tabular}{l | c | c | c}
        \hline  
    	\hline & \Gone\ & \Gtwo\ & \Gthr\ \\
    	\hline my \ activation & sigmoid & - & - \\
    	\hline loss \ function & binary \ crossentropy & huber\_loss & huber\_loss \\
    	\hline  
    	\hline  
    \end{tabular}
    \label{table:my activation and loss}
\end{table}

\begin{figure}
    \centering 
    \includegraphics[height=10cm,width=1\linewidth]{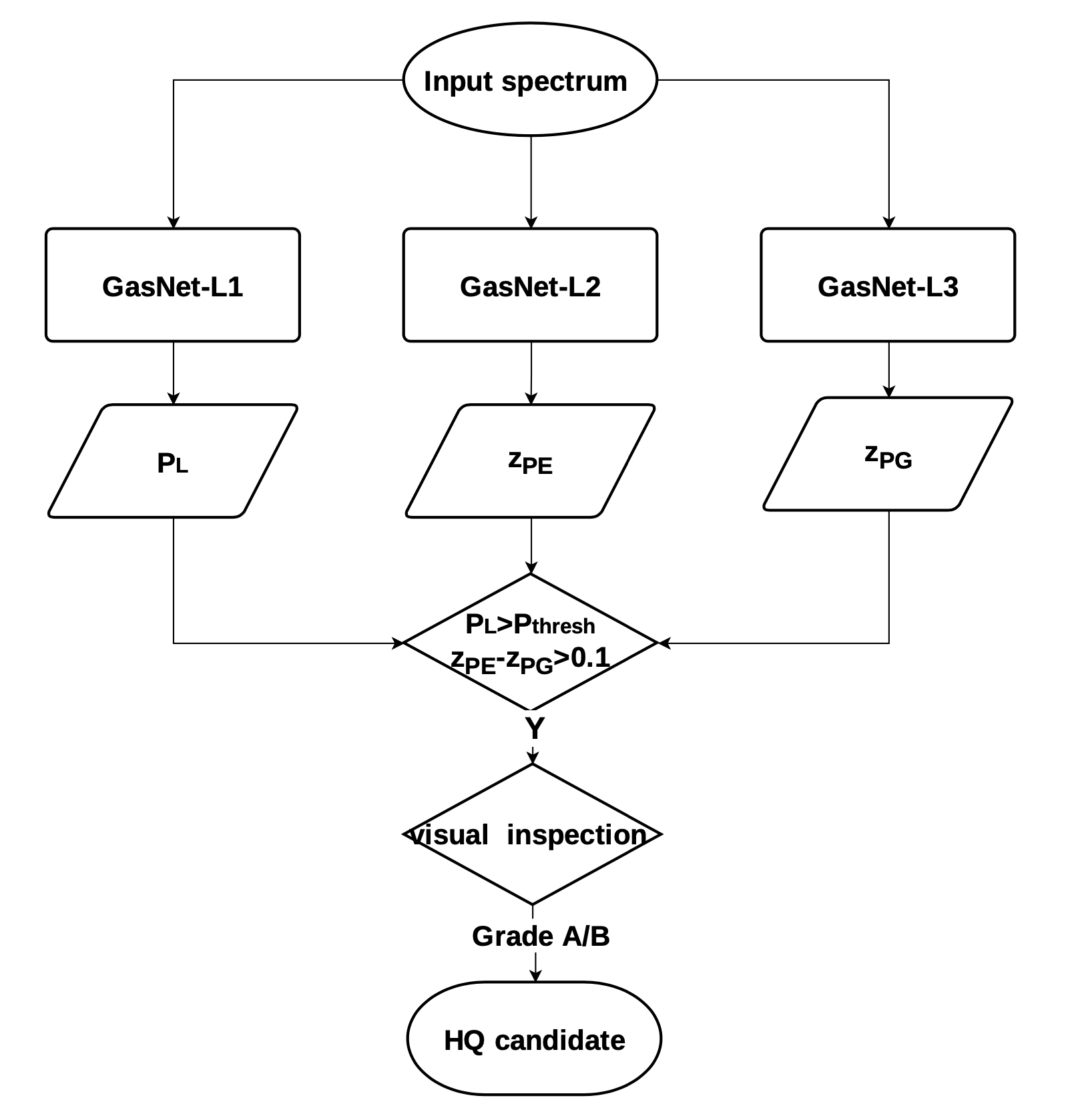}
    \caption{Flow-chart describing the process to obtain the HQ candidates combining the output of the three \Gs. The final step is the visual inspection of the candidates selected using the probability criterion, $P_L > P_{\rm thresh}$, combined with the presence of background emission lines, $z_{PE} > z_{PG} +0.1$.}
    \label{whole_CNN_model}
\end{figure}

From Fig. \ref{CNN_model_detail} we see that the CNNs all accept a 1D spectrum (i.e. a vector of wavelength and fluxes) as input and produce as predicted parameters, either a probability ($P_L$ for \Gone) or a redshift (i.e. $z_{PE}$ and $z_{PG}$ for \Gtwo\ and \Gthr, respectively). 

\subsection{Decomposing a complex CNN model}
\label{sec:dcompose of CNN}
To conclude this section, we briefly discuss the choice 
to combine the outcome of three CNNs to improve the accuracy of the identification of high-quality (HQ) candidates and minimize the chance of false detection. 

This task involves two steps: 1) the identification of different kinds of features that can suggest the presence of a lensing event, i.e. the coexistence of absorption and emissions lines from different objects along the line-of-sight, and 2) the verification that 
(some of) the emission lines come from the background system. This is a complex classification task
that can be more efficiently performed by 
combining different CNNs with different specializations.
Indeed, \Gone\ is designed to identify 
a specific series of emission lines at a higher redshift overlying 
a lower redshift spectrum \Zhong{characterized} either by a continuum plus absorption lines typical of ETGs or continuum plus emission lines from LTGs. Even if the training sample is made of real galaxy spectra, where the simulated emission lines from mock background sources are randomly 
redshifted 
with respect to the main galaxy (see \S\ref{sec: training data}), \Gone\ 
can only give a probability of the coexistence of a lens and a source at different redshifts, but cannot predict by how much the emissions of the source are misplaced. Since this process can be uncertain, we cannot exclude that 
\Gone\ can confuse a lensing event with other ``local'' emission processes (e.g. active galactic nuclei, 
ongoing star-formation, gas outflows,
etc.), and \Zhong{vice versa}. On the other hand, the \Gtwo\ and \Gthr\ are able to predict the redshift of the tentative source and lens, independently, 
meaning that they cannot predict, individually, if there is another object at a different redshift, compatibly with a lensing event.

Only using the outputs of these three GaSNets together, we can both give a ``high probability'' that there are two different systems contributing to the spectrum and establish that the closer one is a galaxy, with redshift $z_{PG}$, and the background one is a fainter line emitter, with redshift $z_{PE}$. 
In particular, to qualify a spectrum as a candidate, we use the following conditions: 1) $P_L > P_{\rm thresh}$, and 2) $z_{PE} > z_{PG}$, where $P_{\rm thresh}$ is an appropriate lower probability threshold that will be chosen later to define the high-probability candidates that will be further visually investigated to assemble the list of candidates to pass to the visual inspection, which finally produces the HQ candidate list. 
The full process for the selection and grading of the HQ candidates is schematized in Fig. \ref{whole_CNN_model}.

\begin{figure}

\hspace{-0.5cm}
            \centering
            \includegraphics[width=1\linewidth]{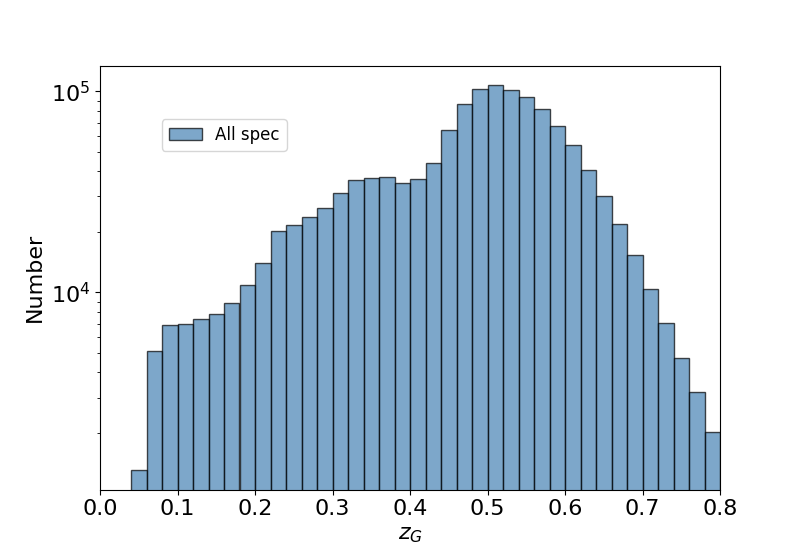}
        \includegraphics[width=1\linewidth]{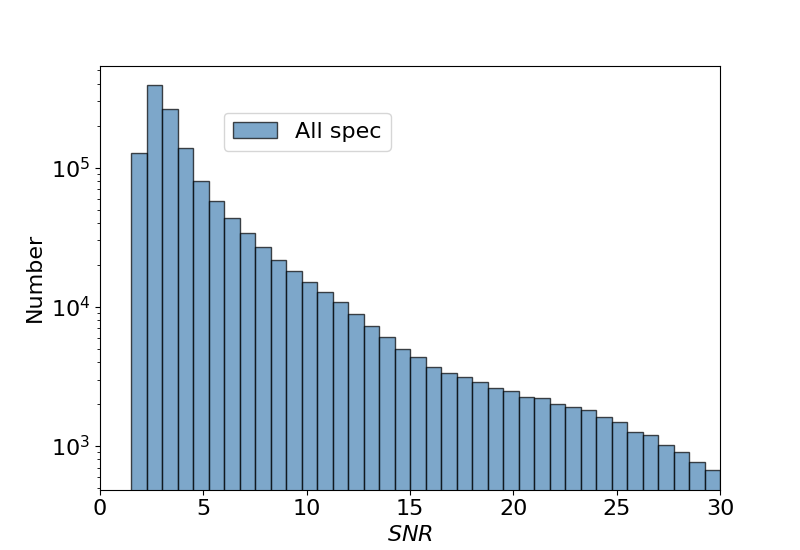}
    
    \caption{The redshift and SNR distribution of the DR16-predictive sample (see text for details).
    }    
    \label{fig:z_dis_and_SNR}   
\end{figure}

\section{ \Zhong {Data} }
\label{sec: training data}
The construction of the training sample is a critical step of any supervised ML algorithm. Indeed, to avoid biased predictions and fictitious performances, the training samples need to be
as close as possible to real observations.
In our \Zhong{case,} we build our training sample starting from real spectra from BOSS, over which we simulate the presence of emission lines. 
Here below, we first introduce the dataset we use for our analysis. Then, we describe the way we have constructed the training set.
This is  
constituted by two samples. 

First, the {\it negative} sample, which represents a catalog of galaxy spectra with no background sources \Zhong{blended in}. As mentioned earlier, we do not make any selection of galaxy types and we include ETGs and LTGs. 

Second, the so-called {\it positive} sample, \Zhong{which represents} a simulated sample of spectra that emulates the presence of emission lines from a background source. This is made of the same galaxy spectra of the negative sample, but with the addition of artificial emission lines, redshifted with respect to the ``foreground'' galaxies.

\subsection{Data selection and predictive sample}
\label{sec:data}
The Sloan Digital Sky Survey (SDSS, see \citealt{2000AJ....120.1579Y}) 
has observed over 10\,000 deg$^2$ of the sky, performing multi-band photometry and spectroscopy \citep{1999cs........7009S}. 
In 2009, before the start of the Baryon Oscillation Spectroscopic Survey \citep[BOSS,][]{2009astro2010S.314S},
in the third stage of the project (SDSS-III), 
the spectrograph operating the observations has been upgraded. 
Compared to SDSS-I/II, the number of fibers was increased from 640 to 1000, and the fiber diameter has been reduced from 3$''$ to 2$''$
\citep{2012ApJS..203...21A}.  
The extended version of the BOSS survey, eBOSS \citep{2016AJ....151...44D}, has overall
produced spectra for around 2.6 million galaxies, in the wavelength range 361–1014 nm. These are 
publicly available through the latest data release 16 (DR16, \citealt{2020ApJS..249....3A}). This is the dataset we use in this work\footnote{For convenience we will address this as eBOSS or DR16.}, over which we operate a series of selections to ensure the quality of the spectra to \Zhong{analyze}.
In particular, we select only: 1) plates labeled as ``good'' quality, 2) ``Object'' flags labeled as ``galaxy'', 3) spectroscopic redshift between $0.05-0.8$, 4) spectra with SNR$>2$, 5) wavelength range 3700–9200\AA.

\begin{table}
    \caption{Model parameters of equation \ref{eq:line flux function}.} 

    \centering
    \begin{tabular}{l l c l l} 
    \hline
    	\hline & $\lambda_{e,1}$ & $\lambda_{e,2}$ & $h_e$ & $z_{max}$ \\
    	\hline $[OII]^1$ & 3726.2 & 3728.9 & [2,10] & 1.44 \\
    	\hline [OIII] & 4959.0 & n/a & [1,5] & 0.82 \\
    	\hline [OIII] & 5007.0 & n/a &  [1,15] & 0.82 \\
    	\hline $H_\alpha$ & 6562.8 & n/a & [1,15] & 0.39 \\
    	\hline $H_\gamma$ & 4340.5 & n/a & [1,5] & 1.10 \\
    	\hline $H_\beta$ & 4861.3 & n/a & [1,5] & 0.87 \\
    	\hline 
    	\hline
    \end{tabular}
    \label{table:model parameters}
\end{table}

This latter criterion is applied to avoid a rather noisy region of the spectra, at $\lambda>9200$\AA, where the residuals from the telluric line subtraction might be a source of spurious detections. This is a problem we expect to deal \Zhong{with in} future developments, but we wanted to avoid \Zhong{in} this first test. We stress, though, that the reduced wavelength range will allow us to train tools that can be straightforwardly applicable to the SDSS-I/II spectra, whose wavelength range is also \Zhong{limited to} 3700-9200\AA. Of course, this makes our tools less sensitive to higher-$z$ systems, as many of the emission lines we want to detect will fall out of the range at redshift \Zhong{z\gsim1.4} (see below). 

Criterion 3 is dictated by the line observability. Indeed, we will assume that typical sources in the SGL events are star-forming galaxies characterized by emission lines as reported in Table \ref{table:model parameters}. Here, for each line, we list the central wavelength(s), the maximum redshift the emission line can reach below 9200\AA,  $z_{\rm max}$, 
and an intensity parameter, $h_e$, that will be used in \S\ref{sec: artificial emission model} for the simulated spectra.
According to this list, for redshift  $z_E\gsim1.4$, all lines would fall out of the eBOSS wavelength range, while with $z_E\lsim1.2$, we can still retain two emission lines, i.e. [OII] and $H_\gamma$. In order to select lenses that are compatible with the visibility of the background lines and with a reasonable lens-source distance to guarantee \Zhong{an} SGL event, we collect spectra in the range $z_G=(0.05 \sim 0.8)$.

Criterion 4, on the other hand, 
is an optimistic lower limit
we have chosen to increase the completeness. We have considered that the emission lines from background sources have \Zhong{a} SNR which is not necessarily correlated to the SNR of the whole \Zhong{spectrum} and, thus, can be seen also in \Zhong{a} noisy galaxy \Zhong{spectrum}.

The final selected sample consists of
$1\,339\,895$ spectra: in the following, we will refer to this as the {\it DR16-predictive} sample. 
In Fig. \ref{fig:z_dis_and_SNR} we show the SNR distribution of the selected spectra 
and the redshift of the central galaxy. 

\subsection{Construction of the negative sample}
\label{sec:neg sample}
The first step to produce our training dataset is the selection of the negative sample. This is chosen \Zhong{to} make the CNN as general as possible, hence assuming that every type of \Zhong{galaxy} can work as a lens, with no particular restrictions in luminosity or color, \Zhong{as} it is typically done to contain the predictive samples in imaging classifiers (see e.g. Petrillo et al. 2019, Li et al. 2020). 

For this purpose, we select 
140\,000 galaxies spectra from the DR16-predictive sample, 
with a wavelength range of 3700-9200 \AA.
We take particular care that the selected spectra uniformly cover, in number, the full $z_G$ range, by counting the spectra in redshift bins of 0.05. This is crucial to avoid any bias in the prediction of the $z_G$ from the poor sampling of one redshift bin with respect to the close ones.

To mimic the presence of emission lines from local processes, for 1/5 of the negative sample, we add artificial emission lines with the same redshift of the galaxy, while the remaining 4/5 of the negative sample is left unchanged. 
In particular, for this simulated ``local emissions'', we use 
the same lines, reported in Table \ref{table:model parameters}, that will be used to simulate the background source emissions, which we expect the \Gs\ to distinguish from the local ones (if any).

\begin{figure}
    {\begin{minipage}[t]{0.45\linewidth}
            \centering
            \includegraphics[height=4cm,width=1\linewidth]{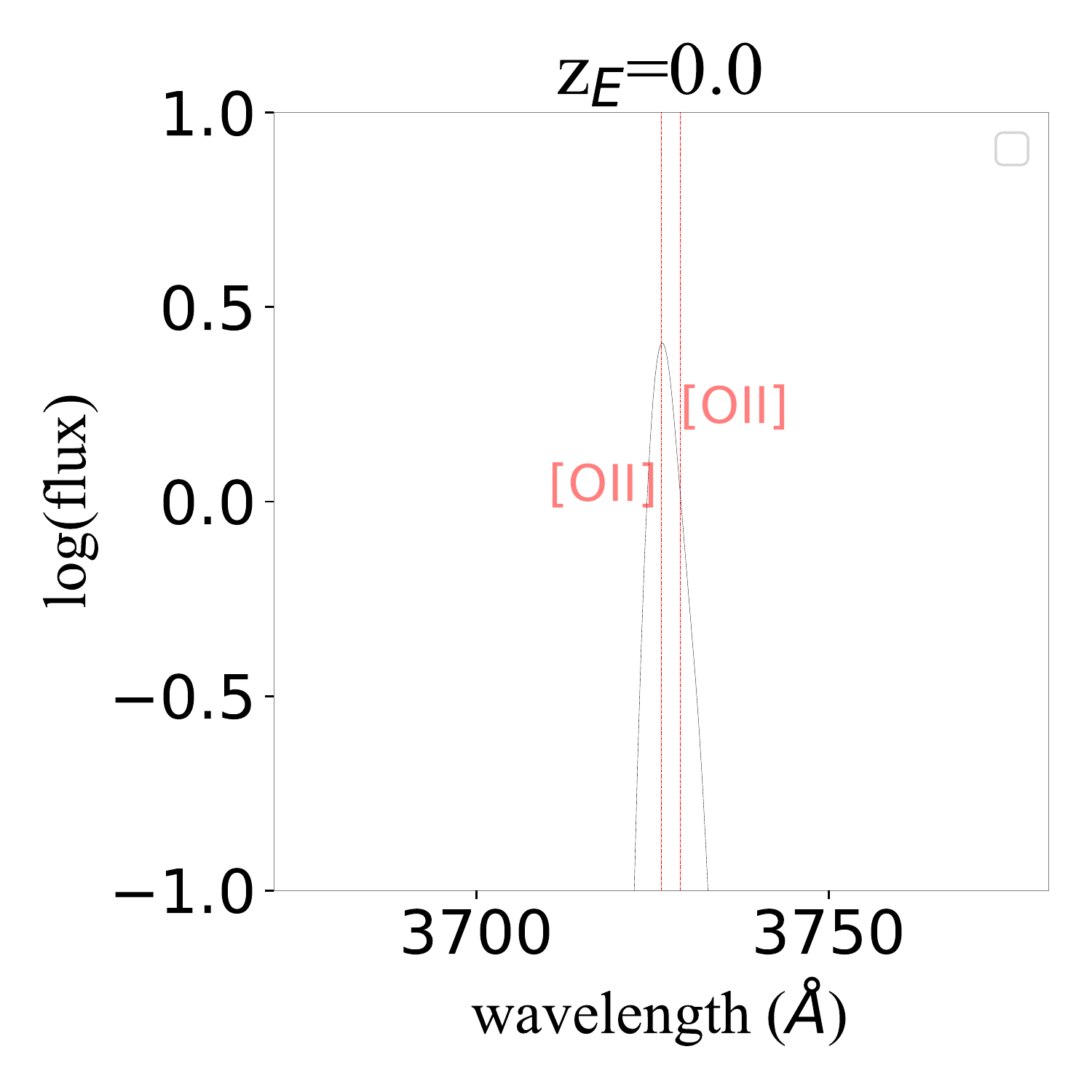}
    \end{minipage}}
    {\begin{minipage}[t]{0.45\linewidth}
        \centering
        \includegraphics[height=4cm,width=1\linewidth]{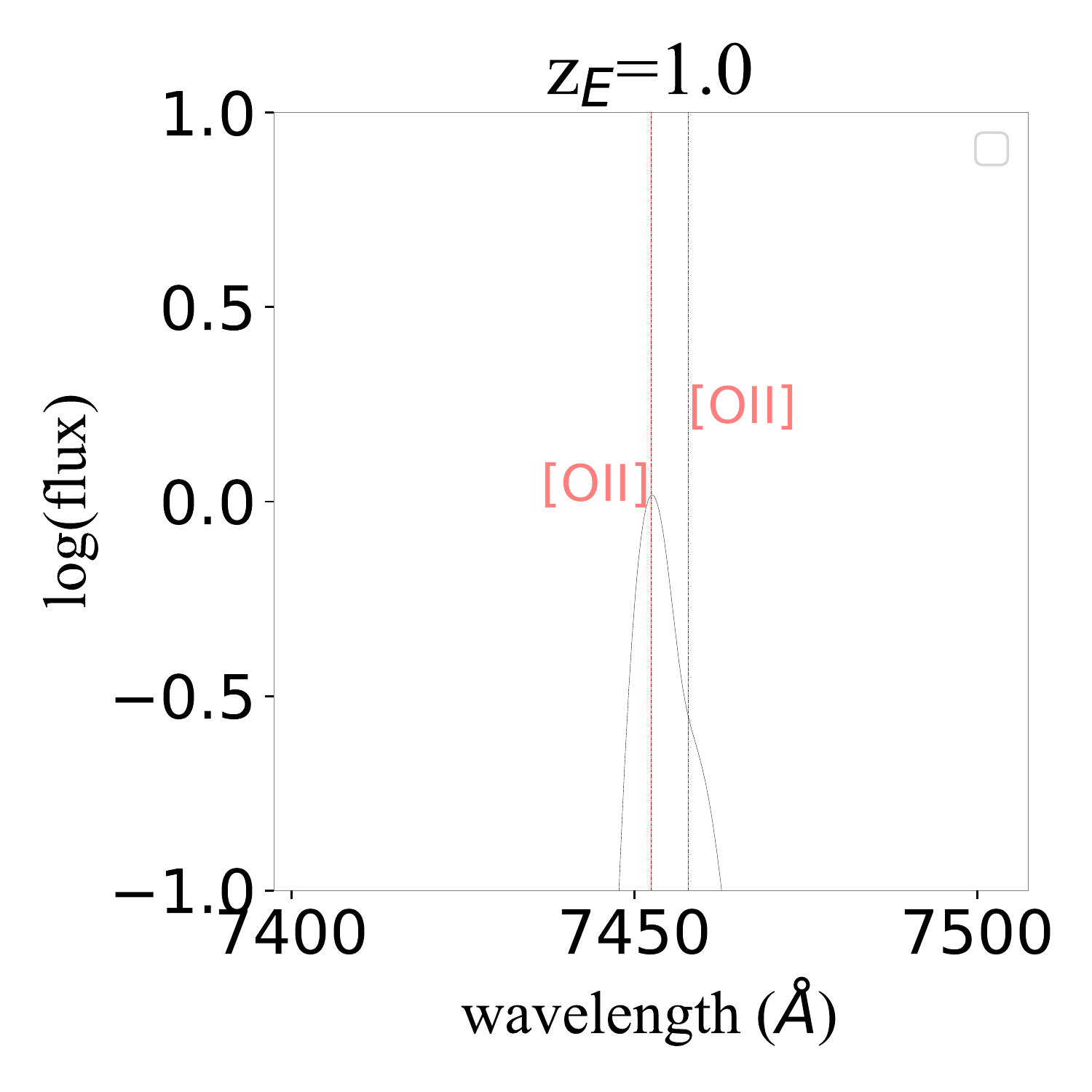}
    \end{minipage}}
    
    {\begin{minipage}[t]{0.45\linewidth}
            \centering
            \includegraphics[height=4cm,width=1\linewidth]{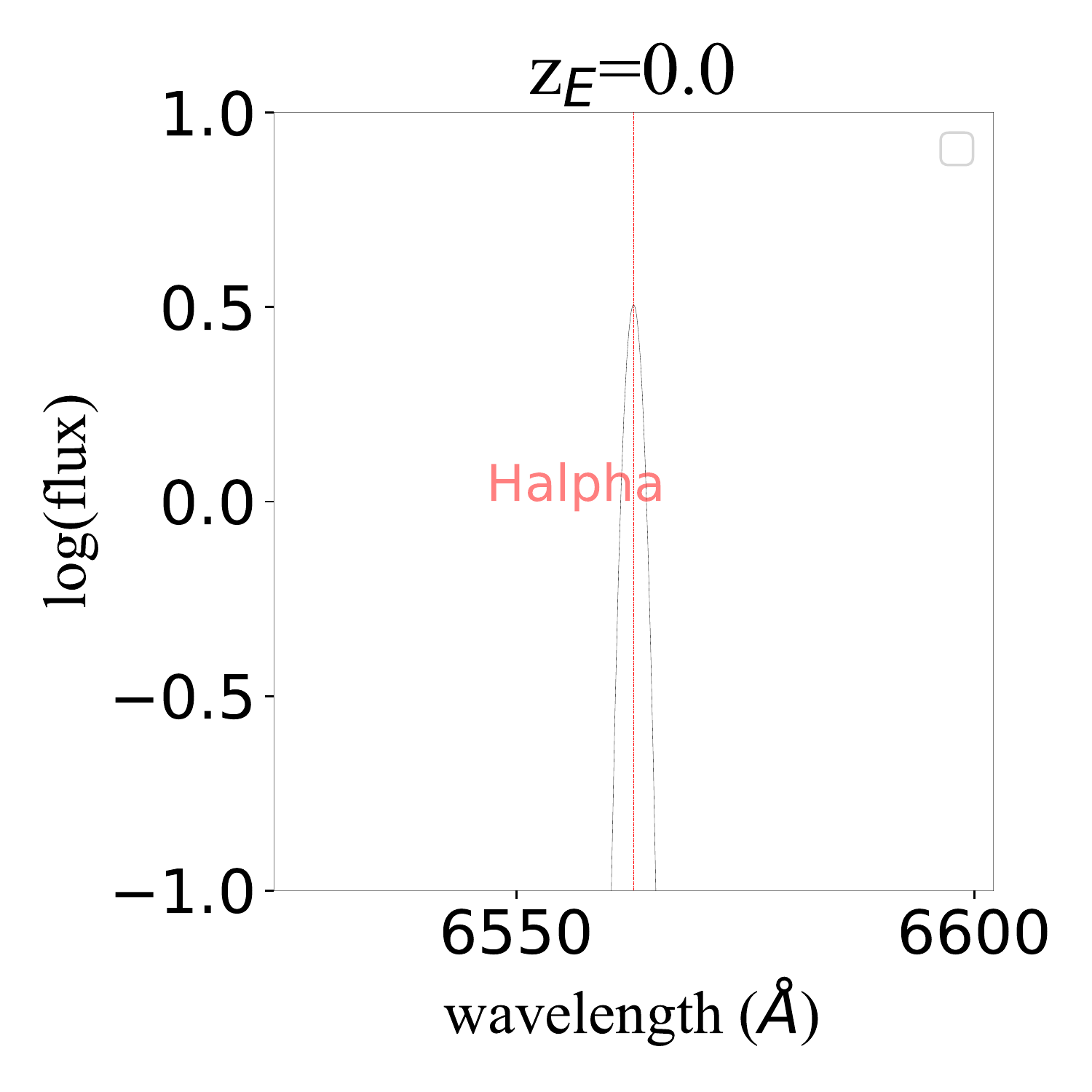}
    \end{minipage}}
    {\begin{minipage}[t]{0.45\linewidth}
        \centering
        \includegraphics[height=4cm,width=1\linewidth]{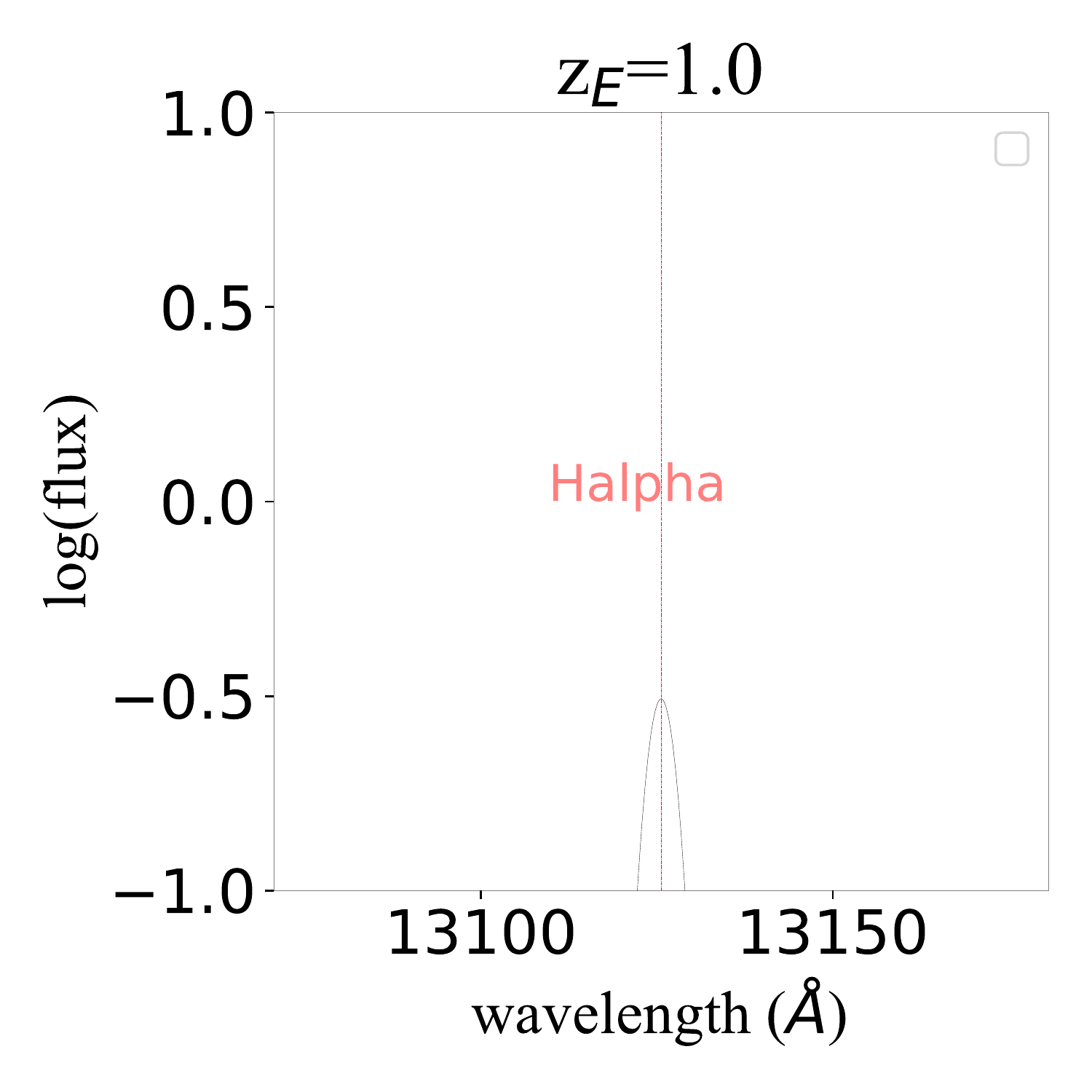}
    \end{minipage}}
    
    \caption{Detail of the line profiles of the simulated emissions. Artificial Oxygen, OII (3727.1\AA\ and 3729.9\AA) lines (top) and Halpha (6562.8\AA) lines (bottom), as observed at different redshifts. The line fluxes are redshifted and dimmed according to Eqs. 4-6. }
    \label{fig:Oxgen}
\end{figure}

\begin{figure}
    {\begin{minipage}[t]{1.0\linewidth}
            \hspace{-0.5cm}
           \includegraphics[width=1.08\textwidth]{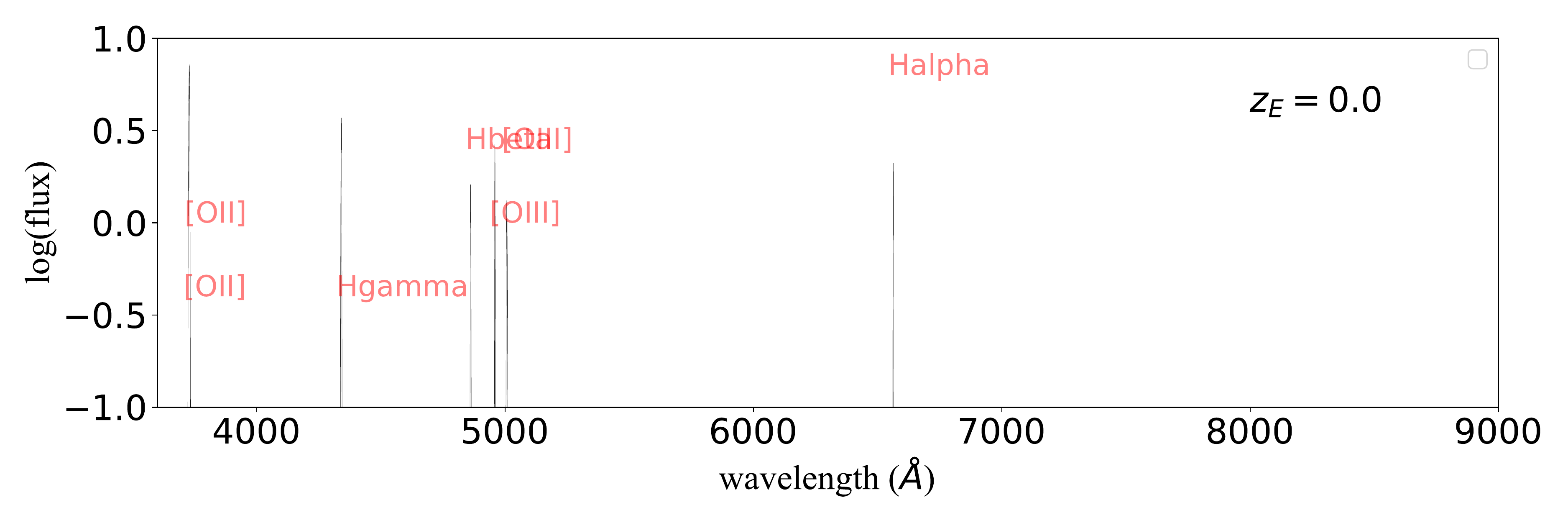}
    \end{minipage}}
    {\begin{minipage}[t]{1.0\linewidth}
            \hspace{-0.5cm}
        \includegraphics[width=1.08\textwidth]{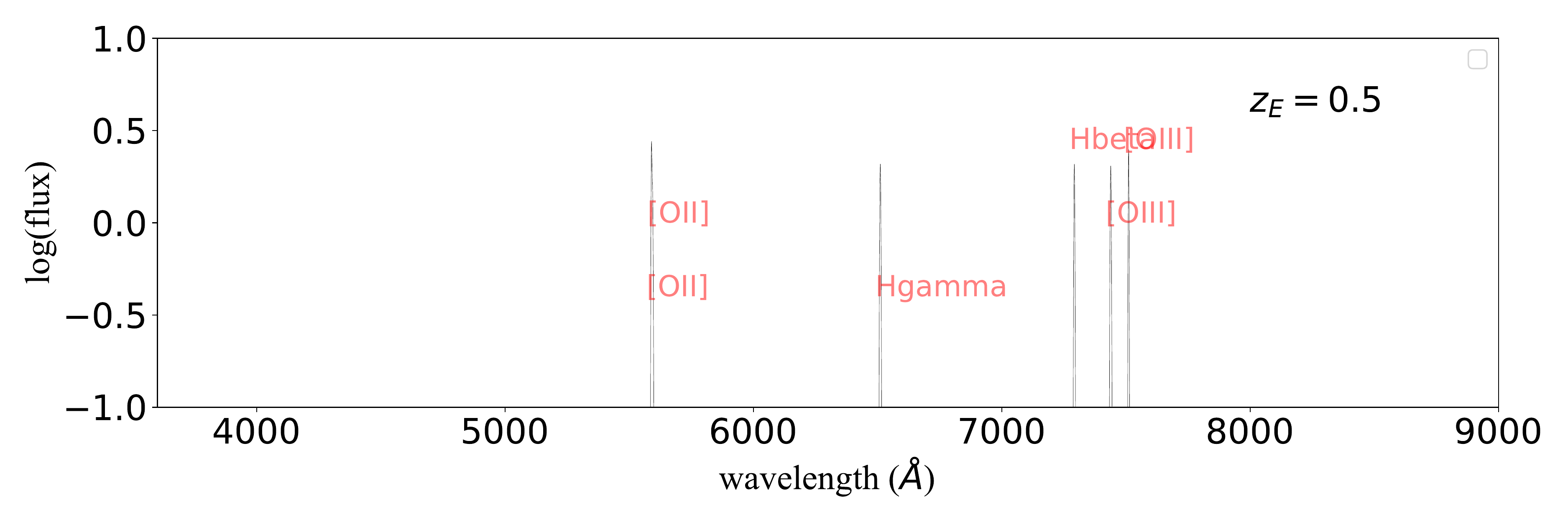}
    \end{minipage}}
    {\begin{minipage}[t]{1.0\linewidth}
            \hspace{-0.5cm}
        \includegraphics[width=1.08\textwidth]{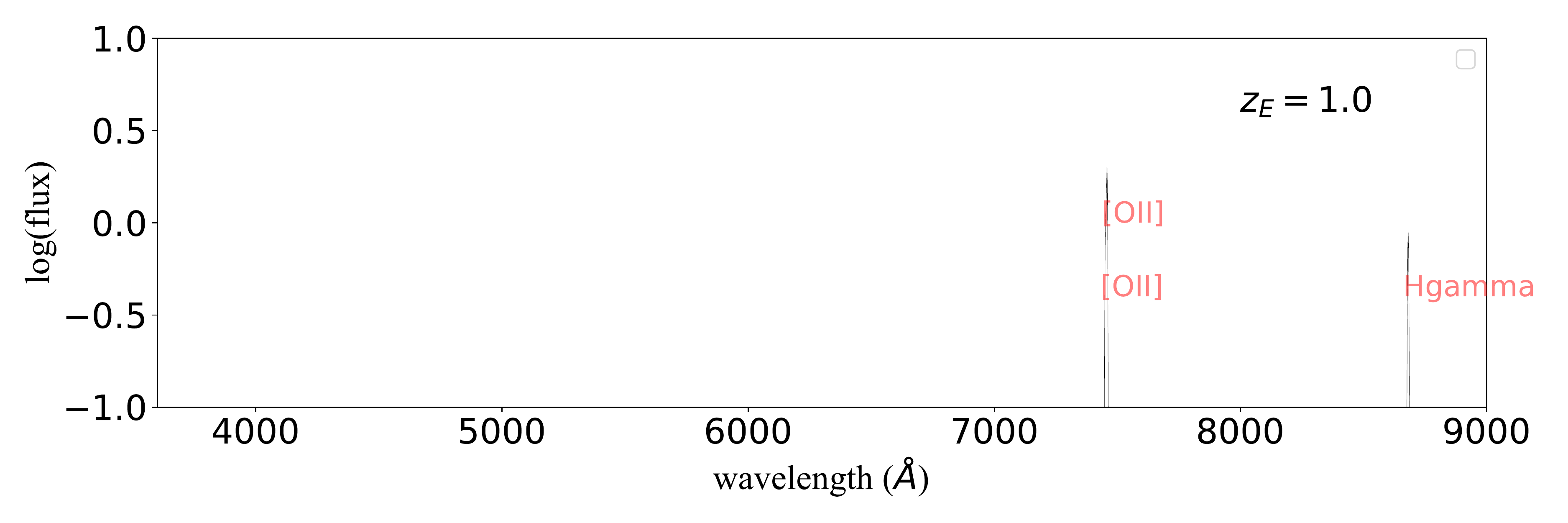}
    \end{minipage}}
    \caption{Full spectrum of artificial emission lines added to the negative sample (see Table \ref{table:model parameters}  and text for details). The line fluxes are redshifted and dimmed according to Eqs. 4-6.}
    \label{fig:all emi}
\end{figure}

\subsection{Artificial emission line model}
\label{sec: artificial emission model}
In this section, we give more details about the  
artificial emission lines we want to add to the original eBOSS spectra to emulate both some local and higher-$z$ emissions in the negative and positive samples, to be used for the CNN training.
Following Li+19, we use a 1-dimensional ``double gaussian'' profile,
defined as:
\begin{equation}\label{eq:line flux function}
F(\lambda) = h_1 \exp\{-\frac{(\lambda - \lambda_{e,1})^2}{2 \sigma_1 ^2}\} + h_2 \exp\{-\frac{(\lambda - \lambda_{e,2})^2}{2 \sigma_2 ^2}\}
\end{equation}
where $F$ is the flux,
$\lambda_{e,1}$ and $\lambda_{e,2}$ are the central 
wavelengths of the emission lines\Zhong{, and}
$h_1$, $h_2$, $\sigma_1$, $\sigma_2$ are the four model parameters. The $\lambda_{e,1}$, $\lambda_{e,2}$, $h_1$, and $h_2$ are listed in Table \ref{table:model parameters}.
These parameters are further defined to satisfy the following conditions:
\begin{equation}
\label{eq:line condition}
    \begin{cases}
     \sigma_1 \in [0.8,1.6]  \\
     \sigma_2 \in \sigma_1 + [0.5 \sigma_1 ,1 \sigma_1 ]  \\
     h_1 =  \frac{h_e}{(1+z_E)^2} \\
     h_2 =  \frac{h_e}{4(1+z_E)^2} \rm ~if ~ \lambda_{e,2}\neq n/a; & =0~ \rm otherwise  \\
    \end{cases}
\end{equation}
where the $\sigma_1$ is uniformly selected in the interval $[0.8,1.6]$,
the amplitude parameter, $h_e$, is given in Table \ref{table:model parameters}, and $z_E$ is the redshift of \Zhong{the emission} line we want to simulate, assumed to be uniform in the range [$z_G+0.1$, 1.2]\footnote{As we will discuss in \S\ref{sec:build positive sample}, since $z_G$ is assumed to be also uniform for the positive sample, this condition produces a final $z_E$ distribution which is pseudo log-normal with a cut at 1.2. For the negative sample, discussed in \S\ref{sec:neg sample}, this condition produces a $z_E$ distribution that follows the distribution of the $z_G$ as in Fig. \ref{fig:z_dis_and_SNR}.}

The $\sigma_1$ range is determined under the assumption
that the emission lines from sources are enlarged by rotation. Hence, the line broadening in wavelength can be written as $\Delta \lambda \approx 2 \lambda_0 v_r/c $, where $\lambda_0$ is the central wavelength of the emission line, $v_r$ is the max velocity along the line-of-sight, and $c$ is the speed of light. 
Then, we can approximate $2 \sigma_1 \approx \Delta \lambda $, which, for a rotation $v_r \sim 100$ km/s and $\lambda_0 \approx 370$nm, gives $\sigma_1 \approx 1.2$\AA. Taking into account a larger wavelength range and rotation spectrum, we can reasonably make $\sigma_1$ \Zhong{vary} over a further $\pm0.4$\AA\ range, i.e. the one we have assumed in Eqs. \ref{eq:line condition}. 
Finally, we remark that the absolute amplitude of the emission lines, depending on $h_e$, is not of major importance, as the final SNR of the line strongly depends on the continuum of the spectrum the lines are added to. On the other hand, two other important features are 1) the relative distance of the line central wavelengths ($\lambda_{e,1}$ and $\lambda_{e,2}$) and 2) their relative full width half maximum (FWHM), connected to the $\sigma_1$ and $\sigma_2$ parameters. 

 \begin{figure}
    \centering 
    {\begin{minipage}[t]{1.0\linewidth}
            \centering
            \includegraphics[width=1\textwidth]{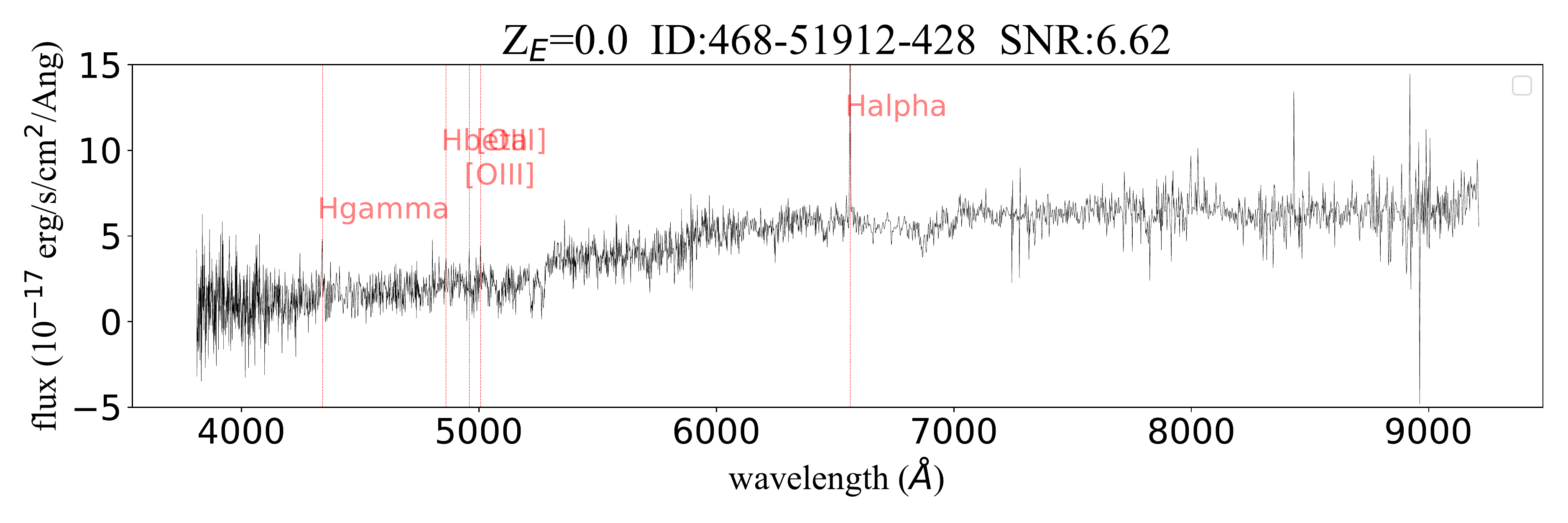}
    \end{minipage}}
    {\begin{minipage}[t]{1.0\linewidth}
        \centering
        \includegraphics[width=1\textwidth]{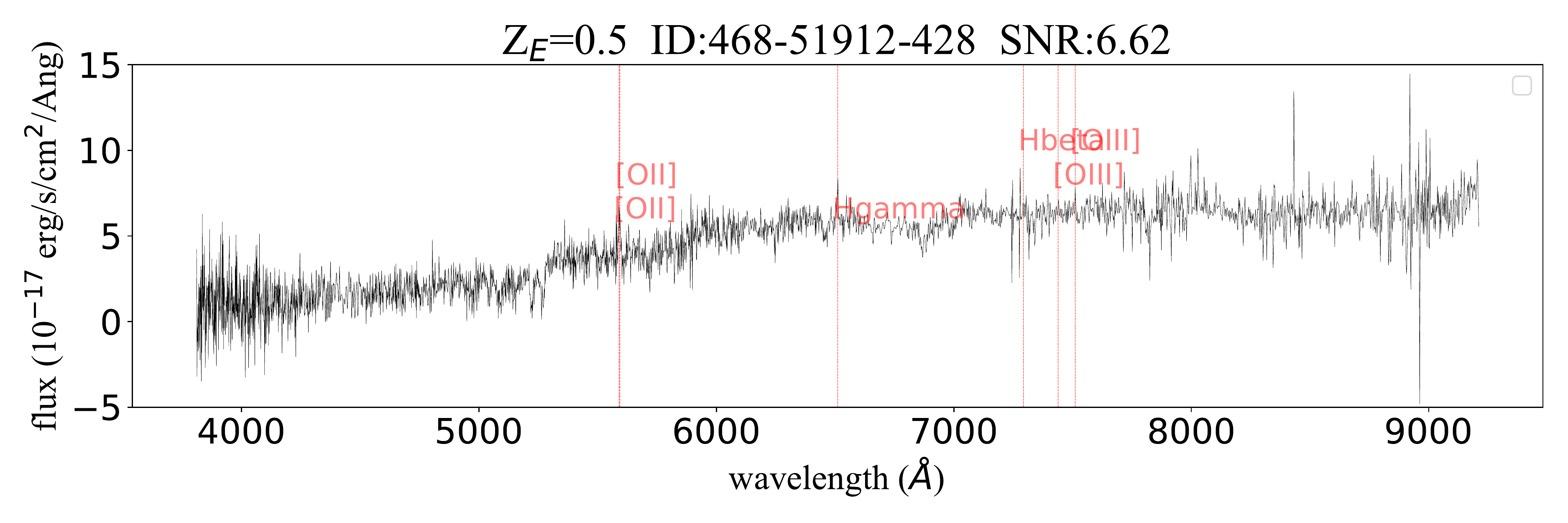}
    \end{minipage}}
    {\begin{minipage}[t]{1.0\linewidth}
        \centering
        \includegraphics[width=1\textwidth]{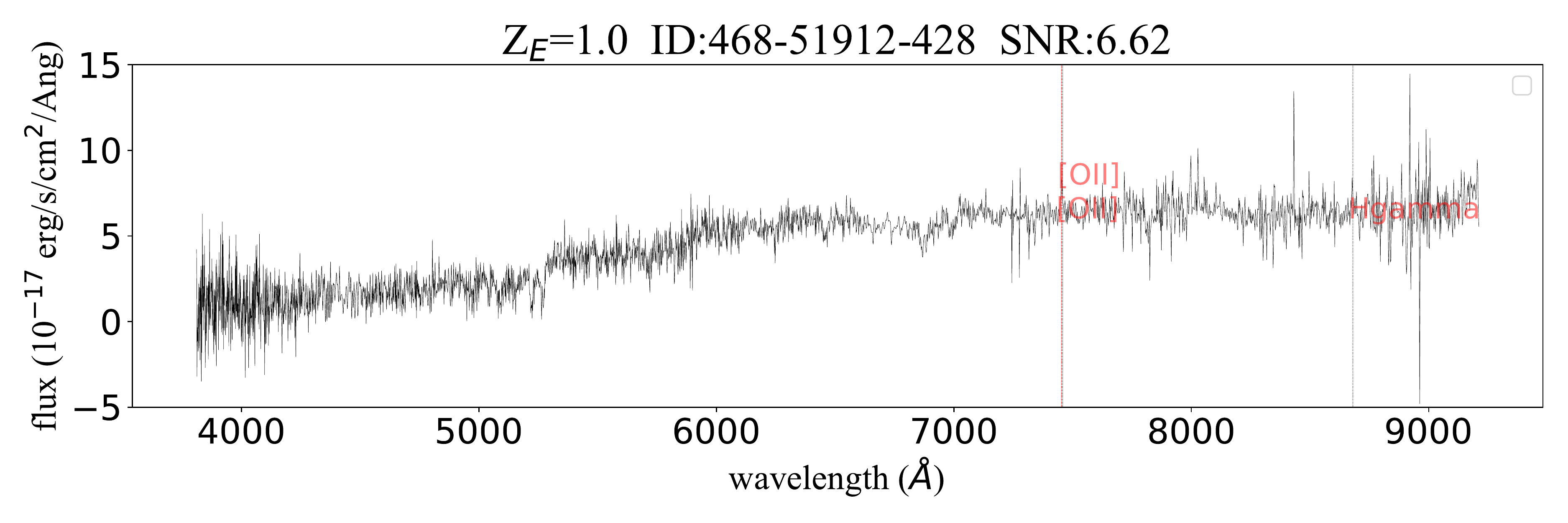}
    \end{minipage}}
    \caption{The spectra after \Zhong{adding} artificial emission \Zhong{lines}. We add the mock emission lines in several different redshift \Zhong{positions} ($z_E$) in \Zhong{the real}  spectrum, in order to simulate the positive sample.}
    \label{fig:simulate galaxy spec}     
\end{figure}

Simulated lines are first randomly generated at $z_E$=0 and then randomly redshifted to $z_E>z_G+0.1$, where $z_G$ is the redshift of the negative spectrum from which the positive is generated (see \S\ref{sec:build positive sample}). 

The flux at the redshift $z_E$ is then defined according to the standard equation:
\begin{equation}
    F_{z_E}(\lambda) = F\left(\frac{\lambda}{1+z_E}\right)
\end{equation}

where the function $F$ is the rest frame emission line flux function (Eq. \ref{eq:line flux function}).

The central wavelength of $F_{Z}$, $\lambda_{cz}$, is defined as
\begin{equation}
\frac{\lambda_{cz}}{1+z_E} = \lambda_{c0} \ \to \ \lambda_{cz} = (1+z_E)\lambda_{c0}.
\end{equation}

The interval of $\lambda$ is equal
\begin{equation}
\frac{\mathrm{d}\lambda'}{1+z_E} =  \mathrm{d}\lambda  \ \to \ \mathrm{d}\lambda' = (1+z_E)\mathrm{d}\lambda
\end{equation}

where $\lambda_{c0}$ is the central wavelength of the rest frame. According to the equations above, $\lambda_{c0}$ shifted to $(1+z_E)\lambda_{c0}$, and the interval of $\lambda$ in \Zhong{the rest} frame will broaden to $(1+z_E)\mathrm{d}\lambda$.

Figs. \ref{fig:Oxgen}
and \ref{fig:all emi} show how typical simulated emission lines from Table \ref{table:model parameters} are simulated according to the random parameters from Eqs. \ref{eq:line condition} and shifted to 0.5 and 1 redshifts.

\subsection{Simulating the positive sample}
\label{sec:build positive sample}
\Zhong{The next} step is to build a positive sample by adding simulated emission lines to the negative sample.
As anticipated, we use the same lines as in Table \ref{table:model parameters}, this time with lines redshifted to $z_E>z_G+0.1$, with the condition that $z_E\lsim 1.2$.

\begin{figure}  
    \centering 
    {\begin{minipage}[t]{1\linewidth}
            \centering
            \includegraphics[width=1.01
            \textwidth]{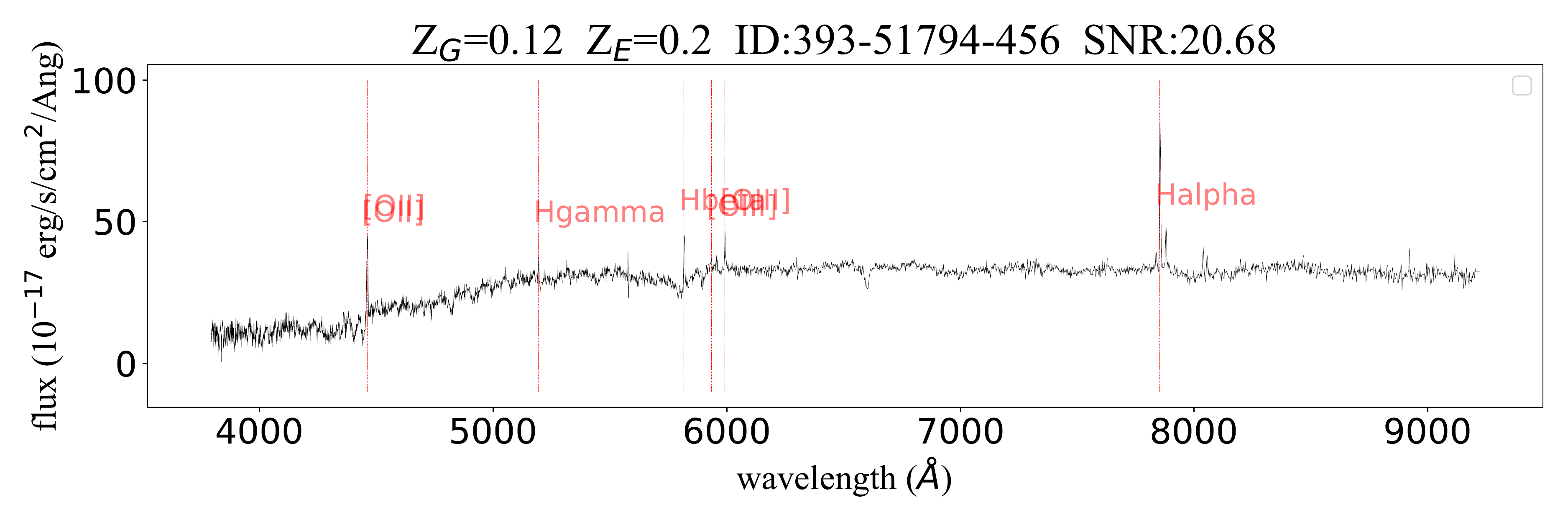}
    \end{minipage}}
    {\begin{minipage}[t]{1\linewidth}
        \centering
        \includegraphics[width=1.01\textwidth]{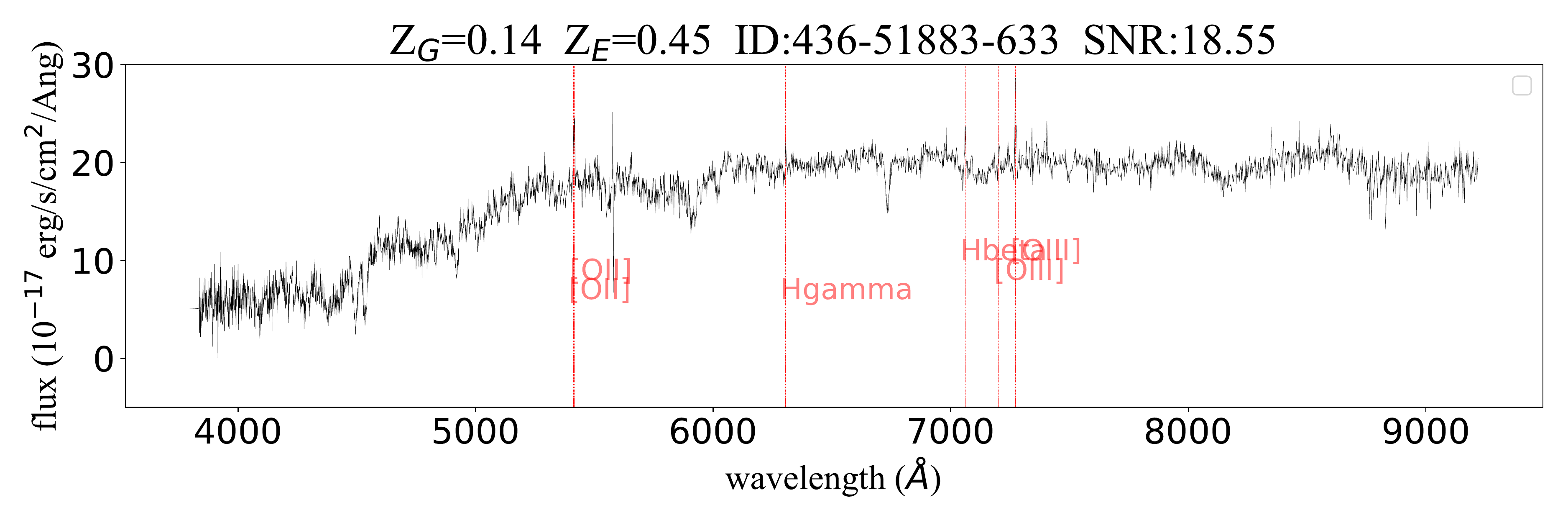}
    \end{minipage}}
    {\begin{minipage}[t]{1.01\linewidth}
        \centering
        \includegraphics[width=1\textwidth]{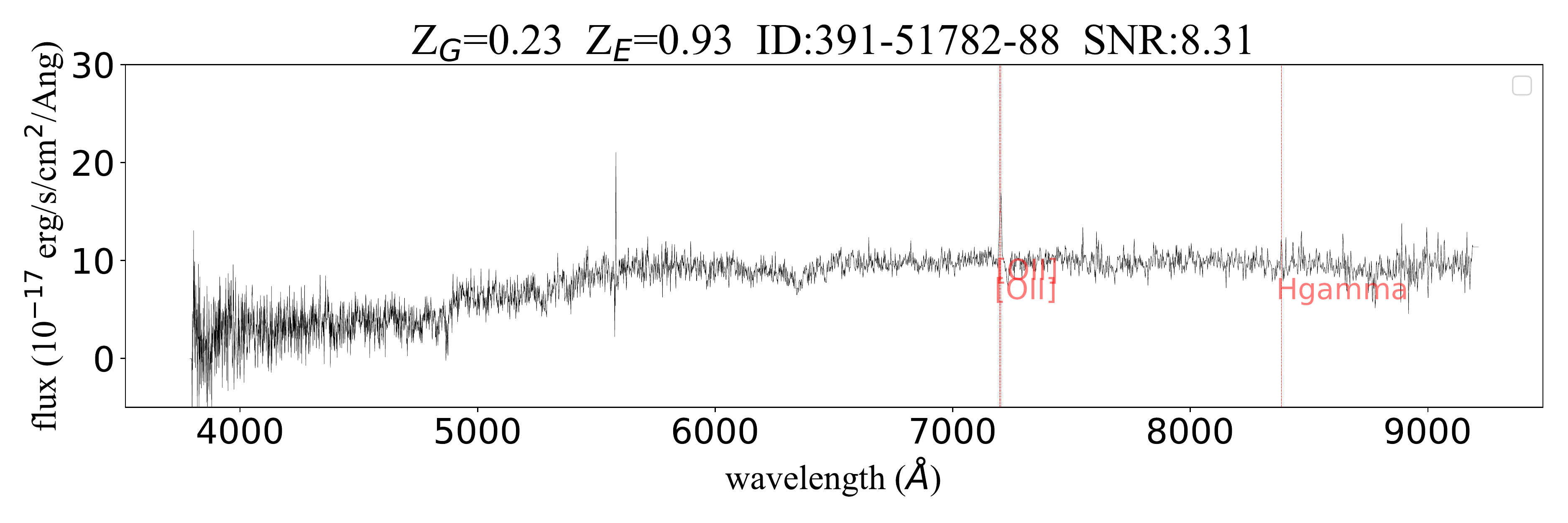}
    \end{minipage}}
    \caption{Spectra of confirmed gravitational lenses in \citet{2008ApJ...682..964B}, \Zhong{the} red line is the location of the identified emission lines.}
    \label{fig:lens spec}
\end{figure}

In Fig. \ref{fig:simulate galaxy spec} we show three simulated positive spectra
for a single moderate SNR ($\sim 7$) negative spectrum (SDSS-468-51912). Here, we 
have marked the location of the simulated emission lines at different redshifts, on top of the continuum of the real eBOSS galaxy spectrum. 
Looking at these spectra, one can visually figure \Zhong{out what are the} major challenges to \Zhong{identify} the ``ground truth'' emission lines in them. First, the SNR of the lines, as this depends not only on the $h_1$ and $h_2$ but also on the intrinsic spectrum noise. 
Second, the contamination from residual sky lines, e.g. at $\lambda>8000$ \AA. Third, \Zhong{the} effect of the source redshift, which can shift most of the relevant emission lines from Table \ref{table:model parameters} out of the spectral range (at $\lambda  > 9200$\AA, e.g. in Fig. \ref{fig:simulate galaxy spec}-c). This latter issue could be in principle solved by including more emission lines in our reference catalog. 
We will consider this option for the next developments of \Gs. However, we stress here that adding more lines, which in \Zhong{most of} the cases have much lower SNRs in real galaxies, might introduce more uncertainties in the predictions of the \Gtwo, as they might be easier confused with random noise, especially in low-SNR spectra. 


As a comparison with real lensing events, in Fig. \ref{fig:lens spec} we report some spectra of confirmed lenses from \citet[][see also \S\ref{sec:confirmed_lens}]{2008ApJ...682..964B}. Here
the locations of the emission lines of the background sources are marked, again, as red vertical lines. 
In particular, in this \Zhong{figure, }
we show spectra with different SNRs to visualize the impact of the spectra quality on the recognisability of the lines. In SDSS-393-51794 all lines \Zhong{are visible} and show a pattern similar to the simulated lines in Fig. \ref{fig:all emi}. In SDSS-436-51883, despite the \Zhong{spectrum's SNR being} comparable to the one above, the lower signal of the background lines makes some of them embedded in the noise, although some others still stick out rather clearly. Here, the number of visible lines is reduced by the higher redshift of the source ($z_E=0.452$). Finally, in SDSS-391-51782, the redshift of the source ($z_E=0.931$) \Zhong{permits the} observations of only two lines, which are yet rather easy to spot because of the decent SNR of the spectrum and the high signal of the lines. Overall, 
these examples show the kind of features the CNN needs to be trained on identifying in the spectra and the impact of the spectra quality and SNR of the background emission on the final 
line detection and redshift determination. 

Similarly, these examples provide textbook cases of HQ candidates we will visually grade among the high probability candidates provided from the GaSNets (see \S\ref{subsec:visual}).  

\subsection{Confirmed lenses from previous spectroscopic searches} \label{sec:confirmed_lens}
As anticipated in the previous section, we also collect candidate/confirmed lenses from previous spectroscopic searches in
SDSS/BOSS, using standard techniques, to be used as a real test sample for our deep learning tools. In \Zhong{particular,} we have collected 131 objects from \cite{2008ApJ...682..964B}, 45 from \cite{2012ApJ...744...41B}\Zhong{, and} 118 from \citet{2017ApJ...851...48S}, that have secure confirmation based on HST follow-up.
This ``test sample'' made of real systems is useful for two main purposes: 1) to measure the {\it completeness} of our tool, by checking how many of these lenses are recovered by \Gone; 2) to test how {\it accurate} the \Gtwo\ and \Gthr\ are in determining the $z_E$ and $z_G$, respectively. 
We will also compare our final catalog of HQ candidates vs. the latest highly complete sample of spectroscopic selected candidates in eBOSS from T+21. This will allow us to check the presence of candidates missed by standard techniques, and \Zhong{compare} the different approaches.

\begin{figure}
    \centering 
    \includegraphics[height=6cm,width=1\linewidth]{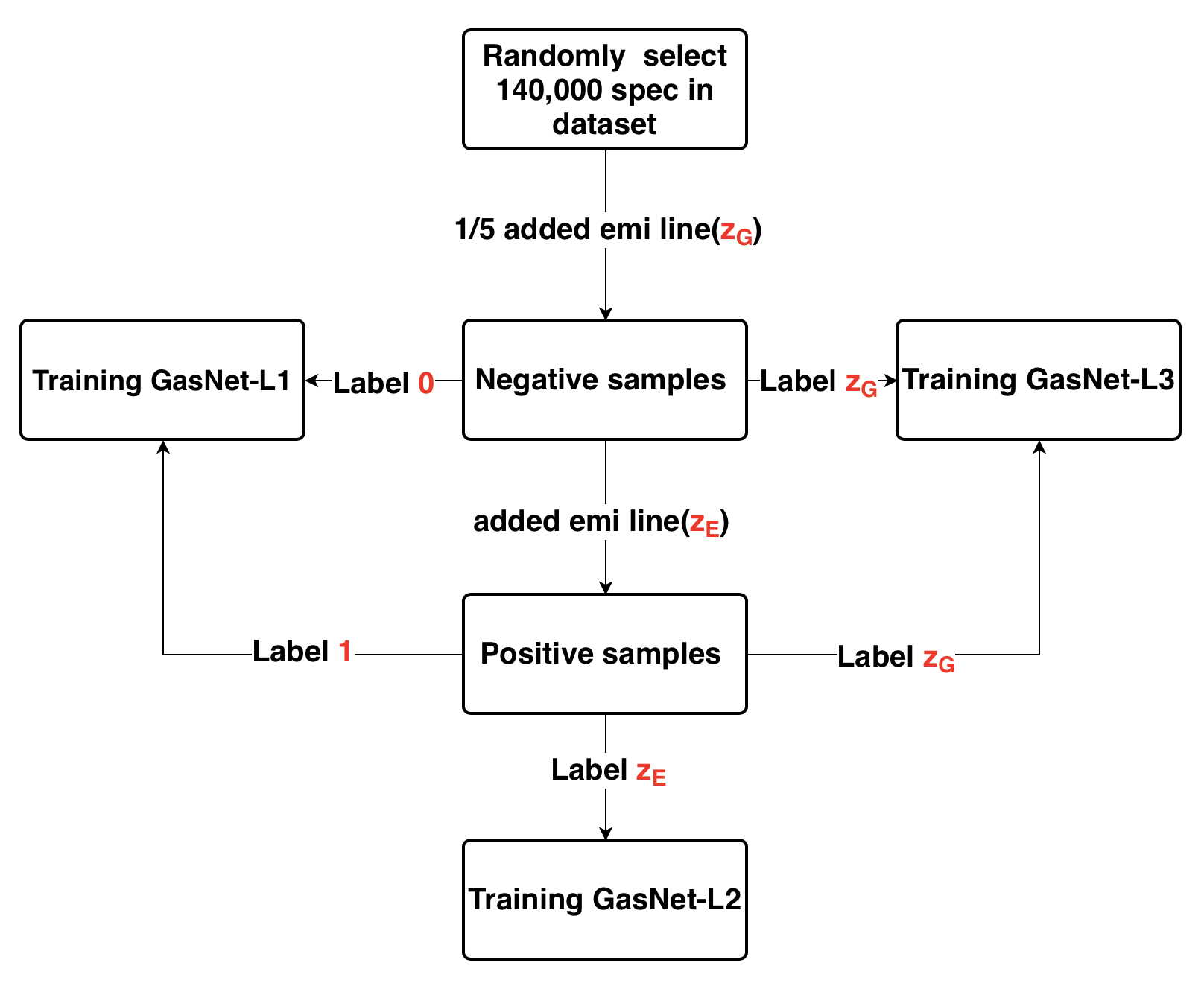}
    \caption{Summary of the training data building process. After \Zhong{constructing} the negative and positive samples, they are labeled before they are fed into the training process of the three \Gs. In this \Zhong{scheme,} we illustrate the steps made to add the label to the two training samples.} \label{fig:build data}
\end{figure}

\section{ \Zhong{ Implementation} }
\label{sec:test of CNN}
{To proceed with the construction of the training and test samples, we \Zhong{collect} 140\,000 positives and the same number of negatives. These samples are further split into the 3 datasets:  100\,000 for training, 20\,000 for validation\Zhong{, and} 20\,000 for testing. The first two samples are used to train the \Gs\ and evaluate how well the model predicts the ground truth targets based on the unseen data during the training process. The last sample is used to qualify the final performance of the \Gs. Finally, we also test the performance against real candidates from literature, as discussed in \S\ref{sec:confirmed_lens}.}

\begin{figure*}[t]
    \centering 
    {\begin{minipage}[t]{0.32\linewidth}
            \centering
            \includegraphics[width=1.0\linewidth]{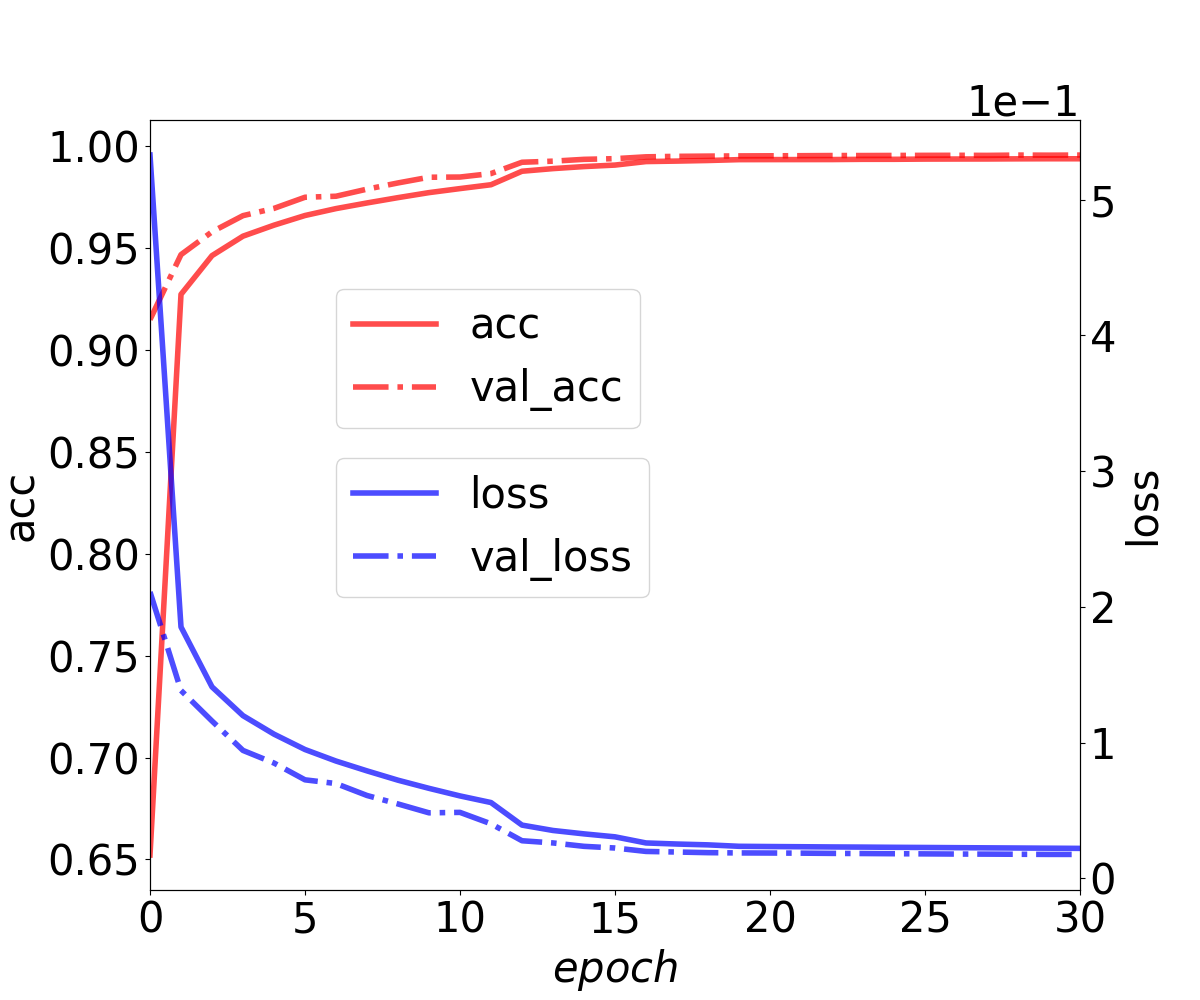}
    \end{minipage}}
    {\begin{minipage}[t]{0.32\linewidth}
        \centering
        \includegraphics[width=1.0\linewidth]{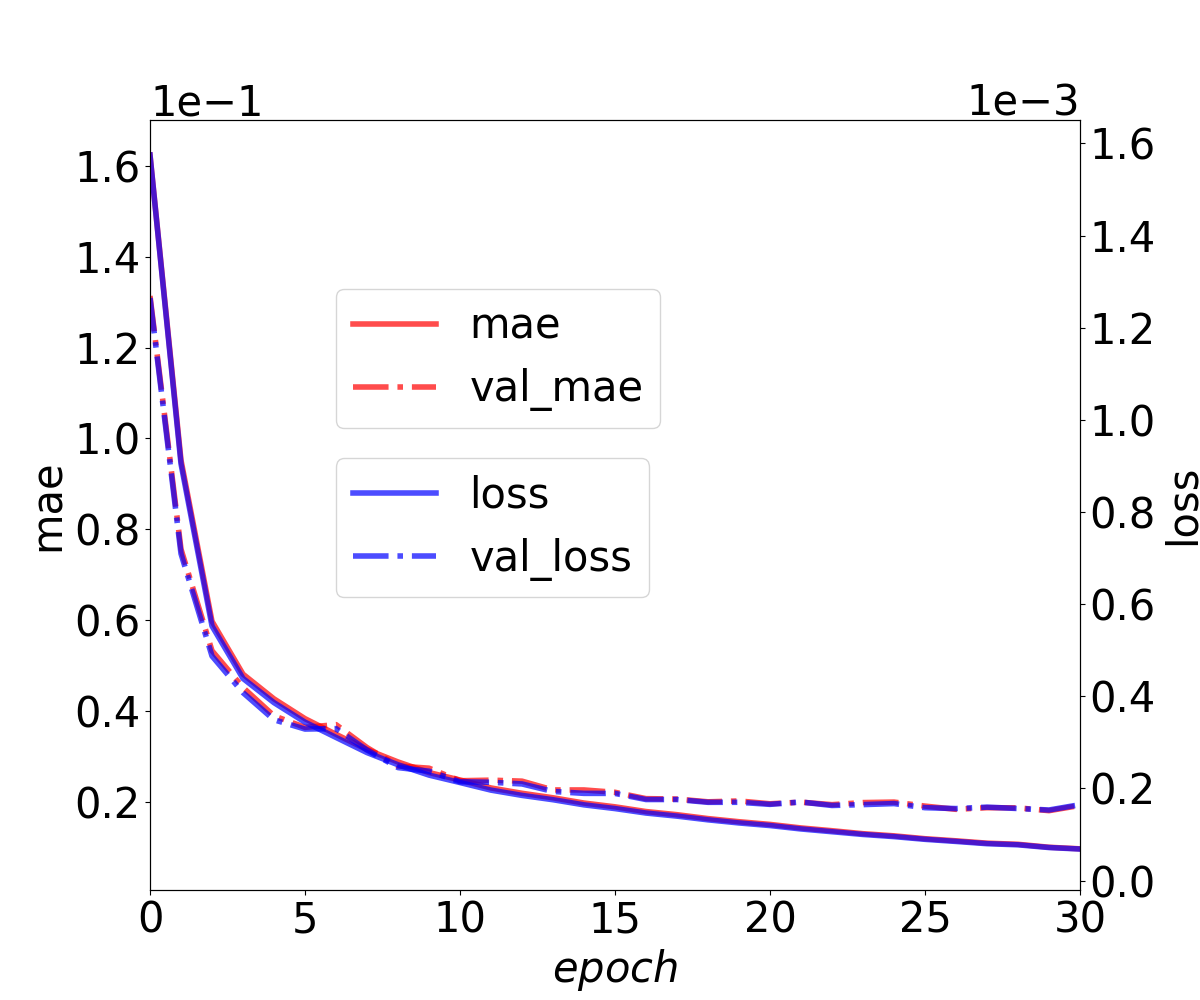}
    \end{minipage}}
    {\begin{minipage}[t]{0.32\linewidth}
            \centering
            \includegraphics[width=1.0\linewidth]{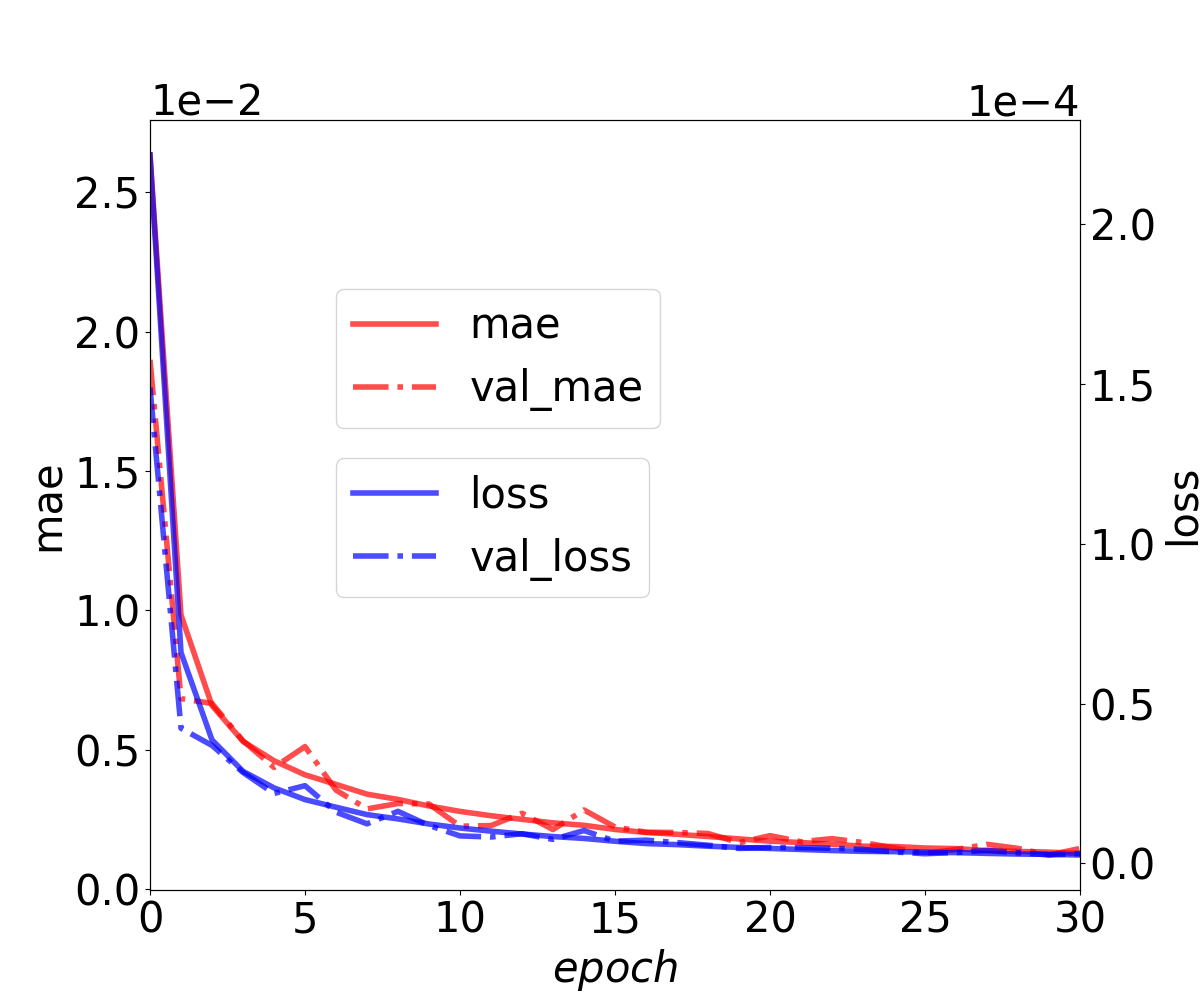}
    \end{minipage}}
    
    \caption{GasNets training results. Left panel: accuracy and the loss of GasNets-L1, the training and evaluation curves both converge to \Zhong{the same} point, and show high accuracy. Middle panel: MAE and loss of GasNets-L2, \Zhong{which show a worse but yet acceptable convergence, with a reasonably low MAE}. Right panel: MAE and loss of GasNets-L2, which well \Zhong{converge to the} same values, because $z_G$ is easier to predict than $z_E$. }
    
    \label{fig:CNN1}
    \label{fig:CNN2}
    \label{fig:CNN3} 
\end{figure*}

\begin{figure}
    \centering 
    {\begin{minipage}{0.95\linewidth}
            \centering
            \includegraphics[width=1\linewidth]{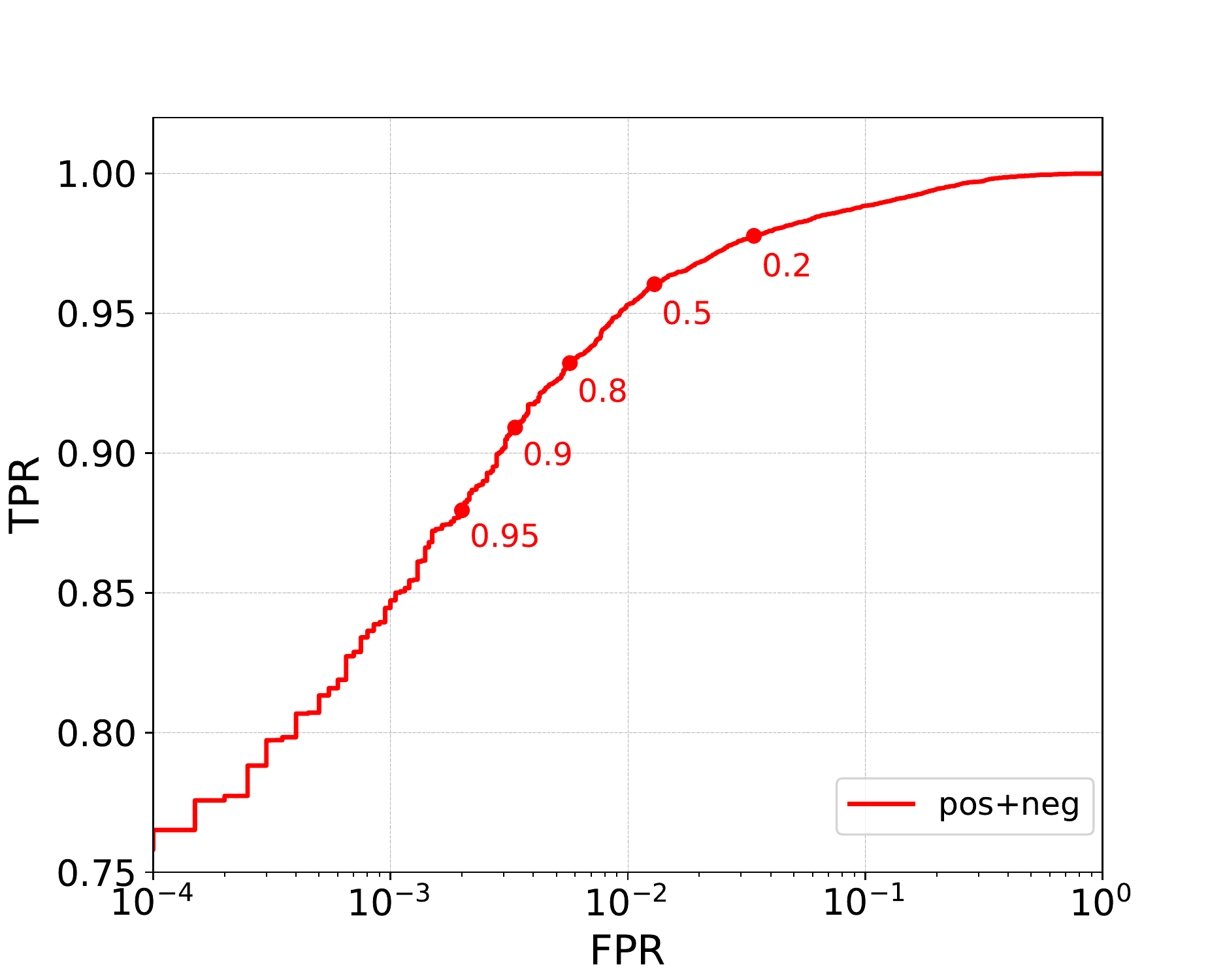}
    \end{minipage}}
    \caption{ROC curve of training samples where the true-positive rate (TPR) is plotted against the false-positive rate (FPR) (see text for details).}
    \label{fig:ROCcuve}
\end{figure}

\subsection{Training the Networks}
\label{sec:Training the Networks}
According to the description in \S\ref{sec:galnets} and the tasks they are expected to fulfill, during the training, the \Gs\ are fed 
with the training spectra to produce accurate predictions of the ``target'' quantities. For \Gone\, the inputs are the spectra of the positives and negatives as well as their labels to give as output the probabilities ($P_L$) to be lens candidates. For \Gtwo, the inputs are the simulated positive spectra with their labels, while the outputs are the predicted redshifts of the emission lines $z_E$. For \Gthr, the inputs are the labeled spectra of positives and negatives and the output are the redshifts of the foreground spectra ($z_G$).
The full process of the training sample building and \Zhong{labeling} is summarized in Fig. \ref{fig:build data}.

\begin{table}[t]
 \caption{Statistical properties of the predicted parameters}
 \label{table:Statistical properties}
 \begin{center}
  \footnotesize
 \begin{tabular}{c c c c c c c c}
 \hline
 \hline
    Sample & var. & $R^2$ & Out. fract. & NMAD & MAE &MSE  \\
  \midrule
    Test   & $z_{PE}$ & $0.941$ & $0.0121$ & $0.0029$ & $0.0164$ & $0.0032$ \\  
    Test   & $z_{PG}$ & $0.998$ & $0.0003$ & $0.0009$ & $0.0017$ &$0.0001$\\
    Real    & $z_{PE}$ & $0.770$ & $0.0840$ & $0.0062$ & $0.0535$ & $0.0173$\\
    Real    & $z_{PG}$ & $0.989$ & $0.0047$ & $0.0012$ & $0.0033$ & $0.0004$\\
    All     & $z_{PG}$ & $0.988$ & $0.0012$ & $0.0009$ & $0.0020$ & $0.0002$\\
 \hline
 \hline
 \end{tabular}
  \end{center}
\end{table}

Regarding the training step, for \Gone\ and \Gthr\ we use the 120\,000 positive (traning+validation data) and 120\,000 negative samples, i.e. a total of 240\,000 spectra. Since \Gtwo\ only predicts the $z_E$, in this \Zhong{case} the training+validation sample \Zhong{is} made by 120\,000 spectra from the positive sample only. 
For each GaSNet, we use the training data to train 30 epochs with a learning rate of 0.0001 and use validation data to evaluate the performance. We have found that this produces rather stable validation results. During the training process, we optimize the 3 \Gs\ with the Adam optimizer (\citealt{Friedman99+huberloss}).

\subsection{Testing on simulation data}
\label{sec: test_simul}

\begin{figure*}
    \centering 
    \vspace{-0.3cm}
    {\begin{minipage}[t]{0.32\linewidth}
            \centering
            \includegraphics[width=1.1\linewidth]{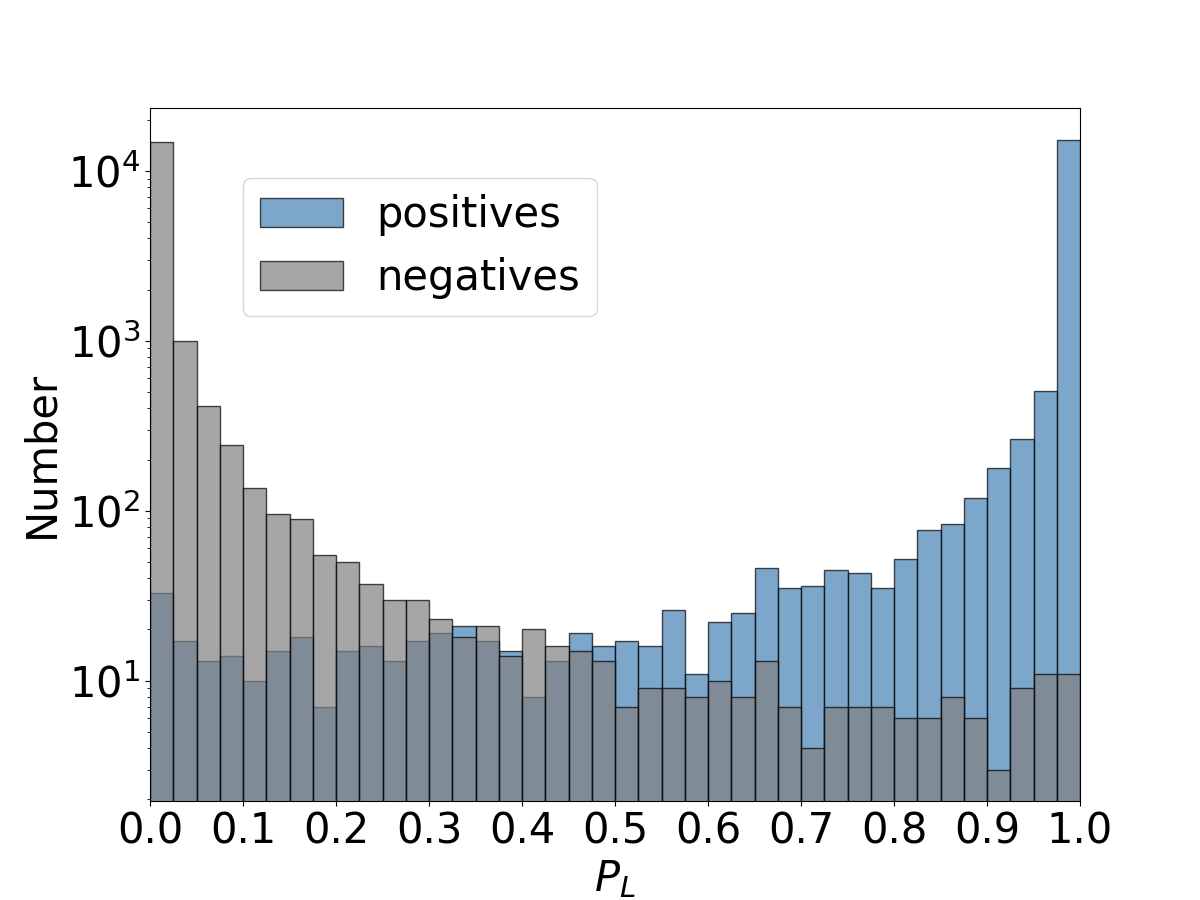}
    \end{minipage}}
    {\begin{minipage}[t]{0.32\linewidth}
        \centering
        \includegraphics[width=1.1\linewidth]{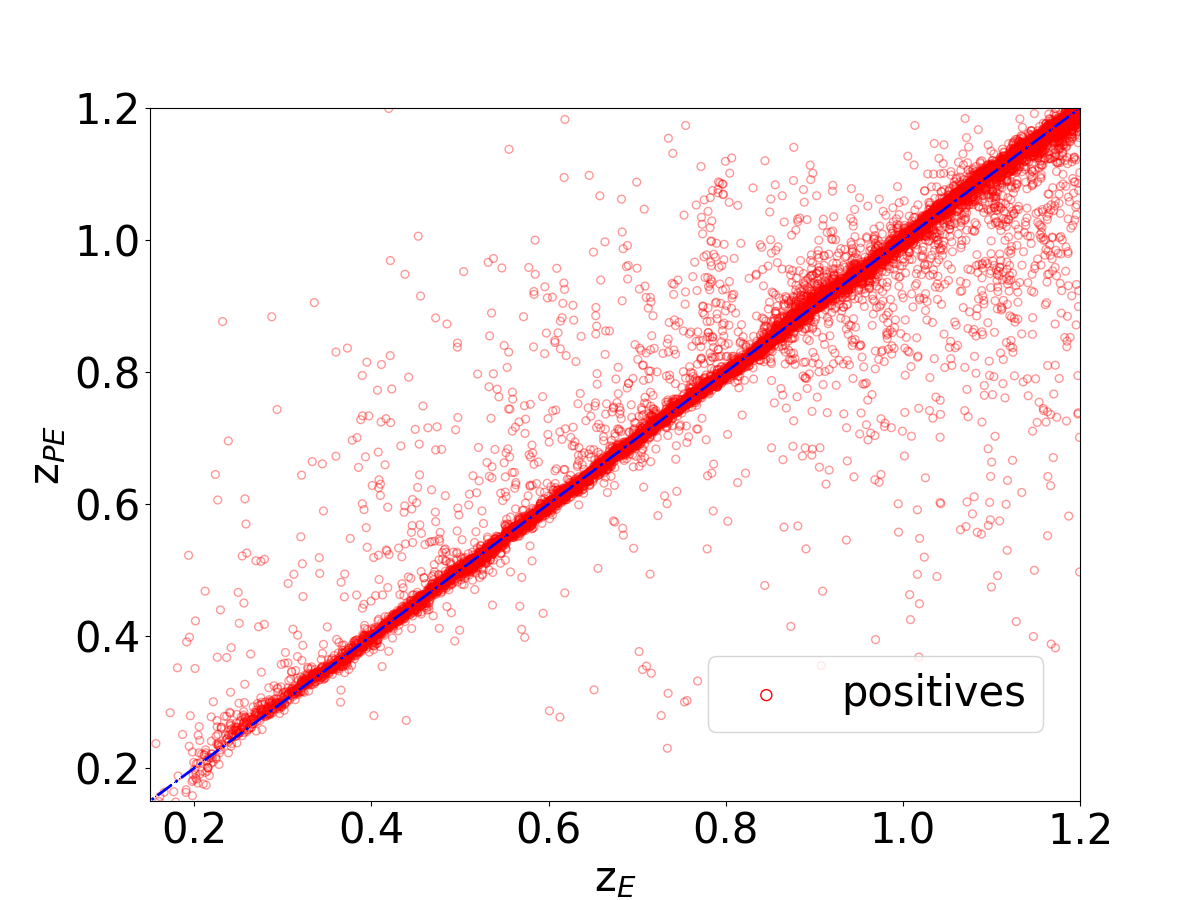}
    \end{minipage}}
    {\begin{minipage}[t]{0.32\linewidth}
            \centering
            \includegraphics[width=1.1\linewidth]{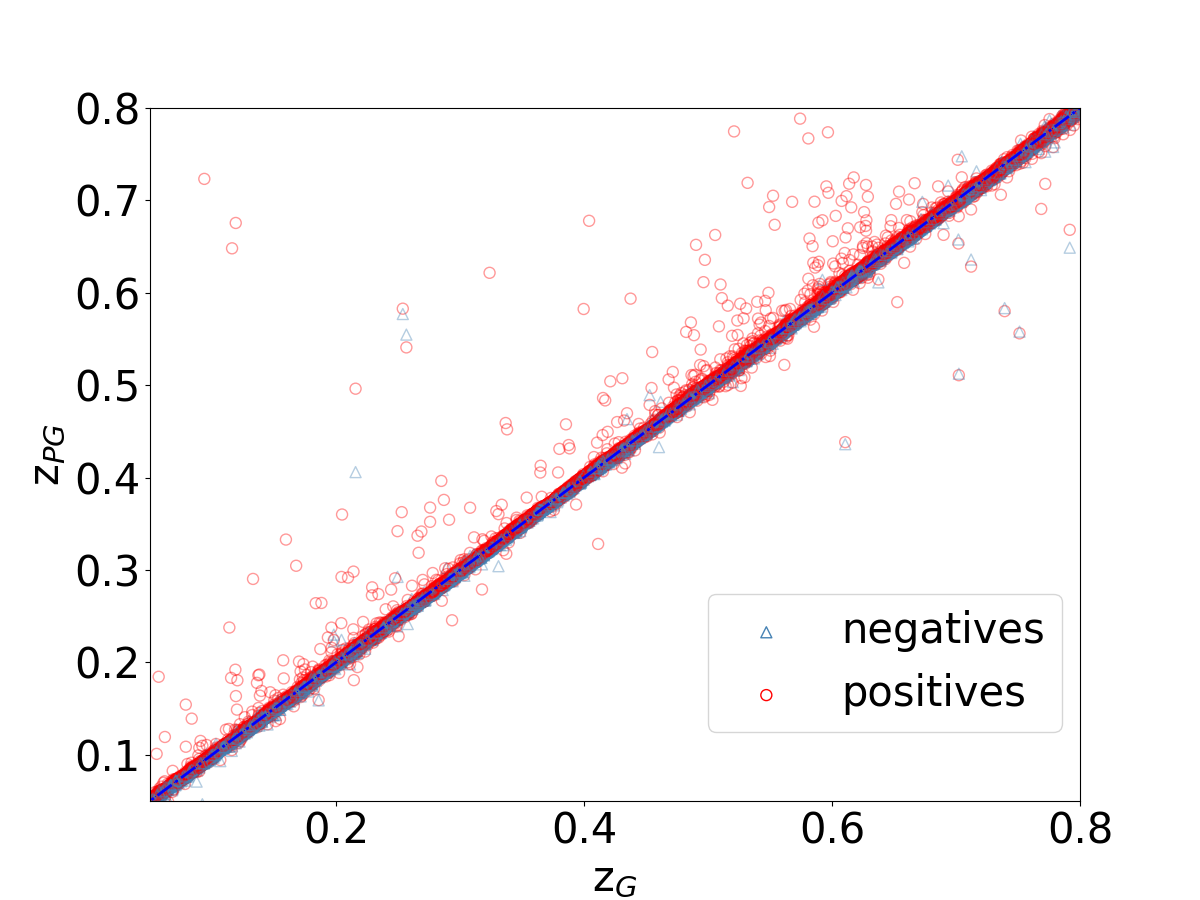}
    \end{minipage}}
    
    \caption{\Gs\ results on the test sample. Left: the $P_L$ distribution from \Gone; Center: predicted emission line redshifts from \Gtwo, $z_{PE}$, vs. ground truth values, $z_{E}$; Right: predicted lens galaxy redshifts from \Gthr, $z_{PG}$, vs. ground truth values, $z_{G}$.}
    \label{fig:pos_test}
    \label{fig:neg_test}
\end{figure*}

After training, we first test the \Gs' performances using the simulated ``test'' spectra. As anticipated, the test sample is made \Zhong{of} 20\,000 positive and 20\,000 negative samples for \Gone\ and \Gthr\ and 20\,000 positive samples for \Gtwo\ .


In Fig. \ref{fig:CNN1} we first show the results of the training run for the three \Gs\ to have a first evaluation of their performances. 
In particular, we plot the first 30 training epochs. The solid lines in Fig. \ref{fig:CNN1} represent the accuracy reached 
on the training data as the average deviation of the predictions from the ground truth (loss). The dot-dashed lines represent the same quantity on the test data. 

For each GaSNet, we set a different evaluation function: for \Gone, 
being a classifier giving a probability as output, we use the ``accuracy" (acc) as loss variable;
for \Gtwo\ and \Gthr\, as they predict the $z_E$ and $z_G$, we set the mean absolute error (MAE) as loss variable.
From Fig. \ref{fig:CNN1}, \Gone\ and \Gthr\ both show good convergence at about the same epoch toward the end of the training, while \Gtwo\ shows a larger loss because \Zhong{of the} degeneracy between noise and emission lines (see comment above). 
One possibility to improve this result might be the adoption of some spectra pre-processing, e.g. via smoothing. However, this would imply an incursion on the data characterization that is beyond the purposes of this paper, and we rather plan to address \Zhong{this in} next analyses. Here, we just stress that the accuracy reached by \Gtwo\ is high enough to \Zhong{separate} the background emission lines from the foreground spectral features in lens candidates, hence more than sufficient for its actual purposes (see also \S\ref{sec:Slightly shift}).

In Table \ref{table:Statistical properties}, we report some statistical estimators to measure the \Zhong{\Gs'} performances. Besides the standard MAE and MSE, we add other three estimators.

First, the R-squared ($R^2$) is used to evaluate the linear relationship between prediction and true values. It is defined as 
$$
R^2 = 1 - \frac{\sum_{i}^{} (z_P-z_T)^2}{\sum_{i}(z_T-\Bar{z_T})^2}
$$
where $z_P$ is the predicted value and $z_T$ is the true value, and $\Bar{z_T}$ is the average value of $z_T$. The closer the $R^2$ is to 1, the \Zhong{better} the prediction. In Table \ref{table:Statistical properties} we see that for the test sample, $R^2$ is close to 1 for both $z_{PG}$ and $z_{PE}$, meaning that both \Gtwo\ and \Gthr\ are expected to produce accurate results.

Second, the outlier fraction, which is defined as the fraction of predicted redshifts scattering more than 15\% from the true values:
$$
\delta Z = \frac{\left |  z_P - z_T \right | }{1+z_T} > 0.15.
$$
For the test sample, in Table \ref{table:Statistical properties} we show that the outlier fractions are $\lsim1\%$, implying a very small fraction of anomalous predictions.

Third, the normalized median absolute deviation (NMAD), which is defined as:
$$
    NMAD = 1.4826 \times  median(\left |\delta Z - median(\delta Z)\right |).
$$
It gives the absolute deviation of the predicted value from the central value of $\delta Z$. As seen in Table \ref{table:Statistical properties}, the NMAD for the test sample is close to zero, meaning again a very small deviation from the true values, i.e. very accurate predictions.

In Fig. \ref{fig:ROCcuve} we also show the receiver operating characteristic (ROC) curve 
where we plot the true-positive rate (TPR) against the false-positive rate (FPR). The TPR is the fraction of lenses that are correctly classified with respect to the total number of ``ground truth'' lenses, while the FPR is the fraction of \Zhong{non-lenses} that are misclassified as lenses with respect to the total number of \Zhong{non-lenses}.
The ROC curve can be used to decide the probability threshold to adopt as a trade-off between true detection and contaminants from \Zhong{false positives}. In the same figure, we report the TPR-FPR for different $P_L$s. We can see \Zhong{that} for a $P_L=0.95$, we almost reach 90\% completeness with a negligible false positive rate. We stress here that this result \Zhong{derived} from simulated spectra \Zhong{is} in rather ideal conditions. Hence, both the TPR and, most of all, the FPR might be just an upper and lower limit, respectively, as compared to the real cases. However, the $P_L=0.95$ occurs before the slope of the ROC becomes flatter, meaning that the gain in the number of true detections, at lower thresholds, increases at the cost of a larger number of contaminants.
We will come back to these results later \Zhong{when} we will discuss the threshold to adopt to select HQ candidates in real data. 

Finally, in Fig. \ref{fig:pos_test} we detail the results obtained for the test sample. On the left \Zhong{panel,} we show the distribution of the $P_L$ from \Gone\ for both the negative and the positive samples. As expected the former \Zhong{tends} to cluster more toward a peak at $P_L=1$, but with a rather long tail toward the $P_L=0$, meaning that, statistically, there \Zhong{is a significant fraction of true positives to which \Gone\ has given a low probability.}
We have checked these latter cases and found no correlation with the overall SNR of the spectra. Instead, we have found a correlation of the low $P_L$ objects with the $z_E$, in the sense that the larger the $z_E$, the bigger \Zhong{the} number of \Zhong{the} object with $P_L<0.5$. This suggests that either the lower number of lines or the intrinsically lower SNR of the lines suppress the $P_L$ and makes the classification of the lensing event more difficult at higher-$z$.

In the central panel of the same figure, we show the output of the \Gtwo\ by comparing the predicted $z_{PE}$, against the ground truth values, $z_E$. 
Overall, the majority of the predicted values are tightly distributed around the one-to-one relation, as also quantified by the large $R^2$ values found in Table \ref{table:Statistical properties} (Test/$z_{PE}$). \Zhong{Numerous predictions scatter} quite largely from the perfect correlation, because of the degeneracy of noise and background emission lines, as mentioned above. However, these are statistically irrelevant as the estimated outlier fraction in \Zhong{Table} \ref{table:Statistical properties} is close to 1\%.

Finally, in the right panel of Fig. \ref{fig:pos_test}, we show the predicted $z_{PG}$ from \Gthr\ against the ground truth values, $z_G$. In this case, the correlation is quite perfect and the outlier fraction is negligible ($<0.1\%$ see Table \ref{table:Statistical properties} -- Test/$z_{PG}$), both for the positive and the negative sample. Indeed, for this latter test, we have also input the negative sample to check the performance of \Gthr, as a pure automatic spectroscopic redshift tool, in absence of artificial emission lines. This shows that the ability of \Gthr\ to predict the galaxy redshift is not driven by the emission lines, easier to spot, but \Zhong{by} the overall features of the spectrum (i.e. continuum and absorption/emission lines).

\subsection{Test on HST confirmed samples}
\label{sec:real_data}
Previous analyses of the SDSS/BOSS spectra have brought to the collection of 
294 
strong lenses candidates: 
131 from SLACS, 45 from BELLS\Zhong{, and} 118 from S4TM (see \S\ref{sec:intro}). 
Being candidates based on spectroscopic features, these samples contain both real lenses and contaminants. Indeed, space imaging follow-ups have confirmed 70/131 SLACS candidates (the Grade-A objects in Table 4 of \citealt{2008ApJ...682..964B}), 25/45 BELLS candidates (Grade-A objects in Table 3 of \citealt{2012ApJ...744...41B})\Zhong{, and} 40/118 S4TM (Grade-A in Table 1 of \citealt{2017ApJ...851...48S}).
As also commented in \S\ref{sec:intro}
these correspond to an average confirmation rate of 46\%. Note, though, that the HST samples often tend to optimize the confirmation rate by pre-selecting targets with low-resolution imaging (see e.g.  \citealt{2004AJ....127.1860B}; \citealt{2016ApJ...824...86S}), hence this can be considered an optimistic upper limit estimate.
These are the main statistical samples that have been systematically \Zhong{followed up} to collect space imaging confirmations of spectroscopically selected SGL candidates, using optical lines. As such, these represent the most secure sample to check our results against.
These data can be used for two main purposes: 1) to compare the classification of the \Gs\ against human selection and help us \Zhong{set} a reasonable threshold to optimize the chance of finding real lenses with the minimal contamination from false positives; 2) to forecast the success rate we might expect from our set-up \Zhong{since} we have a reference sample of ``candidates'' and ``confirmed'' events. 
Being this literature sample far from complete (see \S\ref{sec:challange}), it cannot be fully used to draw firm conclusions about the completeness of the \Gs, \Zhong{however,} this is the only sample we can use to benchmark the \Gs' performances, with a necessary grain of salt.
On the other hand, the large sample from T+21, having no space observations cannot be used for the same purpose \Zhong{as} the ones above. As anticipated, we will use it for an {\it a posteriori} test to assess the differences (if any) \Zhong{between} standard and deep learning approaches.

\begin{figure*}
    \centering 
    \vspace{-0.3cm}
    {\begin{minipage}[t]{0.33\linewidth}
            \includegraphics[width=1.07\linewidth]{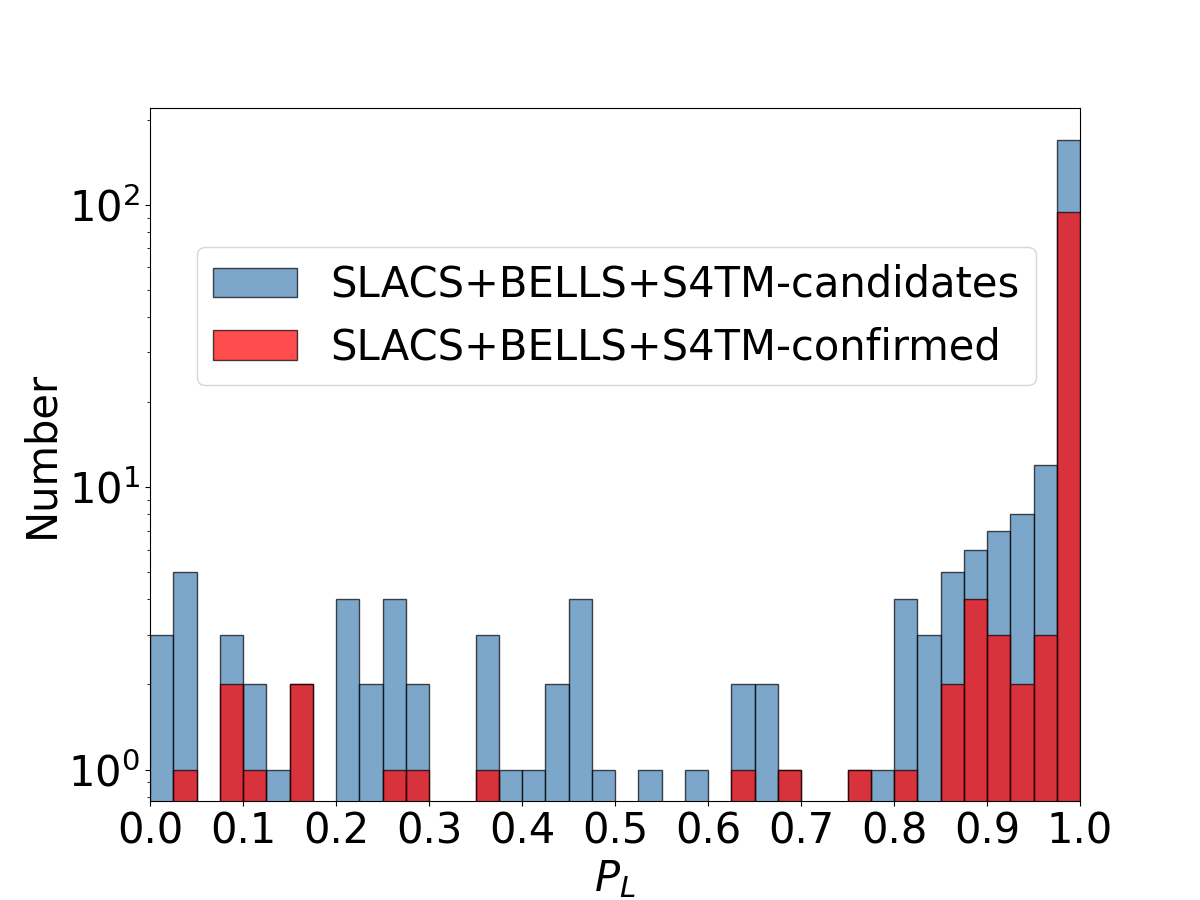}
    \end{minipage}}
    {\begin{minipage}[t]{0.33\linewidth}
        \includegraphics[width=1.07\linewidth]{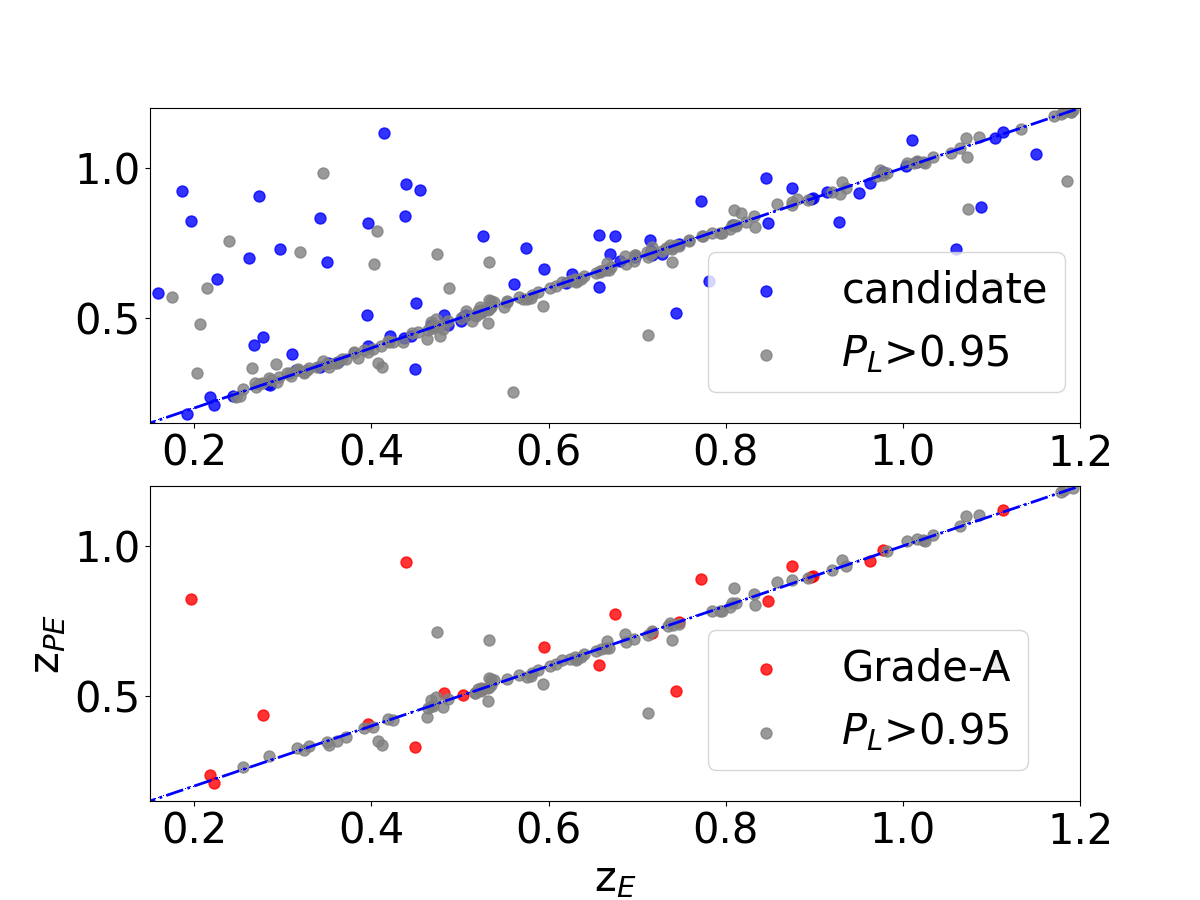}
    \end{minipage}}
    {\begin{minipage}[t]{0.33\linewidth}
            \includegraphics[width=1.07\linewidth]{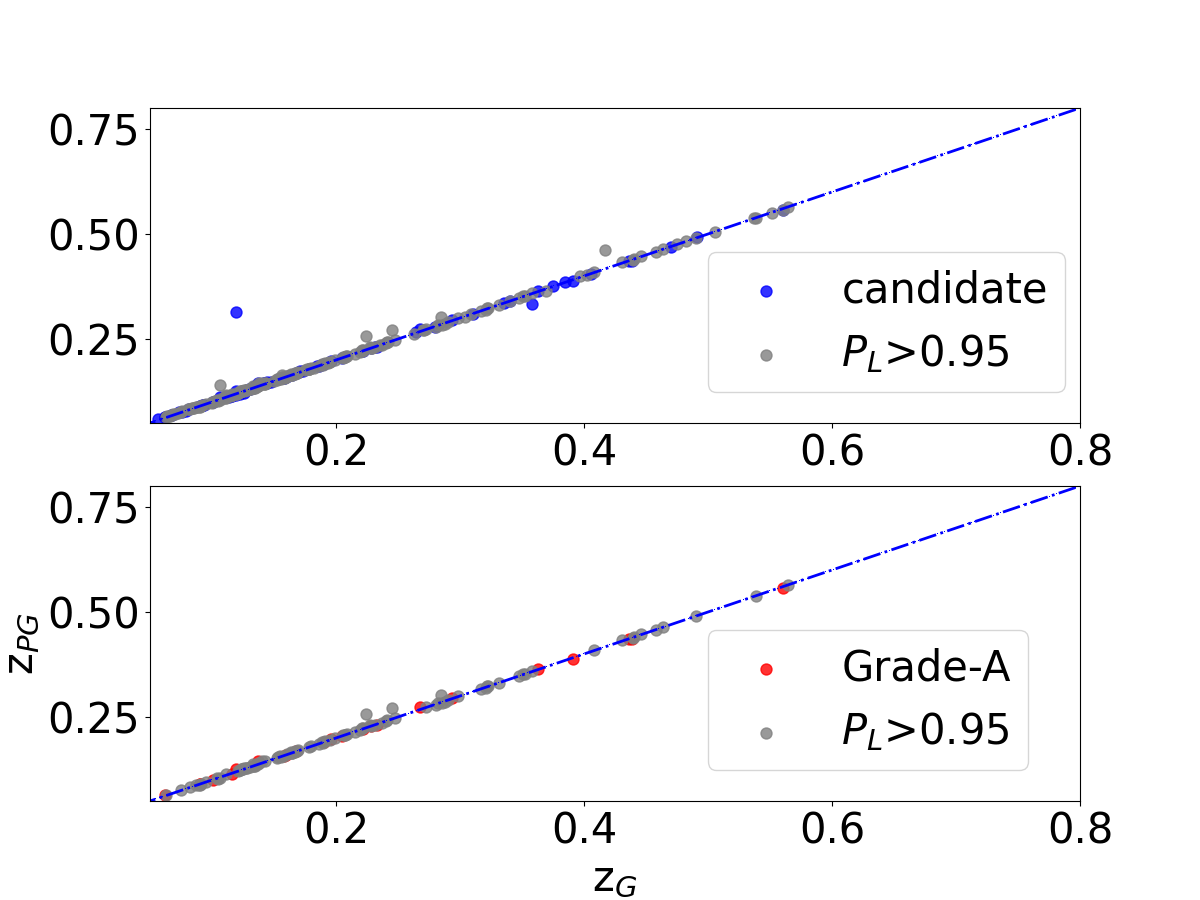}
    \end{minipage}}
    
    \caption{Results of the \Gs\ applied to spectra of strong gravitational lens candidates (in blue) and HST-confirmed events (in red) from SLACS, BELLS\Zhong{, and} S4TM samples (see text for details). Left: the $P_L$ distribution from \Gone; Center: predicted emission line redshifts from \Gtwo, $z_{PE}$, vs. literature redshifts, $z_{E}$, for the candidate objects (top) and the HST confirmed (bottom); Right: predicted lens redshifts from \Gthr, $z_{PG}$, vs. literature redshifts, $z_{G}$, for the candidate objects (top) and the HST confirmed (bottom).}
    \label{fig:real_len}
\end{figure*}
\begin{figure*}
\vspace{-0.45cm}
    \centering 
    {\begin{minipage}{0.33\linewidth}
            \centering
            \includegraphics[width=1.07\linewidth]{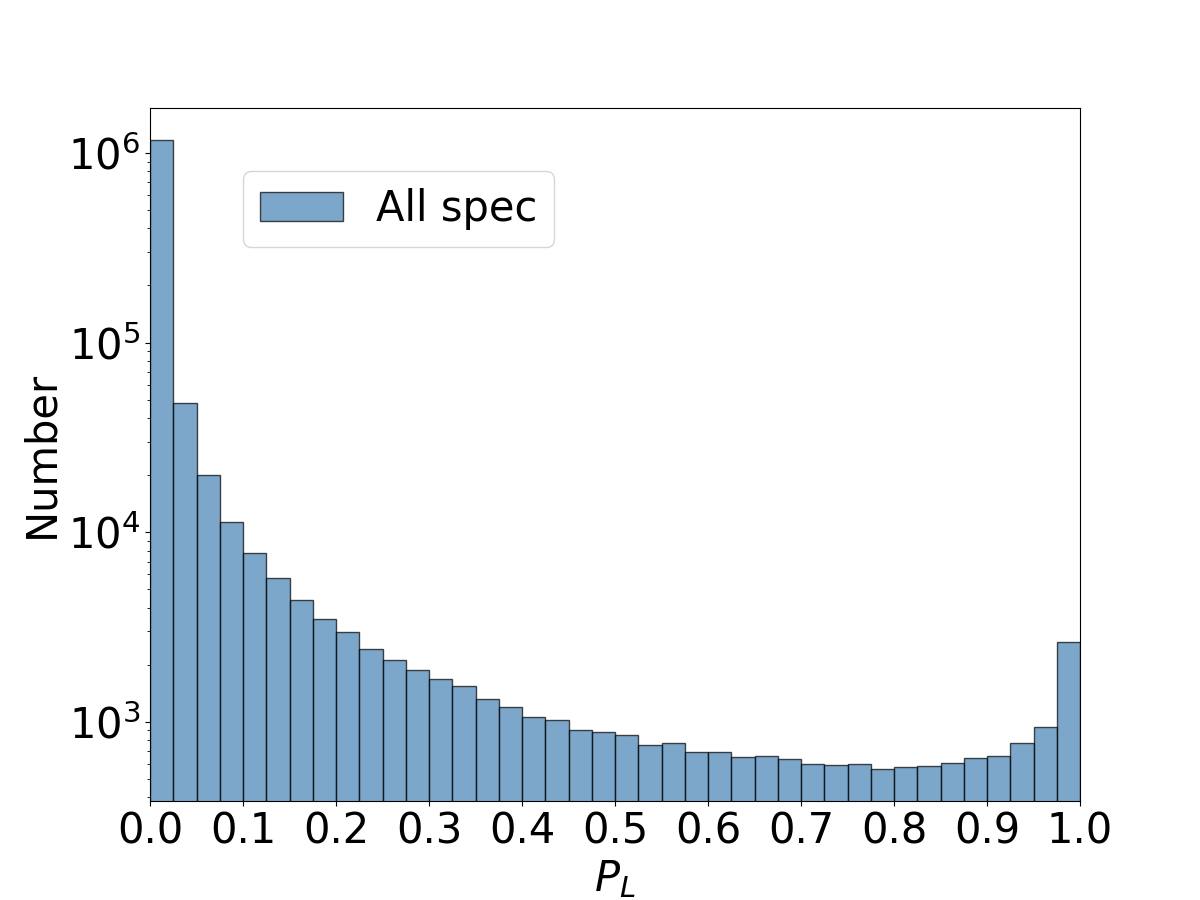}
    \end{minipage}}
    {\begin{minipage}{0.33\linewidth}
        \centering
        \includegraphics[width=1.07\linewidth]{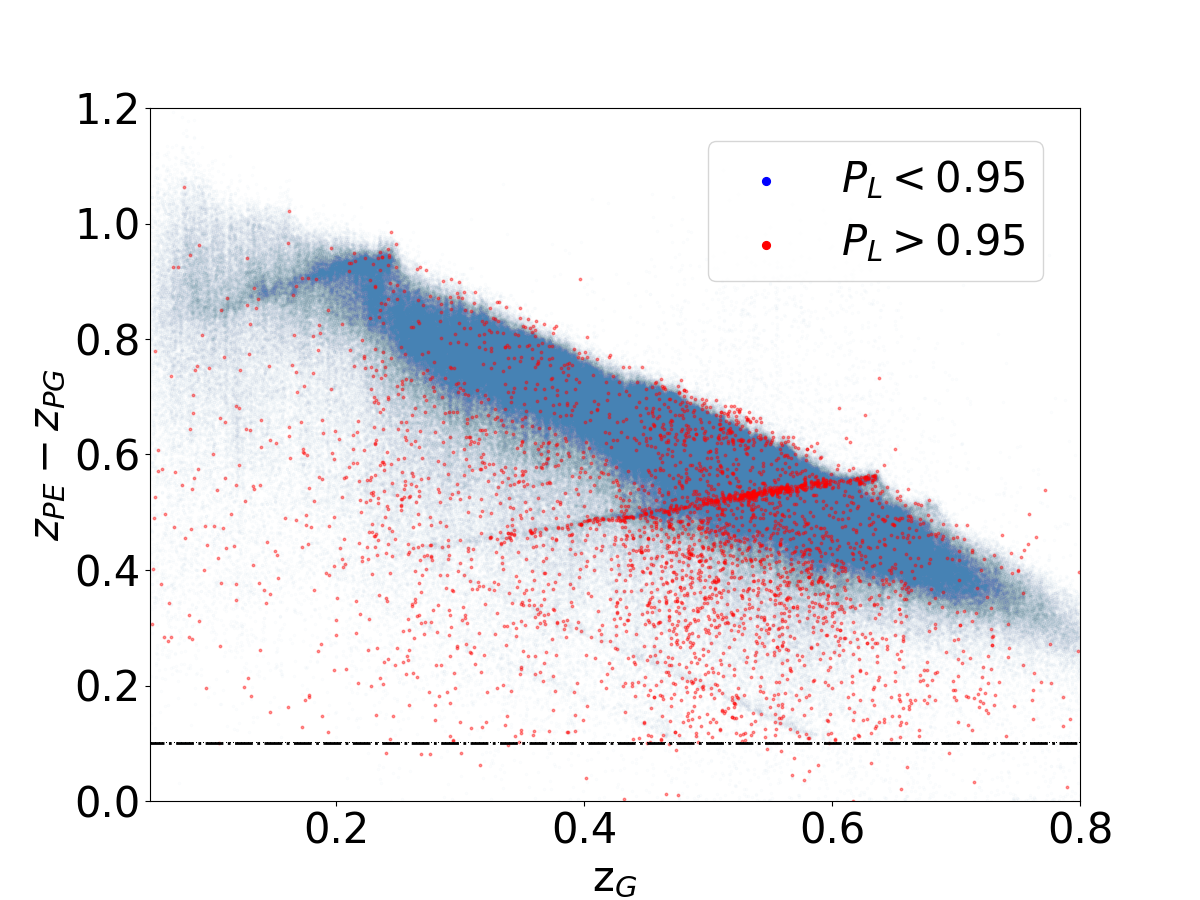}
    \end{minipage}}
   {\begin{minipage}{0.33\linewidth}
            \centering

            \includegraphics[width=1.07\linewidth]{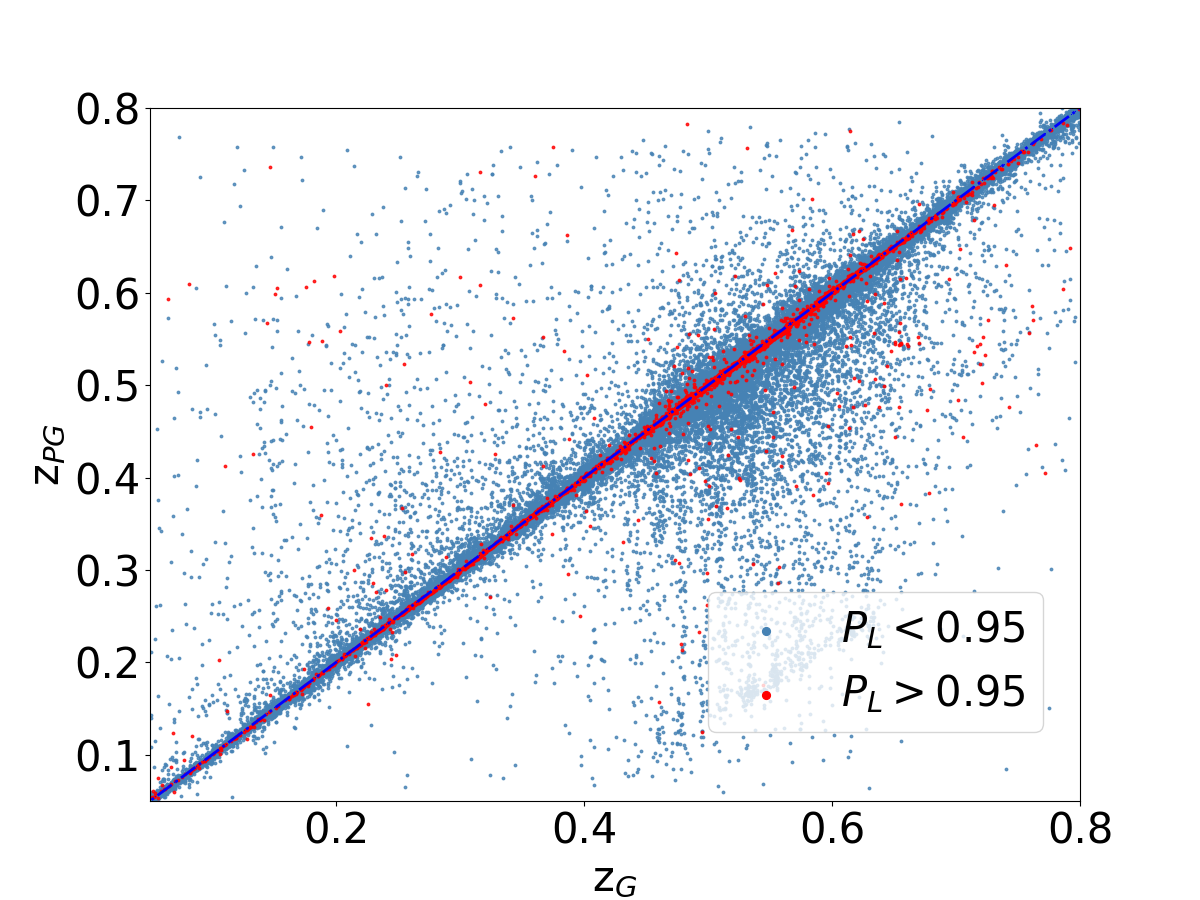}
    \end{minipage}}
    \caption{Predictions of the three \Gs\ on the DR16-predictive sample ($\sim 1.3M$ objects): the $P_L$ from \Gone\ (left), the $z_{PE}-z_{PG}$ (center) and the $z_{PG}$ (right), from \Gtwo\ and \Gthr\ outputs,  vs. $z_G$, the galaxy redshift from eBOSS/DR16 catalog (see text for the details). In the center and right panels we show the $P_L>0.95$ sample (in red) and the $P_L<0.95$ in blue.
    }
    \label{fig:all_spec}  
\end{figure*}

To proceed with the test of the HST confirmed catalogs against \Gs, we first select the literature spectra that are located in the predictive range of our CNNs (i.e. $0.05<z_G<0.8$ and $0.15<z_E<1.2$). These are 264/294 candidates and 121/135 confirmed objects.
In Fig. \ref{fig:real_len} we show the probability predicted from the \Gone\ (left panel), the redshift of the source predicted from \Gtwo\ (central panel), and the redshift of the lens galaxies predicted by the \Gthr\ (right panel), for the candidates and confirmed literature objects face-to-face.

In particular, we see that 
\Gone\ predicts high probabilities for most of the lenses: e.g., 69\% of the candidates and 80\%  of the confirmed objects have $P_L>0.95$, which becomes 81\% of the candidates and 90\% of the confirmed objects for $P_L>0.8$.

More importantly, the ratio of the confirmed/candidates increases dramatically from $0.8<P_L<0.95$ to $P_L>0.95$, as we have 12/33, i.e. 36\% for the former and 97/182, i.e. 53\%, for the latter,
vs. the overall 46\% estimated for the full sample (see above). On the other hand, for $P_L<0.8$ the confirmation rate drops to 12/49, i.e. 24\%, which is 
too low 
for successful space observations and still anti-economical for lens search in spectra. Indeed, as discussed in \S\ref{sec: test_simul}, at $P_L<0.8$ the FPR becomes prohibitive, producing massive false detections in large samples that should be cleaned with tedious visual inspections.   
Interestingly enough, 
for $P_L>0.95$ the fraction of true SGL events recovered (80\%) is rather close to the TPR ($\sim89\%$) predicted by the ROC curve (Fig. \ref{fig:ROCcuve}) for an idealized mock population of strong lenses. This means that the performances of the \Gs\ on the real data might be not far from the expectations from simulated data. 

However, In Fig. \ref{fig:real_len} (left) 
a misalignment between the deep learning and human filtered selections is further demonstrated by the fact that
some confirmed lenses
have received a small probability by \Gone. As discussed in the previous section, these are mainly low-SNR emission line spectra or higher-$z$ systems that, even if accounted \Zhong{for in} the training sample, are difficult to be highly scored by \Gone\ but might have been picked by the human eye with higher confidence. Hence, we conclude that a $P_L=0.95$ threshold is very likely to produce effective completeness higher than the 80\% obtained above over a complete and unbiased true SGL sample. 

\begin{figure*}  
    \centering 
            \centering
            \includegraphics[width=1\textwidth]{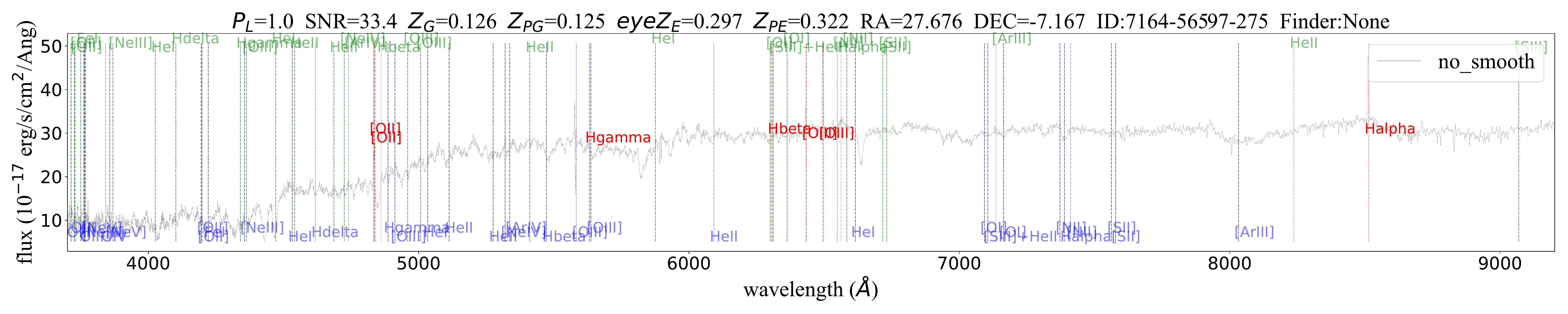}
        \includegraphics[width=1\textwidth]{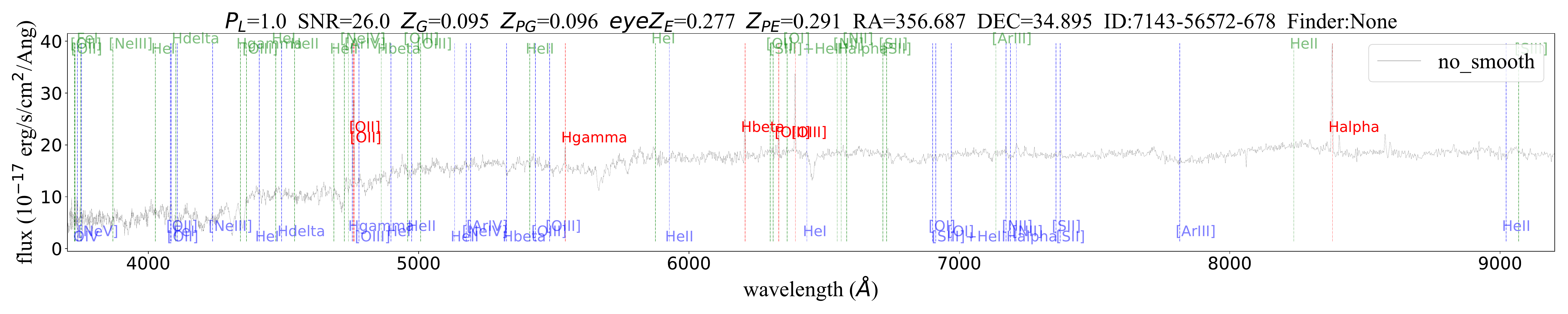}
        \includegraphics[width=1\textwidth]{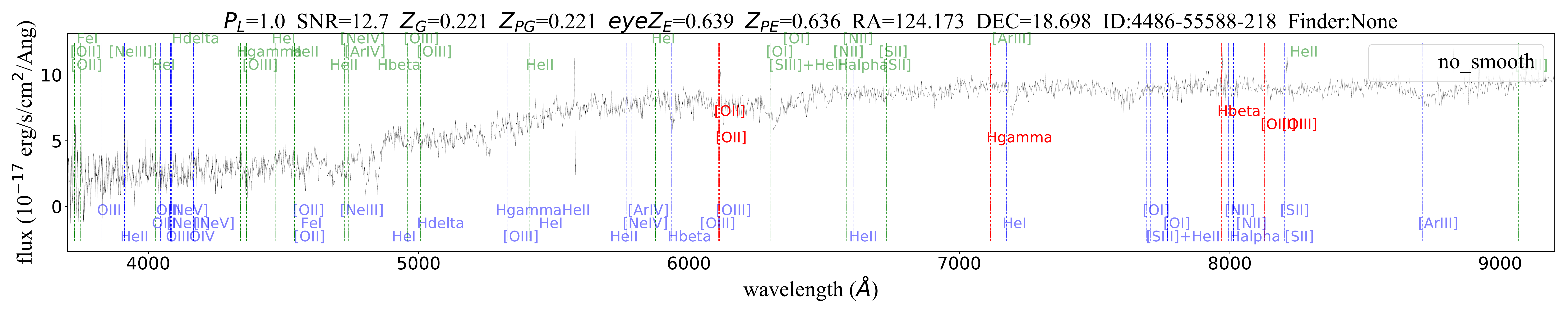}
\caption{HQ candidate spectra. The SDSS/BOSS spectra are plotted with highlighted their main spectral features. Red vertical lines indicate the emission lines of the background source at the redshift $eye z_E$ (i.e. the one corrected \Zhong{during} the visual inspection). Blue vertical lines indicate the spectral features of the lens at redshift 
$z_G$. Green lines show the location at rest frame ($z=0$) of the  emission lines from sky). On the top of each spectrum from left to \Zhong{right,} we report the probability from \Gone, the average SNR of the spectrum, the \Zhong{cataloged} redshift of the galaxy from SDSS/BOSS, \Zhong{the predicted redshift from \Gthr, the corrected redshift from visual inspection, the predicted redshift from \Gtwo}, and finally, the RA, DEC\Zhong{, and} ID of the target. 
}
    \label{fig:candida_spec}
\end{figure*}

The middle and the right panel of Fig. \ref{fig:real_len} show that both \Gtwo\ and \Gthr\ can make good predictions on the redshift of the emission lines and the lens galaxies.
In general, \Gthr\ performs better than \Gtwo\ (see Table \ref{table:model parameters}), possibly because the spectra of the lens galaxies can provide more information, both from the continuum and the absorption or emission lines, 
while \Gtwo\ relies only on a few emission lines, 
which provide intrinsically less information.
We also see that the confirmed objects generally show a smaller scatter and outlier fraction than the candidates, especially in $z_{PE}$, and also that the highest probability objects show tighter one-to-one predictions. This demonstrates that misclassifications of SGL events might be related to uncertainties on the redshift of the background sources, which tend to be placed further away than sometimes they are, i.e. confusing ``local'' emissions with background ones. However, the chance of such misclassification is reduced for $P_L>0.95$ systems.

All in all,
Fig. \ref{fig:real_len} 
indicates that the $P_L>0.95$ sample is accurate enough to produce reliable lens candidates from the DR16-predictive sample.

\section{Results}
\label{sec:applying_BOSS}
In this section, we apply the trained GaSNets to the DR16-predictive sample, introduced in \S\ref{sec:data}. This is made of 1\,339\,895 galaxy spectra and represents the sample among which we want to find new strong lens candidates and, for them, determine 
the redshift of the background source, $z_E$.

\subsection{Predictions on the eBOSS spectra}
\label{sec:DR16data}
According to the workflow described in Fig. \ref{whole_CNN_model}, the first step to perform is the classification of candidates using GaSNet-L1. In Fig. \ref{fig:all_spec} (left) we report the probability $P_L$ distribution obtained from GaSNet-L1 for the DR16-predictive sample.
From this histogram, we see that 
using a $P_L>0.8$, which, according to the ROC curve, would return almost 95\% of the true lenses, 
would produce a list of about 10\,000 candidates. This is a sample hard to handle for two main reasons: 1) it is \Zhong{time-consuming} to visually inspect and 2) it is foreseen to be severely contaminated from false detections. 
This latter case has been confirmed by randomly inspecting $100$ candidates with $0.8<P_L<0.95$ to find that about 90\%
are very poor candidates. 
{On the other hand, choosing $P_L>0.95$, which, for the true lens cases, allowed to recover \Zhong{of} $80\%$ of the confirmed lens known in SDSS/BOSS, would produce a more manageable sample of $\sim4000$ candidates. 
Hence, at the cost of some acceptable incompleteness, for this first test, we decide to adopt a more conservative approach and search for 
high-quality candidates among the ones with $P_L > 0.95$. }
We can now look into the predictions of the \Gtwo\ and \Gthr\ \Zhong{to} finalize the sample to visually inspect. 
In Fig. \ref{fig:all_spec} (center) we report the redshift gap between the lens and the source, $\Delta Z=z_{PE}-z_{PG}$ as a function of the lens redshift $z_G$ for the full predictive sample. Here we highlight the objects with $P_L>0.95$, from all the other spectra in the predictive sample. We can distinguish a few features: 1) the upper limit imposed \Zhong{on} the $z_E$ produces a zone of avoidance \Zhong{on} the up-right side of the image; 2) there is a crowded sequence of high $P_L$ in the box defined by $z_G$=[0.5,0.6] and $\Delta z$=[0.4,0.6]. This is due to the presence of \Zhong{rather} redundant residual emission lines from sky subtraction in the SDSS pipeline at $\lambda\sim5600$\AA\ (see Fig. \ref{fig:candida_spec}) that is very often ignored by \Gtwo\ but that in many cases is confused as a real emission. As we will see later, this sequence is easily filtered out by the visual inspection, but it has to be better accounted \Zhong{for in} the training sample \Zhong{to} reduce its impact in future analyses.

A similar effect is produced by the residual sky lines at $\lambda>8000$\AA, which also produce a sequence of spurious $z_E$ predictions (see $z_G\sim0.2$ and $\Delta Z\sim0.9$). These have a small $P_L$, according to \Gone, and thus they do not bother, as they are excluded by the following analysis. 

Overall, the $P_L>0.95$ sample looks rather unbiased, as seen by the $z_G$ estimates from \Gthr\ in the right panel of Fig. \ref{fig:all_spec}, where the predicted $z_{PG}$ is extremely tightly correlated to the eBOSS catalog values (see also the statistical estimators in Table \ref{table:Statistical properties}). 

However, before proceeding with the visual inspection of the background emissions estimated by the \Gtwo, \Zhong{to} minimize the heterogeneity in the human grading, we pre-select the spectra that show an average SNR, computed at the expected positions of the reference lines from Table \ref{table:model parameters}, 
$\langle$SNR$_{\rm lines}\rangle$, 
to be larger than one. 
This further selection gives us 931 potential candidates \Zhong{pass} to the visual inspection.

\subsection{Visual inspection of spectra}
\label{subsec:visual}
The 931 candidates are visually inspected \Zhong{by} the three authors, according to
an ABCD ranking scheme, being A=``sure positive'', B=``maybe positive'', C=``maybe not a positive'' and D=``sure negative''. To combine the human grading with the $P_L$, we have turned the ranking above \Zhong{into} a score according to the conversion A=10, B=7, C=3, D=0 (see also Li+21). We finally select the spectra for which we have obtained an average score $\ge$7, as the final high-quality candidate sample. This is made of 497 objects in total. 

Some spectra of this ``high quality'' sample are plotted in Fig. \ref{fig:candida_spec}. Here we clearly see the emission lines, marked as red vertical lines, from background lensed star-forming galaxies. 

During the visual inspection process, besides grading, we also check that 
the predicted \Zhong{values}, $z_{PE}$ and $z_{PG}$\Zhong{, given} by GaSNets\Zhong{, are} perfectly aligned with visible spectral features. This is not often the case as the prediction process has some intrinsic uncertainty. For instance, the two GaSNets need to interpolate across a grid of training spectra that have been shifted with a coarse sampling (i.e. 0.05 in redshift, see Sect. \ref{sec:neg sample}). However, other sources of errors are possibly causing even more significant shifts, as we will discuss in more detail in Sect. \ref{sec:Slightly shift}. Using an interactive GUI developed by one of us (ZF), we then determine by eye the needed shift to obtain a perfect visual alignment and a ``corrected'' redshift for 
the $z_{PE}$, assuming the $z_{G}$ from the eBOSS catalog as an unbiased estimate of the main galaxy redshift.

\begin{figure}
\vspace{-0.3cm}
    \centering 
        \includegraphics[width=1\linewidth]{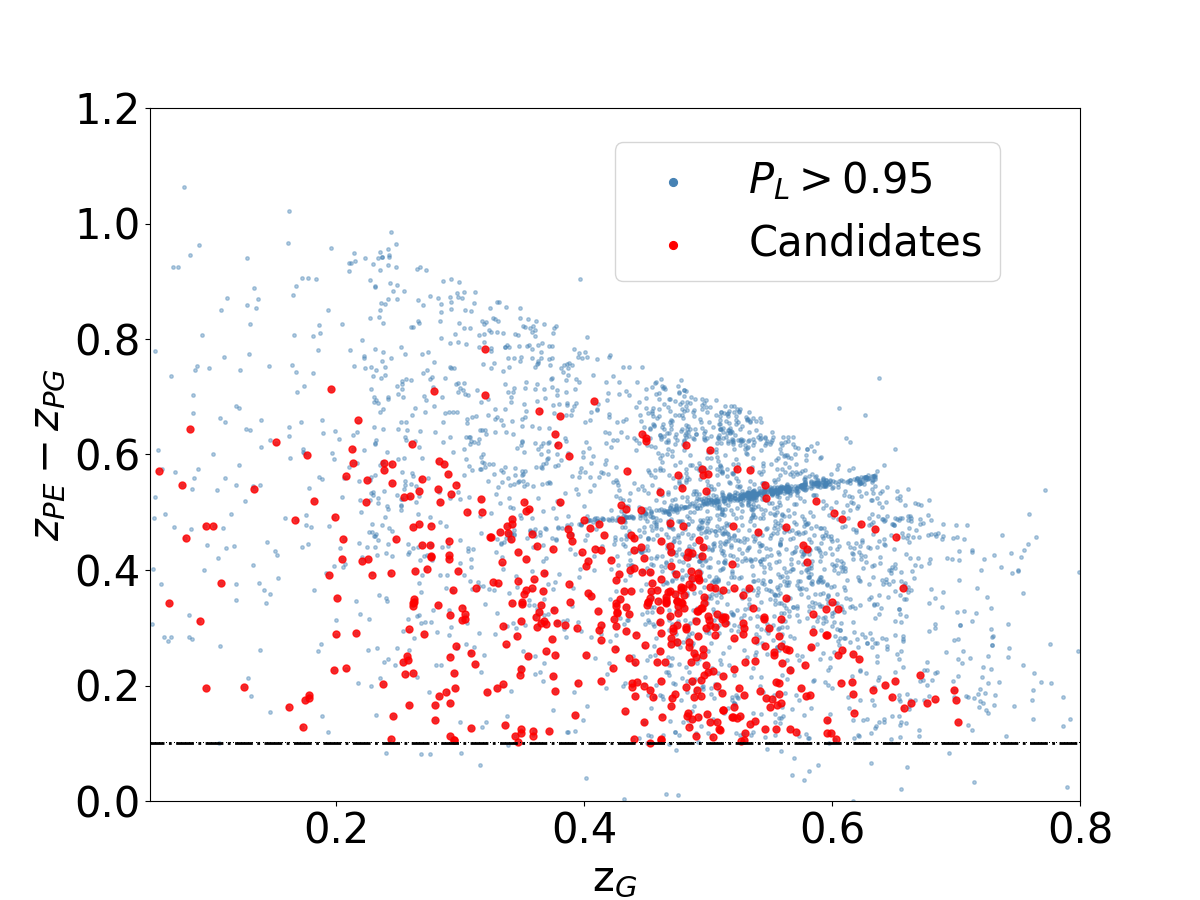}
    \caption{$ z_{PE} -  z_{PG}$ vs. $z_{PG}$ distribution of visual inspection 497 good potential candidates and all $P_L>0.95$ spectra}
    \label{fig:candidate_distr}
\end{figure}

Finally, to qualify a spectrum as a 
lensed galaxies \Zhong{candidate} we check that 1) the emission lines do not belong to the sky lines (green lines in Fig. \ref{fig:candida_spec}) and 2) that the identified emission lines, i.e. red lines in Fig. \ref{fig:candida_spec}, having redshift $z_{PE}$ from \Gtwo, do not correspond to any line from the galaxy (i.e. blue lines in Fig. \ref{fig:candida_spec} at redshift $z_{PG}$ from \Gthr). In other words, the $\Delta Z=z_{PE}-z_{PG}$ has to be larger than 0.1, as shown in Fig. \ref{fig:candidate_distr}, where it is plotted as a function of the estimated $z_{PG}$.
Here we also see that the $\Delta z$ 
is decreasing with the $z_{PG}$ because the further the lenses, the smaller the difference in redshift with the background source.
From Fig. \ref{fig:candidate_distr} it is clear that this is mainly 
a selection effect due to our condition \Zhong{on} the $z_{PE}<1.2$, however, since the \Zhong{high-quality} candidates do not cluster toward the upper bound of the zone of avoidance, we conclude that the candidate distribution becomes incomplete when the $z_{PE}\sim1.2$. This is consistent with the correlation of the low $P_L$ with the higher-$z_{PE}$ we have discussed in \S\ref{sec: test_simul}. An encouraging feature, in the same figure, is that the combination of the $\langle$SNR$_{\rm lines}\rangle>1$ and the visual inspection, allows us to drop the stripe of spurious detection from residual sky lines discussed in \S\ref{sec:DR16data}. 


\subsection{Deep learning vs. traditional methods}
\label{sec:T+21}
We end this section by comparing our HQ catalog, based on deep learning, with the catalog of 1551 candidates selected with the \Zhong{rest-frame} optical bands from T+21, using traditional selection methods. They used the complete eBOSS/DR16 database and applied the standard spectroscopic detection method introduced in the eBOSS Emission-Line Lens Survey (BELLS) and added Gaussian fit information, grading, additional inspection observables, and additional inspection methods to improve the BELLS selection method. 
They used a total of 2 million objects with no selection on the redshift of the lenses.
Furthermore, they used a larger database of reference lines, including also [NII]a/b and [SII]a/b: these are best suited for low-redshift detections being all placed at $\lambda>6500$\AA, leaving the only [OII] doublet as a feature for the identification of background sources at $z\gsim1.2$.
As such, their predictive sample is wider in the parameter space than the DR16-predictive we have adopted. For a proper comparison, we have selected the T+21 candidates that fall in the \Gs\ predictive space (i.e. $z_{G}=0.05-0.8$, spectra SNR$>2$, $z_E\lsim1.2$, $z_E=z_G+0.1$) and finally obtain 778 ``compatible'' candidates ($\sim 50\%$ of the original sample). We have checked the excluded 773 and found that 739 detections are, indeed, based on a single line (generally in spectra with SNR$>2$) and 29/5 are based on 2/3 lines (all with spectra SNR$<2$), according to the T+21 catalog. Hence, the majority of these ``known candidates'' would have been missed anyways in our HQ catalog because of the conservative selection in the number of lines to use for the classification, either in the deep learning training or visual ranking.   

We have, then, matched the compatible 778 candidates with our HQ sample of 497 entries and, surprisingly, we have found a match for only 68 \Zhong{objects.}

The positive note is that {\it \Gs\ have found $\sim 430$ new HQ candidates that have been missed by standard techniques}. The negative note is that the \Gs\ seem to have missed $710$ candidates from \Zhong{T+21.}

{Is this true?} To answer this question we need to first check how many of these objects are lost by the \Gs\ according to the criteria imposed \Zhong{on} their outputs, i.e. they do not fall in the criteria $P_L>0.95$ and $z_E-z_G>0.1$. These are 327, i.e. $42\%$ of the compatible sample. This is larger than the fraction of lost objects found in the test against the real systems in \S\ref{sec:real_data} (i.e. $100-69=31$\% of ``candidates'' and $100-80=20$\% confirmed ones, having $P_L<0.95$). One explanation of this excess of lost objects with low $P_L$ can be that these are mainly optimistic candidates in T+21, for which the \Gs\ have given low reliability. To confirm this we have checked that 215/327 are single line detections, according to T+21, \Zhong{and} only 87/327 have scored A+ or A in their check against low-resolution imaging\footnote{As we will comment later, the image quality of the low-resolution DES imaging used by T+21 does not consent a firm classification, except for very clear features. Hence, we have conservatively assumed the A+ and A scores sufficient to preliminary quantify the confirmation rate.}. Hence, we can fairly conclude that this sample of lost candidates is overall low-valuable, having a tiny (albeit insecure) confirmation rate. This also implies that the fraction of lost SGL ``real'' events in our catalog is in line with the one estimated in \S\ref{sec:real_data}, reported above (i.e. 20\%). 

Going to the remaining lost candidates ($710-327=383$), in Fig. \ref{fig:missed_plot} we show the spectra (not line) SNR 
vs. the estimated redshift of the background lines from T+21, color-coded by the number of detected lines. 
From this figure, we observe that:
\begin{figure}
\vspace{+0.3cm}
    \centering 
        \includegraphics[width=1.05\linewidth]{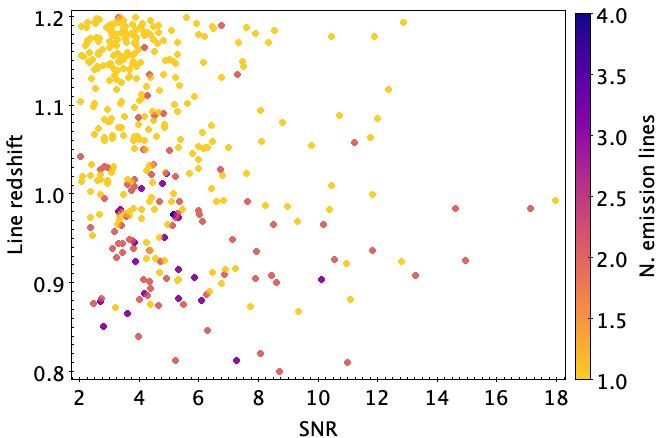}
    \caption{Sample of missing candidates in the HQ catalog from \Gs + visual inspection. In this \Zhong{figure,} we show the distribution of the missing candidates in the parameter space adopted for the training of the \Gs\ (i.e. spectra SNR$>2$ and $z_E<1.2$). Each candidate is \Zhong{color-coded} by the number of detected lines in their spectra (according to T+21). Most of the missed candidates are 1-line and did not qualify in our HQ sample.}
    \label{fig:missed_plot}
\end{figure}

\begin{figure*}
\vspace{+0.3cm}
    \centering 
        \includegraphics[width=0.895\linewidth]{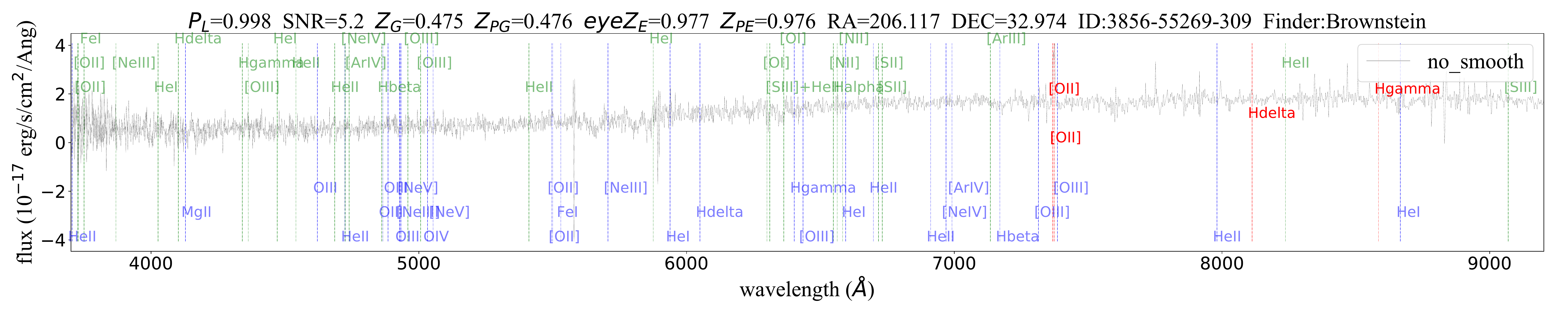}
        \includegraphics[width=0.9\linewidth]{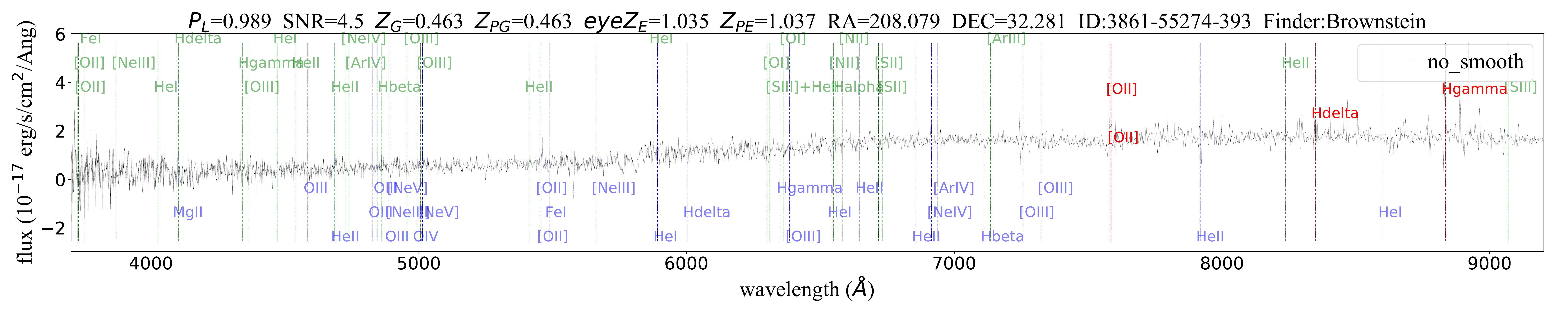}
        \includegraphics[width=0.91\linewidth]{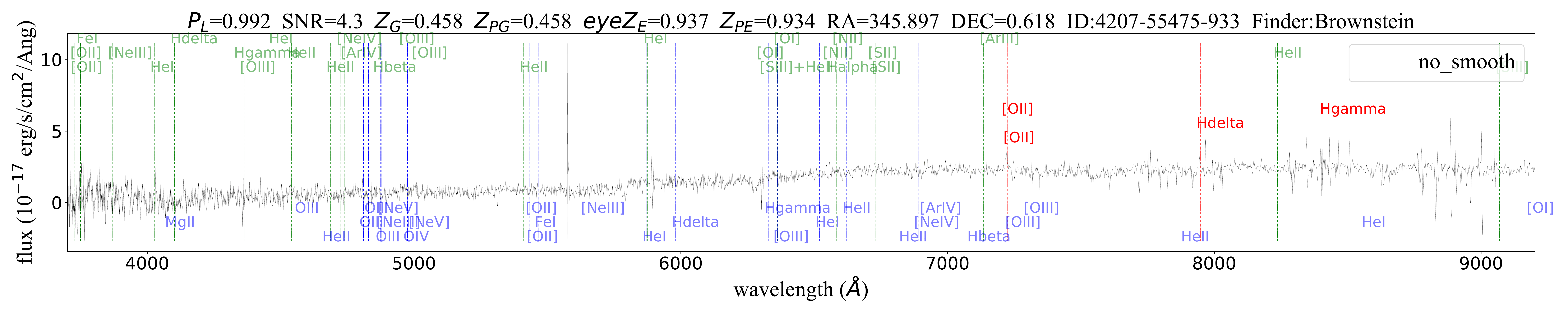}
        \includegraphics[width=0.92\linewidth]{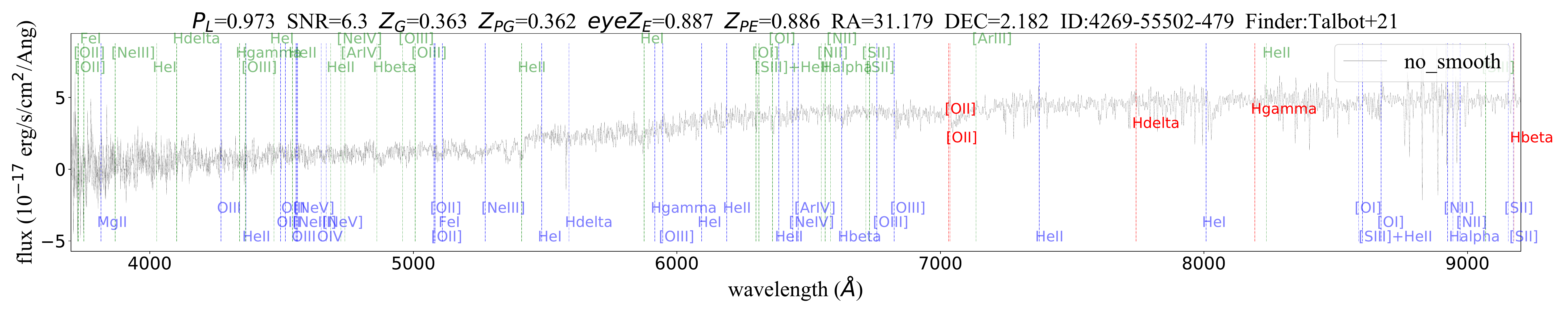}
    \caption{Sample of missing candidates in the HQ catalog from \Gs\ + visual inspection and found in T+21. The red vertical lines are the features identified as multi-lines in T+21, but that have been excluded by us because either too faint or embedded in noisy regions, making them poorly reliable to qualify as HQ candidates. 
}
    \label{fig:missed_spec}
\end{figure*}

1) The majority (286/383, i.e. 75\%) of the missing candidates have 1-line detection, thus they are lost from our HQ catalog because we excluded them in our filtering (both because of the $\langle$SNR$_{\rm lines}\rangle$ or the visual inspection, see \S\ref{sec:DR16data} and \ref{subsec:visual}). According to the T+21 low-resolution grading, 164/286 of the 1-line detections have A or A+ scores, which implies a rather large confirmation rate, $\sim$60\%, if confirmed by higher-quality imaging. This is a sample we can easily intercept with \Gs, by simply releasing the conservative criterion of the 1-line. From Fig. \ref{fig:missed_plot}, we see that above $z~1.05$ we \Zhong{lose} some 2-line candidates, which supports further the conclusion in \S\ref{subsec:visual} that we are incomplete at $z_E\lsim1.2$.

2) The remaining 97 multi-line objects, in Fig. \ref{fig:missed_plot}, majorly concern us, as according to their $P_L$ and number of lines should have been picked by the \Gs\ + visual inspection. First, we have found 10/97 objects classified as quasar or unknown in DR16, so these could not be in our catalog. For all the other 87 we have visually inspected the spectra and found that despite they \Zhong{being} classified as multi-lines in T+21, no line, except the [OII] doublet, had an acceptable SNR. Hence, these are all candidates that have been substantially treated as 1-line from us or given a rather poor visual grade. We give some examples of these spectra in Fig. \ref{fig:missed_spec}. Since 60/97 have received A or A+ scores 
from the low-resolution confirmation in T+21, i.e. 60\%, this is a sample that is likely to be valuable and should not be missed.
However, we need to point out that this sample was not lost by the \Gs\ but by human selection.

\subsection{First catalog of new HQ strong lensing candidates in eBOSS from Deep Learning}
\label{sec:HQ_catalog}
After having subtracted the 68 candidates already found in T+21, we obtain a final catalog of 429 new HQ candidates in eBOSS, the first fully derived using deep learning.
The full catalog 
is reported in Appendix \ref{appendix}. 
This includes information about 1) RA/DEC coordinates; 2) plate ID; 3) MJD (Modified Julian Day), the observation date; 4) the \Gone\ probability, $P_L$; 5) the redshift of the galaxy from the eBOSS catalog; 6) the predicted redshift of the galaxy from \Gthr; 7) the predicted redshift of the background source from \Gtwo; \Zhong{8)} the corrected redshift of the source from the visual inspection (see Sect. \ref{sec:Slightly shift}); the total probability, $P_T=P_L\times0.1$ visual scores, i.e. combining the \Gs\ and human probabilities to be a lens. 

\section{Discussion}
In the previous section, we have presented the final list of 429 new strong galaxy lensing candidates, obtained by applying the three \Gs\ to the latest eBOSS database (DR16), and further cleaning the sample via visual inspection. 
\begin{figure*}
    \centering 
    \includegraphics[width=0.8\linewidth]{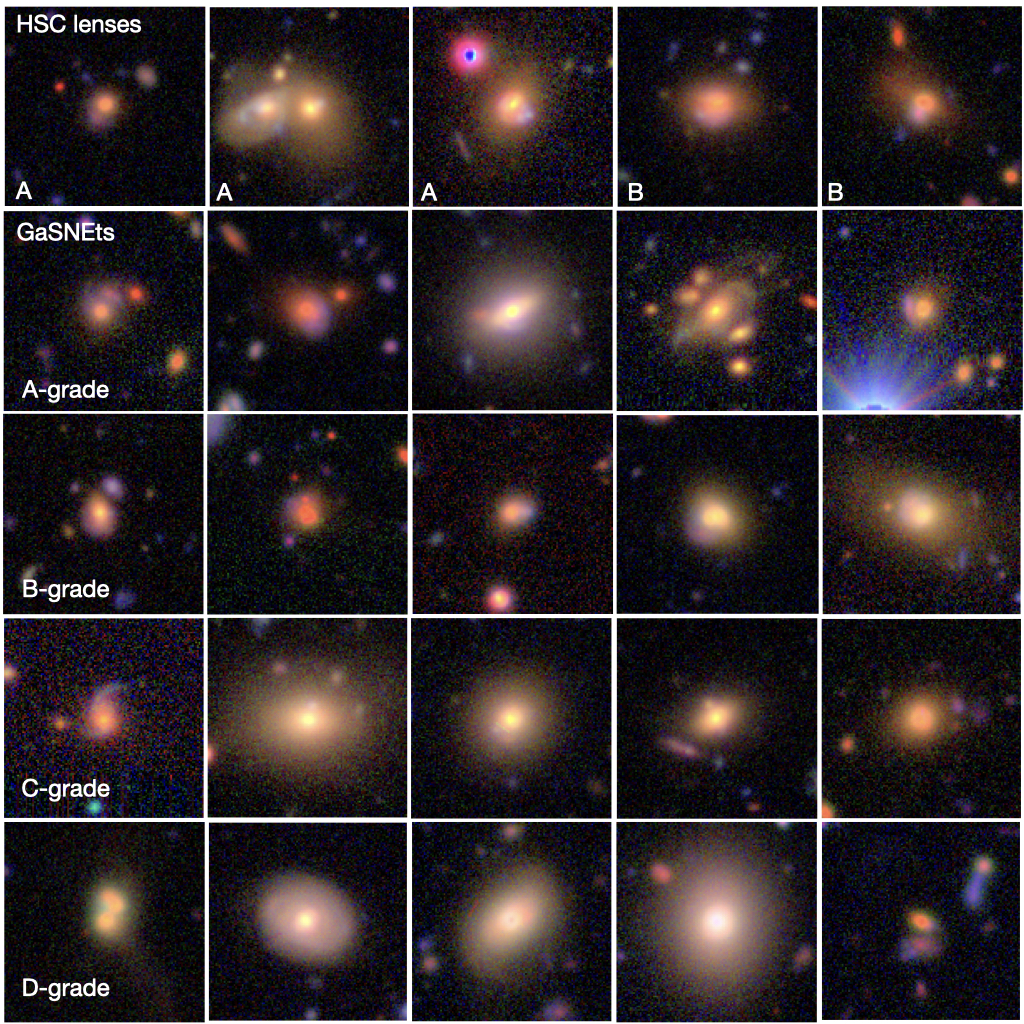}
    \caption{Some examples of \Zhong{ground-based} color cutouts ($20'' \times 20''$) of GaSNet candidates. Top row: match with HSC known candidates, re-graded as in the bottom-left corner. \Gs\ have found them as HQ candidates independently.
    Bottom 4 rows: A, B, C and D ranked HSC counterparts of GaSNet candidates from the HQ sample in \S\ref{sec:HQ_catalog}.}
    \label{fig:HSC_match}
\end{figure*}
Strictly speaking, the \Gs' candidates consist of systems where, in the spectrum of a foreground galaxy, 
we have found emission lines that are incompatible with belonging to the same galaxy. We have assumed, so far, that all these lines come from lensing events. In reality, they can be emitted by other kinds of sources, like overlapping galaxies along the line of sight, outflows in late-type galaxies, interacting systems, etc., although we have set a redshift gap, $\Delta z$, that might have prevented the confusion with some ``local'' phenomena. Hence, to fully assess the new catalog, we need to estimate a fiducial confirmation rate based on space observations or high-quality \Zhong{ground-based} imaging. Such a confirmation rate is important 1) to compare with the one from standard techniques, to see whether Deep Learning can outperform them in terms of reliability of the candidates; 2) to check whether the large spectroscopically selected samples accumulated \Zhong{so far}, are compatible with expected numbers of SGL events from theoretical predictions (see e.g. \ref{sec:challange}), or we might expect to find more events with more refined tools.

Besides the confirmation rate, in this section, we also discuss
the possibility to use the \Gtwo\ and \Gthr\ as automatic tools for redshift estimates and spectra classification.
We will conclude this discussion with some perspective \Zhong{on} the next improvements of the \Gs.

\subsection{Confirmation rate via ground based imaging}
\label{sec:confir_rate}
To properly derive a fiducial confirmation rate for the 429 HQ candidates in \S\ref{sec:HQ_catalog}, we have checked the HST archive observations to look for serendipitous matches with our newly discovered candidates \Zhong{but} found no matches. Hence, the only remaining check we can perform is inside archive observations from the ground.
There are three datasets potentially useful for the test: 1) DECaLS\footnote{https://portal.nersc.gov/cfs/cosmo/data/legacysurvey/dr7/};
2) KiDS\footnote{https://kids.strw.leidenuniv.nl/DR4/access.php} and 3) HSC\footnote{https://hsc-release.mtk.nao.ac.jp/das\_cutout/pdr3/}. We have found 279 matches with DECaLS, 16 with KiDS\Zhong{, and} 63 with HSC, however: 1) the quality of the DECaLS {\it grz} color images from the public data is rather poorer than other surveys and made the identification of the lensing features extremely uncertain (see Appendix \ref{app:B}); 2) the number of KiDS matches is too small to have a fair statistics and we decided to leave the few convincing candidates for future analyses; 3) the HSC sample is the one \Zhong{with} sufficient large statistics, image quality\Zhong{, and} uniformity to make a fair estimate of the fraction of convincing lenses without strong biases.

Looking \Zhong{at} this latter sample, we find that 7 candidates have corrupted color images or are too close to some bright source to be used with sufficient confidence. Hence, we finally inspect 56 systems. Of these, our HQ candidates match 8 known lens candidates from HSC \footnote{http://www-utap.phys.s.u-tokyo.ac.jp/~oguri/sugohi/} (e.g., \citealt{2018PASJ...70S..29S, 2019A&A...630A..71S}), although they are all C-graded by the imaging only in their catalogs.
We have visually inspected them again and, applying the ABCD scheme as in \S\ref{subsec:visual} and taking into account the spectroscopic evidence, we have reclassified 3 of them with A-grade and 5 with B-grade.

Of the remaining 49 matches,
we have classified 7 candidates
as A-grade and 17 as B-grade systems. 
Taking the A-grade as {\it bona fide} confirmed lenses and weighting the B-grade ones by a 0.5 factor to account that they \Zhong{may be} not lenses, 
we conclude that the lens confirmation rate is 21/56 or 38\%, which is lower than the confirmation rate estimated in \S\ref{sec:real_data} using space imaging. 

In Fig. \ref{fig:HSC_match} we show a gallery of the ``confirmed'' lens and, as a comparison, the ``unconfirmed'' ones (i.e. the ones C- and D-graded). In the first row, {we report some of the lenses previously found in the HSC imaging and confirmed and re-graded by us,} in the second and third rows some examples of new \Gs' confirmed lenses with A-grade, 
and B-grade, respectively. In the final two rows the unconfirmed C and D cases. \Zhong{These clearly show} the variety of potential contaminants, including arc-like features of unclear nature, blue/faint background galaxies similar to other objects in the field-of-view, interacting systems\Zhong{, and} large late-type or lenticular galaxies. In these latter examples, especially the large galaxies, if we exclude the cases where it is likely that the background emissions found in the spectra come from unlensed faint background systems as they can be seen in field-of-view, it is difficult to identify any other potential high-$z$ emitters. This leaves the nature of these emissions unresolved. In principle we cannot exclude that, given the small area covered by the fibers in eBOSS ($2''$, see also Fig. \ref{fig:HSC_match}) there is some very low separation arc, embedded in the bright foreground galaxy light, remaining undetected in the seeing-confused images from HSC. In this case, we can argue that the confirmation rates estimated above (38\%) might represent a lower limit.

If this conclusion is correct, we can attempt to derive a prediction of the total number of true SGL events in eBOSS, based on the current candidates from T+21 and this work. Put together they are 1551+429=1980. Assuming a pessimistic confirmation rate of 38\%, they make 
$752$
real SGL events, {while for a more optimistic $46\%$ conformation rate of SLACS+BELLS+S4TM, it makes $911$ real SGL.} If we add the other candidates found in BOSS from BELLS (25) and BELLS GALLERY (17\footnote{Note that more can be still found on their sample of remaining 155 candidates remaining unconfirmed. Assuming $\sim 50\%$ confirmation rate they can be $\sim70$.}) we reach 794 and 953 real SGL, which nicely bracket the expected number we have estimated in \S\ref{sec:challange} for BOSS ($\sim920$). This suggests that we have possibly reached the full completeness of the lens \Zhong{population accessible by} the largest spectroscopic database currently available.

\subsection{Statistical errors of \Gtwo\ and \Gthr}
\label{sec:Slightly shift}

\Gtwo\ and \Gthr\ are two CNNs that can perform the generic task to estimate the redshift of given features in 1D spectra. As such, they can be applied to spectroscopic databases regardless of the specific task of looking for strong gravitational lenses. 

Certainly, the search for lenses requires a much lower accuracy in the $z_{PG}$ and $z_{PE}$, because the only condition to ring the bell for potential events is $\Delta z = z_{PE}-z_{PG}>0.1$, which is rather higher than typical spectroscopic redshift errors based on the human measurements. However, this condition is physically meaningful if 
$\Delta z$ is larger than the combination of the typical errors on $z_{PE}$ from \Gtwo\ and $z_{PG}$ from \Gthr, which also include the uncertainties that a deep learning process might introduce (activation, loss, training, etc.). 

Hence, if on one hand, the assessment of the ``bias'' and typical ``statistical errors'' of the two \Gs\ (L2 and L3) is needed to validate the pre-condition for the HQ candidates, on the other hand, they can also  quantify 
the accuracy of the individual CNN as ``automatic tools'' for redshift measurements. In this latter case, we can possibly require the typical errors to be of the order of $<1\%$, and systematics smaller than this precision. At the same time, we should expect a negligible fraction of outliers/catastrophic events.

\begin{figure}
    \centering 
            \includegraphics[width=1\linewidth]{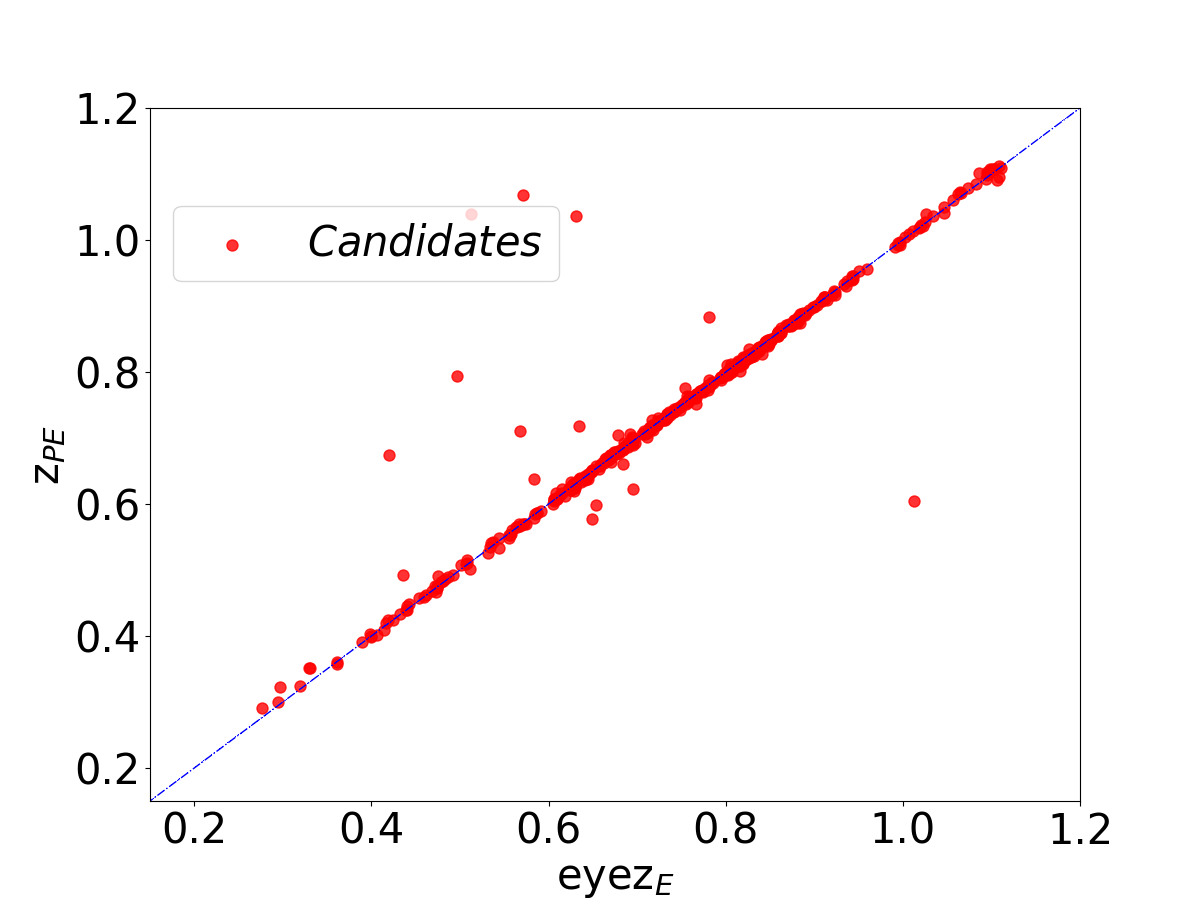}
    \caption{$Eye z_E$ vs. $z_{PE}$ and $eye z_E$, $z_{PE}$ distribution of visual inspected 429 HQ candidates.}
\label{fig:ZE_accuracy}
\end{figure}

As mentioned in \S\ref{subsec:visual}, during the visual inspection we had the chance to check the accuracy of the $z_{PE}$ estimates 
and correct them by hand. This process is not error-free itself, as the resulting $eyez_E$ is a combination of a subjective identification of the line center and the accuracy in the line alignment by eye.
However, we can confidently use these corrections, together with the nominal $z_G$ given in the SDSS catalogs, to compute the scatter of the \Gtwo\ and \Gthr\ predictions and derive systematics and statistical errors for $z_{PE}$ and $z_{PG}$.
For the $z_{PG}$ we can use all the galaxies in the predictive catalog as shown in  Fig. \ref{fig:all_spec} (right), for which $z_G$ is known, to determine the $\delta _{z_{PG}}=z_{PG}-z_{G}$. 
The scatter in this case is $\sigma(\delta_{z_{PG}})=0.015$\Zhong{. The} outliers, defined as the spectra for which the $(|z_{PG}-z_{G}|)/(1+z_{G})>$0.15, are about 0.113\%.

For the $z_{PE}$ we can use the spectra that we have visually inspected and for which we have collected the average $eyez_E$ estimated by the three of us. These are shown against the $z_{PE}$ in Fig. \ref{fig:ZE_accuracy} and used to estimate the $\delta_{z_{PE}}=z_{PE}-eyez_{E}$. In this case, we have estimated the scatter $\sigma(\delta_{z_{PE}})=0.046$, while the outliers are about 1.21\%.

In both cases, the scatter and accuracy are reasonably good, and so it is the outlier fraction. This result confirms that the adoption of the $\Delta z>0.1$ is conservative enough to account for the nominal statistical errors of the predicted redshifts. Furthermore, if we consider that the SNR is generally poor for the majority of the emission lines of the background galaxies, then we believe that both \Gs\ (L2 and L3) are a very promising start and can be possibly be already used to automatically provide a first accurate guess of the redshift of galaxies in large surveys, while a more dedicate training would possibly improve the overall accuracy. We will dedicate future analyses to \Zhong{} the \Gs\ on this latter and more applications, including specialized tasks for spectra classification (e.g. starburst galaxies, active galactic nuclei, irregular systems\Zhong{, etc.}).




\begin{figure}
    \centering 
    \vspace{+0.5cm}
    \includegraphics[width=1\linewidth]{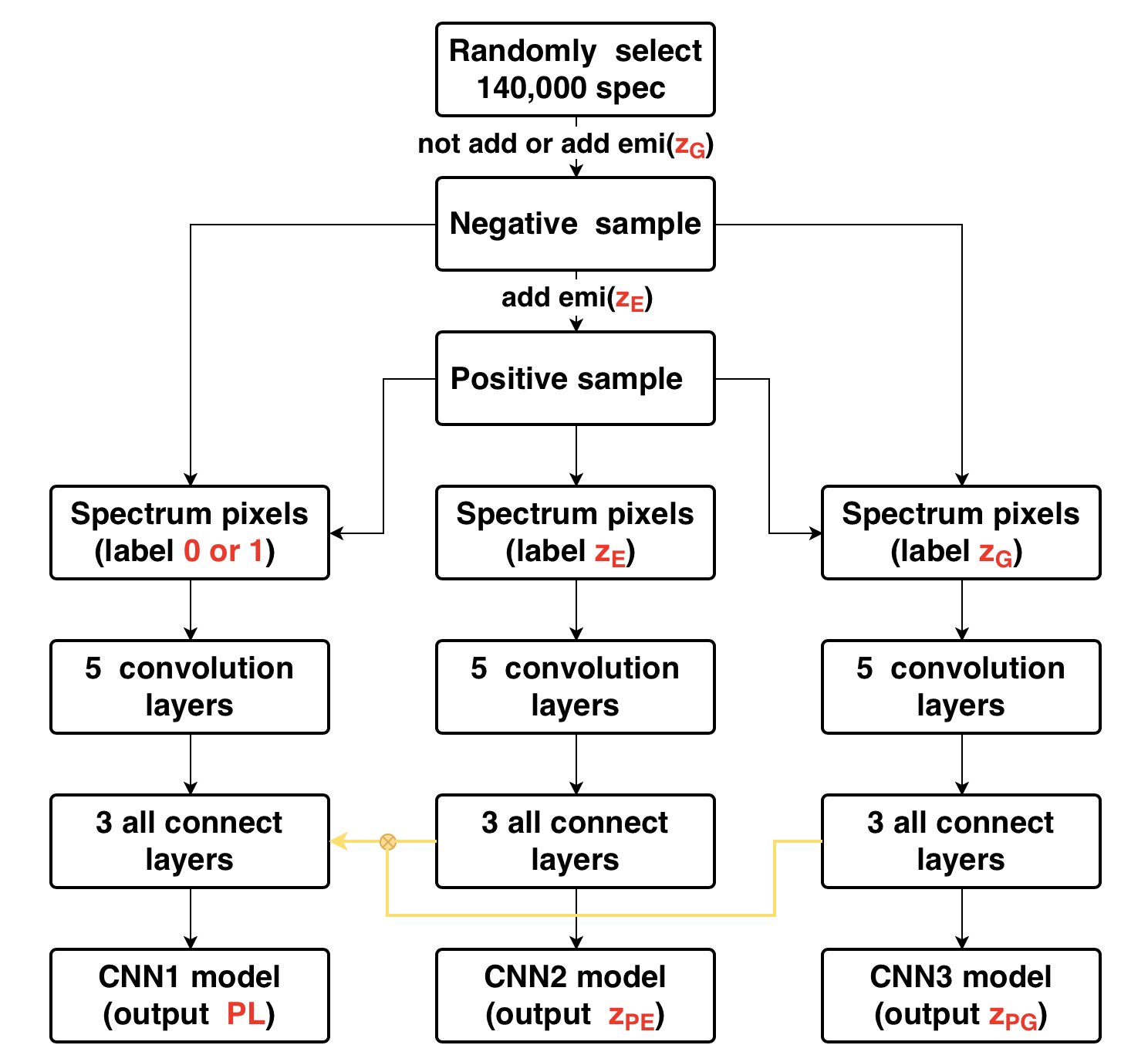}
     \caption{A possible scheme for increasing the interplay between the three \Gs. Here, we foresee \Zhong{input} \Gone\ with the outputs of \Gtwo\ and -L3, as shown \Zhong{in} the yellow line show, to improve the $P_L$.}
\label{fig:new_CNN_model}
\end{figure}

\subsection{Improvements of CNN model}
In this work we have used 
3 independent CNN models and combined their outputs according to some physically meaningful conditions (see Fig. \ref{whole_CNN_model}), to identify strong lens candidates.
In fact, because of 
the physics of the SGL, which involves the position of the source and lens with respect to the observer, the properties of the projected potential, etc., the 3 outputs of the \Gs\ are not fully independent. Rather, 
they must be connected via the ray-tracing equation of the SGL. For instance, one can define a more meaningful probability for spectra to have caught a lens candidate, \Zhong{by looking} at the relative distance of the $z_{PG}$ and $z_{PE}$, or at the absolute value of $z_{PG}$ (e.g. giving a lower $P_L$ if the galaxy is a very low redshift)\Zhong{, etc}. 
One possible future development is to  
connect different individual CNN networks (just like the neurons in our brain), for example, as in Fig. \ref{fig:new_CNN_model}, to make a more educated probability for a spectrum to be an SGL system.

In this figure, we 
suggest \Zhong{using} the prediction of $z_{PG}$ as conditional information for the prediction of $P_L$, \Zhong{then} using the prediction of $z_{PG}$ and $P_L$ as auxiliary information for the prediction of $z_E$.
If, on one hand, this architecture can help to improve the accuracy, the cost to pay is
the model complexity, including a larger correlation among the different branches with some large back-propagation. This would make the overall model more \Zhong{time-consuming} in terms of training and prediction, but likely more accurate and false detection free.

\section{conclusions}

In this paper, we have presented a novel deep learning tool to search for strong gravitational lensing (SGL) events in 1D galaxy spectra. This is the first attempt to use multiple emission lines after Li+19 used Ly$\alpha$ only.

The new algorithm is made of different CNNs, dubbed Galaxy Spectra convolutional neural Networks (GaSNets). These are optimized to work together to provide SGL candidates, but can also perform classification and regression tasks independently. As such, they are extremely suitable for further applications in large databases of tens to hundreds of millions of spectra, \Zhong{like} the ones expected from the next generation spectroscopic surveys (4MOST, DESI, EUCLID, CSST). 

In this paper, we have started by applying these new tools to \Zhong{the} strong lensing search in the eBOSS/DR16 database (\citealt{2020ApJS..249....3A}). To this aim we have introduced: 1) \Gone\ giving to each eBOSS spectrum the probability to be an SGL event ($P_L$); 2) \Gtwo\ estimating the redshift of background sources ($z_{PE}$) from a series of pre-selected emission lines (see Table \ref{table:model parameters}); and 3) \Gthr\ estimating the redshift of the galaxy itself ($z_{PG}$), using the information it learns from the continuous spectrum, including local absorption/emission features. Only working together, the three \Gs\ efficiently pinpoint SGL candidates combining a high $P_L$ with the condition that the $z_{PE}>z_{PG}$, as expected for typical strong lensing configurations.

In particular, by testing the GaSNets on a list of known spectroscopically selected gravitational lenses in SDSS/BOSS (from \citealt{2008ApJ...682..964B}, \citealt{2012ApJ...744...41B}\Zhong{, and} \citealt{2017ApJ...851...48S}) we have found that using a $P_L>0.95$ we can recover about 80\% of the strong lenses confirmed by HST. This very conservative probability threshold provided a reasonable trade-off between \Zhong{significant} completeness and a reasonably small sample to visually inspect, 
with low contamination from false-positive detection. 

Using this set-up, with the condition that $z_{PE}>z_{PG}+0.1$, we have applied the \Gs\ to $\sim 1.3$ \Zhong{million} spectra from the SDSS-DR16, after having imposed some appropriate cuts to guarantee a good spectrum quality and the visibility of at least two emission lines from the putative sources (namely, [OII] and $H_\gamma$), assumed to be star-forming galaxies.   

We have collected $\sim 930$ candidates that have been further cleaned by misclassified SGL events, via visual inspection. The final sample of visual HQ candidates is made of 497 spectroscopic selected objects. This catalog has been {\it a posteriori} compared to the most extended catalog of spectroscopic selected lens candidates from 
\Zhong{T+21} and found an overlap of only 68 candidates, meaning that 429 of our candidates are newly found. On the other hand, we have demonstrated that \Gs\ did not recover the remaining T+21 sample because of the conservative constraints we have adopted for the number of lines to be detected ($>2$). Releasing them, half of the sample from T+21 (i.e. the one for which \Gs\ has $P_L>0.95$) remains under the \Gs\ discovery reach.

For the new HQ \Zhong{catalog,} we provide RA, DEC, the probability, $P_L$, the redshift of the galaxy from the eBOSS catalog, the predicted redshift of the galaxy from \Gthr, the predicted redshift of the background source from \Gtwo, the corrected redshift of the source from the visual inspection, in Appendix \ref{appendix}.

To estimate a tentative confirmation rate of these candidates, we have matched the coordinates with archive HST observations and found no matches. Instead, we have found optical counterparts in DECaLS, KiDS\Zhong{, and} HST observations, but only HSC has provided sufficient statistics and image quality to confidently confirm the first sample of \Gs' candidates. Among these, we have independently confirmed 8 SGL candidates from previous HSC lens imaging searches, thus providing spectroscopic evidence of lensing events, even though for only 3 of them we have found convincing features in the imaging to be ``sure lens''.
Besides these ``known'' lenses, we have found a preliminary optical confirmation of \Zhong{a further} 24 GaSNet HQ candidates, although, also in this case, for 17 of them the HSC images allowed only  a ``maybe lens'' B-grade\Zhong{, and} only 7 have a ``sure lens'' A-grade. {Taking the A-grade as {\it bona fide} lenses and giving a 0.5 weight to the B-grade candidates,
we have estimated a confirmation rate of $38\%$ for our HQ catalog.}

Some examples of the HSC matched are shown in Fig. \ref{fig:HSC_match}, where we also show low-graded imaging of GaSNet candidates. The possible contaminants are higher redshift galaxies, overlapping in the fiber spectra, or maybe local \Zhong{phenomena}  mimicking an SGL event. \Zhong{For example, local gas outflows, with typical velocities of $3 \times 10^3$ kms$^{-1}$, will introduce asymmetric velocity distribution along the ejection direction  \citep{2020A&ARv..28....2V}, which would shift the wavelength of some characteristic emission lines. Among these the [OIII] line could deviate from the $z_G$ by $\sim 50$\AA\ and produce a false positive.}

In this paper, we have demonstrated that Deep Learning represents a very efficient method to search for strong lenses in galaxy spectra.
This can be applied to next generation spectroscopic surveys in a fast and automated way. This first application to the eBOSS database has confirmed that the spectroscopic selection of SGL candidates is complementary to the \Zhong{imaging-based} SGL searches. For instance, of the 32 A/B grade candidates from the \Gs\ matching with HSC imaging, only 8 were found previously on HSC images. This over-performance of the spectroscopic searches with respect to imaging is particularly evident for \Zhong{ground-based} observations, where the typical seeing has no impact on emission lines of background sources in spectra \Zhong{but} makes it hard to resolve low-separation gravitational arcs of the same sources.  

For this first application, we have made conservative choices \Zhong{regarding} 1) the number of features to use for the training of the \Gs; 2) the overall Network architecture, e.g. limiting the interconnections between the three \Gs\; 3) the probability threshold to optimize the sample to visual inspect and keep the false positive under control. These are all directions to consider for future improvements. 
As a final positive note, we have discussed that the \Gthr, in particular, has reached an accuracy and scatter of its predictions, \Zhong{sufficient} to be used to automatically measure galaxy redshifts in large spectroscopic surveys.


\section*{Acknowledgements}
We thank Dr. C. Tortora and Dr. Y. Shu for useful comments \Zhong{on} the manuscripts. RL \Zhong{acknowledges} the science research grants from the China Manned Space Project (No CMS-CSST-2021-B01,CMS-CSST-2021-A01). NRN \Zhong{acknowledges} financial support from the “One hundred top talent program of Sun Yat-sen University” grant N. 71000-18841229.


\section*{Data Availability}
The data that support the findings of this study are available at the URLs provided in the text and the Table in Appendix A. All other data that are not provided in the paper can be requested \Zhong{from} the authors.    



\bibliographystyle{raa}
\bibliography{references} 


\onecolumn
\appendix
\section{HQ catalog from \Gs}
\label{appendix}
In this appendix we list the HQ candidates obtained with the \Gs\ and described in \S\ref{sec:applying_BOSS}. The table content is listed in \S\ref{sec:HQ_catalog}.
\begin{center}

\renewcommand\arraystretch{0.925} 

\begin{tabular}{cccccccccccc}
\toprule
\multicolumn{12}{c}{New HQ candidates from \Gs\ }\\
\toprule
 & RA & DEC & PLATE & MJD & FIBER & $P_L$ & z & $z_{PG}$ & $z_{PE}$ & $mean_{eyeZE}$ & $P_T$ \\ 
 \hline
1&27.6755&-7.1667&7164&56597&275&1.0&0.126&0.125&0.322&0.297&1.0 \\
2&356.6865&34.8954&7143&56572&678&1.0&0.095&0.096&0.291&0.276&1.0 \\
3&124.173&18.6976&4486&55588&218&1.0&0.221&0.221&0.636&0.641&1.0 \\
4&139.3617&31.157&5808&56325&434&1.0&0.239&0.238&0.823&0.82&1.0 \\
5&205.2421&18.6312&5862&56045&110&1.0&0.244&0.244&0.638&0.645&1.0 \\
6&233.5735&3.4663&4805&55715&326&1.0&0.291&0.288&0.707&0.708&1.0 \\
7&22.0923&22.3834&5107&55940&409&1.0&0.277&0.275&0.7&0.695&1.0 \\
8&356.4175&-2.1613&4356&55829&729&1.0&0.294&0.294&0.659&0.658&1.0 \\
9&9.4789&17.8124&6193&56237&298&1.0&0.298&0.297&0.695&0.695&1.0 \\
10&4.8287&26.2703&6276&56269&26&1.0&0.262&0.262&0.6&0.604&1.0 \\
11&244.2444&59.9424&6976&56448&830&1.0&0.257&0.258&0.509&0.507&1.0 \\
12&221.3814&60.9204&6982&56444&516&1.0&0.258&0.257&0.501&0.511&1.0 \\
13&228.0721&14.4225&5486&56030&270&1.0&0.327&0.327&0.706&0.708&1.0 \\
14&143.3092&22.088&5770&56014&842&1.0&0.35&0.35&0.694&0.694&1.0 \\
15&352.9407&29.3635&6581&56540&976&1.0&0.313&0.312&0.68&0.678&1.0 \\
16&202.7383&47.6651&6743&56385&86&1.0&0.349&0.35&0.662&0.662&1.0 \\
17&145.3751&1.865&4736&55631&264&1.0&0.353&0.354&0.719&0.722&1.0 \\
18&212.4583&58.3783&6804&56447&957&1.0&0.36&0.36&0.744&0.742&1.0 \\
19&189.6944&51.0158&6674&56416&262&1.0&0.388&0.387&0.762&0.766&1.0 \\
20&358.2568&32.3241&6498&56565&550&1.0&0.367&0.366&0.729&0.731&1.0 \\
21&215.0974&30.7816&3866&55623&45&1.0&0.466&0.47&0.813&0.814&1.0 \\
22&127.2466&6.7193&4866&55895&335&1.0&0.47&0.471&0.878&0.877&1.0 \\
23&176.7038&31.9488&4614&55604&299&1.0&0.478&0.479&0.845&0.846&1.0 \\
24&322.7388&-1.587&4384&56105&873&1.0&0.506&0.509&0.81&0.814&1.0 \\
25&338.7334&15.1238&5038&56235&231&1.0&0.514&0.513&0.822&0.825&1.0 \\
26&118.3424&11.0724&4511&55602&755&1.0&0.565&0.558&0.82&0.824&1.0 \\
27&189.7042&24.9656&5984&56337&333&1.0&0.559&0.558&0.826&0.828&1.0 \\
28&153.3944&25.7594&6465&56279&537&1.0&0.616&0.619&0.825&0.828&1.0 \\
29&321.1025&1.1421&4193&55476&665&1.0&0.7&0.7&0.875&0.878&1.0 \\
30&358.3876&-8.6495&7166&56602&784&1.0&0.167&0.167&0.653&0.657&1.0 \\
31&253.2719&50.6749&6311&56447&56&1.0&0.095&0.093&0.57&0.575&1.0 \\
32&321.5421&1.9777&5143&55828&640&1.0&0.152&0.151&0.772&0.78&1.0 \\
33&150.7402&6.319&4874&55673&82&1.0&0.225&0.225&0.781&0.781&1.0 \\
34&17.3453&18.7404&5124&55894&36&1.0&0.226&0.226&0.645&0.644&1.0 \\
35&207.8211&-1.7923&4041&55361&522&1.0&0.208&0.208&0.77&0.774&1.0 \\
36&31.4846&-2.3339&4347&55830&156&1.0&0.283&0.281&0.462&0.462&1.0 \\
37&181.6564&37.855&4700&55709&242&1.0&0.263&0.262&0.607&0.609&1.0 \\
38&26.7007&22.0027&5108&55888&811&1.0&0.267&0.266&0.745&0.745&1.0 \\
39&128.9045&8.3751&5285&55946&510&1.0&0.259&0.258&0.555&0.558&1.0 \\
40&42.3078&-2.8227&4342&55531&112&1.0&0.292&0.29&0.459&0.459&1.0 \\
41&18.952&5.1636&4425&55864&376&1.0&0.337&0.336&0.607&0.606&1.0 \\
42&173.7037&54.187&6697&56419&386&1.0&0.302&0.301&0.614&0.613&1.0 \\
43&134.3035&19.7311&5175&55955&804&1.0&0.377&0.379&0.632&0.629&1.0 \\
44&145.78&55.292&5710&56658&116&1.0&0.355&0.356&0.607&0.609&1.0 \\
45&258.3841&42.9416&6062&56091&91&1.0&0.352&0.354&0.713&0.714&1.0 \\
46&242.464&37.8714&5200&56091&151&1.0&0.448&0.448&0.648&0.648&1.0 \\
47&176.8551&35.8301&4646&55622&879&1.0&0.473&0.473&0.658&0.658&1.0 \\
48&199.5047&14.466&5425&56003&820&1.0&0.478&0.481&0.826&0.825&1.0 \\
49&203.1219&19.9543&5861&56069&347&1.0&0.494&0.496&0.831&0.835&1.0 \\
50&189.5541&23.2605&5983&56310&25&1.0&0.48&0.48&0.849&0.85&1.0 \\
\bottomrule 
\end{tabular} 

\newpage 
\renewcommand\arraystretch{1.0}
\begin{tabular}{cccccccccccc}
\toprule
 & RA & DEC & PLATE & MJD & FIBER & $P_L$ & z & $z_{PG}$ & $z_{PE}$ & $mean_{eyeZE}$ & $P_T$ \\ 
\hline
51&248.9654&17.222&4064&55366&77&1.0&0.497&0.496&0.715&0.714&1.0 \\
52&189.5441&46.0467&6635&56370&501&1.0&0.492&0.493&0.822&0.823&1.0 \\
53&219.7836&54.8809&6710&56416&27&1.0&0.461&0.462&0.722&0.722&1.0 \\
54&214.9087&19.692&5895&56046&973&1.0&0.528&0.527&0.883&0.883&1.0 \\
55&331.1801&30.5522&5961&56460&545&1.0&0.538&0.552&0.83&0.832&1.0 \\
56&6.3176&28.2843&6274&56550&455&1.0&0.501&0.502&0.816&0.816&1.0 \\
57&240.6962&7.4048&4893&55709&858&1.0&0.622&0.623&0.869&0.874&1.0 \\
58&115.6977&25.8798&4462&55600&271&1.0&0.441&0.441&0.682&0.684&1.0 \\
59&160.3564&43.6801&4690&55653&424&1.0&0.238&0.236&0.439&0.439&1.0 \\
60&180.9566&23.4359&5969&56069&846&1.0&0.263&0.263&0.612&0.619&1.0 \\
61&177.6002&19.2965&5881&56038&604&1.0&0.289&0.288&0.476&0.474&1.0 \\
62&139.0009&21.6609&5771&56011&574&1.0&0.322&0.322&0.511&0.508&1.0 \\
63&30.0581&3.8889&4270&55511&7&1.0&0.429&0.427&0.776&0.754&1.0 \\
64&133.0317&40.5817&4607&56248&711&1.0&0.447&0.446&0.842&0.844&1.0 \\
65&129.9769&59.5436&5148&56220&583&1.0&0.475&0.474&0.821&0.822&1.0 \\
66&164.8521&11.0051&5356&55979&881&1.0&0.473&0.473&0.832&0.834&1.0 \\
67&189.5441&46.0467&6667&56412&1&1.0&0.491&0.493&0.822&0.822&1.0 \\
68&154.1187&9.2856&5334&55928&953&1.0&0.545&0.544&0.859&0.86&1.0 \\
69&354.6202&27.6904&6516&56571&543&1.0&0.549&0.547&0.846&0.848&1.0 \\
70&118.5745&16.4872&4495&55566&811&1.0&0.607&0.607&0.812&0.814&1.0 \\
71&152.2575&38.1894&4567&55589&235&1.0&0.648&0.648&0.828&0.828&1.0 \\
72&337.6706&-1.1175&4203&55447&368&1.0&0.33&0.33&0.526&0.531&1.0 \\
73&251.7678&29.8492&4189&55679&752&1.0&0.375&0.374&0.59&0.591&1.0 \\
74&137.8548&16.6327&5301&55987&328&1.0&0.394&0.394&0.693&0.698&1.0 \\
75&18.8537&-1.4834&4353&55532&663&1.0&0.438&0.438&0.802&0.805&1.0 \\
76&144.1259&17.0156&5317&56000&579&1.0&0.403&0.404&0.819&0.821&1.0 \\
77&6.512&9.7381&6195&56220&424&1.0&0.246&0.244&0.391&0.39&1.0 \\
78&160.1055&3.5501&4773&55648&290&1.0&0.439&0.438&0.642&0.643&1.0 \\
79&213.0731&46.9419&6751&56368&355&1.0&0.402&0.405&0.657&0.653&1.0 \\
80&215.0974&30.7816&3867&55652&579&1.0&0.466&0.465&0.813&0.814&1.0 \\
81&137.8928&43.2083&4687&56369&409&1.0&0.497&0.495&0.837&0.84&1.0 \\
82&135.5464&1.2383&3816&55272&61&1.0&0.514&0.513&0.831&0.833&1.0 \\
83&143.9865&53.3553&5724&56364&508&1.0&0.198&0.198&0.424&0.424&1.0 \\
84&28.2369&-2.3081&4348&55559&155&1.0&0.534&0.536&0.669&0.665&1.0 \\
85&350.6051&34.2495&7139&56568&687&1.0&0.292&0.29&0.612&0.612&1.0 \\
86&224.9773&-2.6469&4020&55332&247&1.0&0.511&0.509&0.824&0.83&1.0 \\
87&6.2129&-2.6018&4367&55566&263&1.0&0.583&0.584&0.85&0.852&1.0 \\
88&156.6438&37.2146&4559&55597&409&1.0&0.484&0.484&0.636&0.639&1.0 \\
89&185.1209&10.9196&5399&55956&433&1.0&0.507&0.507&0.707&0.712&1.0 \\
90&140.0093&36.1607&4644&55922&855&1.0&0.595&0.594&0.882&0.883&1.0 \\
91&201.2076&5.5061&4839&55703&339&1.0&0.58&0.581&1.018&1.018&1.0 \\
92&184.1385&35.6992&4613&55591&234&0.999&0.347&0.347&0.728&0.73&1.0 \\
93&22.0924&-3.7932&7047&56572&878&0.999&0.395&0.394&0.598&0.654&1.0 \\
94&7.5599&24.7932&6281&56295&145&0.999&0.405&0.402&0.874&0.876&1.0 \\
95&180.5188&55.022&6688&56412&521&0.999&0.512&0.512&0.83&0.832&1.0 \\
96&9.2422&10.4302&5656&55940&264&0.999&0.178&0.178&0.357&0.361&1.0 \\
97&336.5143&28.8703&6586&56485&471&0.999&0.507&0.502&0.825&0.829&1.0 \\
98&39.6814&-5.6274&4399&55811&917&0.999&0.358&0.358&0.726&0.724&1.0 \\
99&333.8121&13.4517&5041&55749&103&0.999&0.605&0.608&0.941&0.944&1.0 \\
100&5.8983&-6.5602&7151&56598&634&0.999&0.292&0.291&0.541&0.534&1.0 \\
\bottomrule 
\end{tabular} 

\newpage
 
\begin{tabular}{cccccccccccc}
\toprule
 & RA & DEC & PLATE & MJD & FIBER & $P_L$ & z & $z_{PG}$ & $z_{PE}$ & $mean_{eyeZE}$ & $P_T$ \\ 
 \hline
101&9.2422&10.4302&6201&56186&264&0.999&0.178&0.178&0.361&0.361&1.0 \\
102&236.1196&19.2834&3937&55352&191&0.999&0.468&0.469&0.831&0.834&1.0 \\
103&135.0759&24.7684&5179&55957&95&0.999&0.476&0.477&0.803&0.807&1.0 \\
104&177.0745&26.108&6411&56331&427&0.999&0.291&0.289&0.74&0.742&1.0 \\
105&134.2845&35.8839&4602&55644&841&0.999&0.538&0.54&0.679&0.675&1.0 \\
106&3.0886&2.5671&4298&55511&759&0.999&0.414&0.413&0.691&0.688&1.0 \\
107&251.1692&50.6684&6311&56447&334&0.999&0.182&0.183&0.702&0.712&1.0 \\
108&255.5705&36.1605&4988&55825&645&0.998&0.552&0.553&0.73&0.731&1.0 \\
109&353.1292&21.7432&6115&56237&979&0.998&0.369&0.368&0.627&0.629&1.0 \\
110&200.5318&26.3394&5996&56096&706&0.998&0.304&0.303&0.626&0.63&1.0 \\
111&1.4434&15.5096&6177&56268&906&0.998&0.295&0.294&0.4&0.4&1.0 \\
112&339.6335&29.3202&6585&56479&719&0.998&0.482&0.482&0.68&0.68&1.0 \\
113&5.6359&-0.3464&4220&55447&481&0.998&0.498&0.5&1.037&0.632&1.0 \\
114&208.6658&51.3517&6740&56401&11&0.998&0.491&0.49&0.683&0.684&1.0 \\
115&337.2294&24.2962&6300&56449&292&0.998&0.309&0.308&0.434&0.432&1.0 \\
116&195.0441&9.6711&5417&55978&646&0.998&0.297&0.297&0.565&0.564&1.0 \\
117&203.5244&53.5181&6757&56416&275&0.998&0.49&0.49&0.739&0.739&1.0 \\
118&243.9984&51.2459&6316&56483&834&0.998&0.359&0.36&0.482&0.478&1.0 \\
119&197.2786&11.4382&5422&55986&379&0.998&0.525&0.524&0.719&0.721&1.0 \\
120&175.9948&45.3268&6645&56398&264&0.997&0.44&0.441&0.548&0.544&1.0 \\
121&185.4293&40.7285&6633&56369&164&0.997&0.216&0.216&0.507&0.501&1.0 \\
122&156.8435&46.1132&6659&56607&299&0.997&0.521&0.52&0.749&0.75&1.0 \\
123&124.7393&29.6162&4447&55542&81&0.997&0.579&0.576&0.757&0.756&1.0 \\
124&137.5782&34.9972&5811&56334&583&0.997&0.52&0.521&0.666&0.666&1.0 \\
125&1.4434&15.5096&6172&56269&69&0.996&0.295&0.294&0.399&0.4&1.0 \\
126&161.1167&13.2743&5348&55978&957&0.996&0.349&0.348&0.472&0.474&1.0 \\
127&7.3882&29.8282&6274&56550&757&0.996&0.496&0.497&0.761&0.767&1.0 \\
128&0.7136&15.5908&6177&56268&744&0.995&0.427&0.428&0.756&0.756&1.0 \\
129&202.3971&16.0299&5434&56033&376&0.995&0.296&0.295&0.491&0.475&1.0 \\
130&340.6461&5.7733&4411&56164&113&0.994&0.436&0.437&0.684&0.685&0.99 \\
131&171.6804&23.9861&6421&56274&577&0.994&0.566&0.564&0.791&0.794&0.99 \\
132&21.4969&19.6116&5135&55862&850&0.993&0.101&0.102&0.579&0.584&0.99 \\
133&256.4372&30.4618&4997&55738&389&0.993&0.553&0.551&0.809&0.814&0.99 \\
134&224.4952&18.075&5902&56042&203&0.993&0.47&0.483&0.914&0.912&0.99 \\
135&172.253&20.7167&5879&56047&630&0.992&0.208&0.209&0.439&0.44&0.99 \\
136&246.2337&28.3755&5007&55710&53&0.99&0.453&0.452&0.553&0.557&0.99 \\
137&188.5778&57.148&6832&56426&591&0.99&0.432&0.433&0.797&0.798&0.99 \\
138&350.4799&34.3269&7139&56568&640&0.99&0.292&0.291&0.403&0.398&0.99 \\
139&185.5842&9.6532&5396&55947&939&0.989&0.465&0.477&0.718&0.72&0.99 \\
140&15.5525&29.6444&6257&56274&28&0.989&0.359&0.359&0.472&0.474&0.99 \\
141&328.2786&25.191&5960&56097&96&0.989&0.201&0.201&0.553&0.555&0.99 \\
142&13.899&11.3643&5706&56165&718&0.988&0.082&0.083&0.727&0.717&0.99 \\
143&20.6702&24.4333&5693&56246&139&0.988&0.542&0.541&0.731&0.724&0.99 \\
144&125.5973&32.6817&4446&55589&756&0.987&0.553&0.553&0.805&0.806&0.99 \\
145&178.7055&4.095&4765&55674&115&0.986&0.443&0.444&0.65&0.649&0.99 \\
146&139.9778&0.8461&3821&55535&955&0.986&0.433&0.432&0.587&0.587&0.99 \\
147&322.6555&-0.5033&4193&55476&21&0.985&0.47&0.471&0.754&0.757&0.98 \\
148&342.6829&31.1553&6508&56535&998&0.982&0.425&0.426&0.69&0.696&0.98 \\
149&221.893&16.5992&5474&56036&341&0.98&0.53&0.53&0.771&0.771&0.98 \\
150&249.4794&33.6617&5188&55803&146&0.98&0.335&0.333&0.535&0.534&0.98 \\
\bottomrule 
\end{tabular} 
 
\newpage
 
\begin{tabular}{cccccccccccc}
\toprule
 & RA & DEC & PLATE & MJD & FIBER & $P_L$ & z & $z_{PG}$ & $z_{PE}$ & $mean_{eyeZE}$ & $P_T$ \\ 
 \hline
151&137.5782&34.9972&4645&55623&95&0.979&0.519&0.522&0.667&0.667&0.98 \\
152&169.6515&11.6851&5368&56001&392&0.979&0.504&0.506&0.617&0.615&0.98 \\
153&17.4289&30.2485&6263&56279&445&0.976&0.385&0.384&0.689&0.69&0.98 \\
154&141.3097&23.3655&5774&56002&235&0.975&0.575&0.573&0.768&0.769&0.98 \\
155&353.682&30.3667&6501&56563&228&0.973&0.079&0.079&0.534&0.544&0.97 \\
156&123.0994&38.6284&3805&55269&423&0.972&0.5&0.502&0.726&0.723&0.97 \\
157&209.103&36.1326&3853&55268&951&0.966&0.555&0.558&0.764&0.756&0.97 \\
158&13.3108&1.9039&4308&55565&439&0.966&0.494&0.493&0.675&0.675&0.97 \\
159&206.6969&4.1721&4785&55659&610&0.966&0.414&0.416&0.623&0.623&0.97 \\
160&225.1194&61.5364&6982&56444&843&0.963&0.506&0.506&0.793&0.797&0.96 \\
161&15.3889&20.9285&5703&56190&896&0.959&0.403&0.402&0.762&0.764&0.96 \\
162&197.0164&28.3228&6483&56341&63&0.953&0.582&0.58&0.764&0.764&0.95 \\
163&136.6796&30.4776&5809&56353&409&0.953&0.512&0.513&0.73&0.73&0.95 \\
164&169.6424&18.8327&5878&56033&391&0.953&0.439&0.439&0.637&0.634&0.95 \\
165&334.2061&5.2761&4319&55507&547&0.952&0.512&0.514&0.671&0.67&0.95 \\
166&150.7027&20.8144&5784&56029&104&0.951&0.372&0.371&0.492&0.492&0.95 \\
167&172.5368&12.4397&5372&55978&709&0.95&0.492&0.491&0.777&0.781&0.95 \\
168&114.5797&22.8478&4470&55587&420&1.0&0.205&0.205&0.624&0.623&0.9 \\
169&22.4308&7.8777&4553&55584&393&1.0&0.437&0.438&0.838&0.838&0.9 \\
170&5.0461&32.0964&7128&56567&584&0.999&0.162&0.162&0.325&0.319&0.9 \\
171&152.2732&40.8465&4562&55570&557&0.999&0.512&0.51&0.668&0.667&0.9 \\
172&317.5483&5.4372&4078&55358&620&0.999&0.414&0.416&0.721&0.721&0.9 \\
173&211.6158&15.8081&5451&56010&815&0.999&0.476&0.475&0.812&0.815&0.9 \\
174&234.0938&51.1223&6722&56431&735&0.999&0.596&0.595&0.736&0.735&0.9 \\
175&208.9611&32.3567&3861&55274&229&0.999&0.487&0.488&0.695&0.696&0.9 \\
176&220.9617&1.3445&4021&55620&106&0.999&0.526&0.527&0.674&0.671&0.9 \\
177&223.3063&6.3302&4857&55711&393&0.999&0.566&0.571&0.696&0.694&0.9 \\
178&8.9715&5.6913&4419&55867&498&0.999&0.365&0.366&0.674&0.669&0.9 \\
179&356.9756&20.0522&6126&56269&845&0.999&0.411&0.412&0.741&0.741&0.9 \\
180&237.5934&52.8266&6715&56449&938&0.999&0.442&0.443&0.731&0.733&0.9 \\
181&16.4021&28.2601&6259&56565&461&0.999&0.486&0.486&0.884&0.781&0.9 \\
182&176.2923&0.7391&3841&56016&957&0.999&0.53&0.524&0.642&0.642&0.9 \\
183&127.3317&54.1993&5156&55925&574&0.999&0.361&0.36&0.678&0.675&0.9 \\
184&216.0698&4.4967&4781&55653&660&0.999&0.309&0.309&0.565&0.563&0.9 \\
185&346.7121&22.9308&6591&56535&991&0.998&0.194&0.195&0.586&0.587&0.9 \\
186&343.2212&15.6879&5040&56243&734&0.998&0.273&0.272&0.674&0.419&0.9 \\
187&191.5569&12.2667&5409&55957&966&0.997&0.633&0.633&0.826&0.827&0.9 \\
188&191.3409&17.2785&5856&56090&427&0.997&0.559&0.557&0.726&0.726&0.9 \\
189&233.1948&-0.6518&4010&55350&236&0.997&0.342&0.343&0.686&0.685&0.9 \\
190&40.7073&-3.6349&7053&56564&664&0.997&0.28&0.278&0.445&0.44&0.9 \\
191&325.0915&4.4258&4083&55443&908&0.996&0.509&0.509&0.633&0.637&0.9 \\
192&146.7165&58.7815&5715&56598&782&0.996&0.349&0.349&0.577&0.649&0.9 \\
193&0.7136&15.5908&6172&56269&232&0.996&0.427&0.428&0.754&0.757&0.9 \\
194&232.7971&52.0507&6722&56431&545&0.995&0.482&0.487&0.784&0.785&0.9 \\
195&332.4037&31.2179&5954&56462&521&0.995&0.481&0.479&0.836&0.836&0.9 \\
196&13.8325&3.9522&4307&55531&699&0.994&0.664&0.666&0.835&0.827&0.9 \\
197&211.0407&40.2749&6630&56358&67&0.993&0.51&0.506&0.629&0.626&0.89 \\
198&220.7873&35.7258&4718&56014&659&0.993&0.557&0.556&0.761&0.764&0.89 \\
199&174.6345&15.131&5378&56011&392&0.993&0.254&0.253&0.493&0.436&0.89 \\
200&161.1167&13.2743&5350&56009&284&0.992&0.349&0.349&0.466&0.473&0.89 \\
\bottomrule 
\end{tabular} 
 
\newpage
 
\begin{tabular}{cccccccccccc}
\toprule
 & RA & DEC & PLATE & MJD & FIBER & $P_L$ & z & $z_{PG}$ & $z_{PE}$ & $mean_{eyeZE}$ & $P_T$ \\ 
 \hline
201&233.7202&30.6619&4722&55735&603&0.992&0.546&0.544&0.669&0.67&0.89 \\
202&23.0487&25.9812&5694&56213&919&0.992&0.494&0.495&0.704&0.678&0.89 \\
203&163.2455&40.0812&4629&55630&852&0.991&0.302&0.303&0.638&0.583&0.89 \\
204&191.086&-2.7746&3793&55214&203&0.991&0.55&0.55&0.713&0.716&0.89 \\
205&238.0017&12.4111&4882&55721&634&0.99&0.107&0.108&0.486&0.483&0.89 \\
206&163.8696&31.3152&6445&56366&883&0.99&0.453&0.451&0.802&0.816&0.89 \\
207&178.0232&21.705&6407&56311&543&0.99&0.618&0.619&0.772&0.775&0.89 \\
208&339.1266&9.5256&5054&56191&495&0.989&0.424&0.425&0.74&0.74&0.89 \\
209&29.6199&1.0778&4234&55478&548&0.988&0.325&0.324&0.781&0.781&0.89 \\
210&176.0852&30.9175&6433&56339&930&0.987&0.076&0.076&0.623&0.695&0.89 \\
211&228.2038&53.9555&6713&56402&396&0.985&0.441&0.44&0.621&0.624&0.89 \\
212&336.7086&5.148&4428&56189&98&0.984&0.09&0.09&0.401&0.406&0.89 \\
213&174.3192&32.0515&4616&55617&283&0.984&0.456&0.461&0.641&0.641&0.88 \\
214&157.7968&28.1283&6456&56339&122&0.983&0.312&0.311&0.549&0.555&0.88 \\
215&233.2035&26.8323&3959&55679&69&0.982&0.523&0.524&0.668&0.664&0.88 \\
216&186.9392&5.4132&4833&55679&266&0.979&0.47&0.47&0.634&0.625&0.88 \\
217&336.3138&6.0182&4428&56189&199&0.979&0.498&0.5&0.736&0.734&0.88 \\
218&174.0321&12.571&5376&55987&258&0.977&0.486&0.487&0.727&0.731&0.88 \\
219&21.6396&14.2556&5140&55836&213&0.976&0.423&0.418&0.648&0.648&0.88 \\
220&252.8827&36.6409&5198&55823&396&0.973&0.603&0.605&0.712&0.718&0.88 \\
221&252.9652&35.5515&5198&55823&372&0.964&0.412&0.414&0.71&0.708&0.87 \\
222&25.5021&31.3093&6601&56335&413&0.961&0.5&0.502&1.068&0.572&0.86 \\
223&118.0482&15.4782&4495&55566&310&0.96&0.282&0.282&0.622&0.623&0.86 \\
224&345.3411&-1.4885&4362&55828&827&0.959&0.434&0.435&0.73&0.734&0.86 \\
225&221.6961&7.9757&4858&55686&786&0.955&0.269&0.268&0.711&0.568&0.86 \\
226&232.7992&10.7432&5493&56009&967&0.954&0.563&0.56&0.823&0.825&0.86 \\
227&23.8523&14.5389&5137&55833&167&0.954&0.449&0.449&0.719&0.72&0.86 \\
228&149.5008&10.1101&5324&55947&843&0.951&0.49&0.491&0.803&0.807&0.86 \\
229&8.8946&9.0665&4540&55863&69&1.0&0.214&0.213&0.799&0.805&0.85 \\
230&204.6975&5.1005&4786&55651&678&1.0&0.271&0.271&0.561&0.559&0.85 \\
231&27.2466&21.0868&5108&55888&99&1.0&0.277&0.276&0.753&0.752&0.85 \\
232&11.5709&25.3449&6286&56301&265&1.0&0.428&0.429&0.822&0.828&0.85 \\
233&190.6792&-0.3908&3793&55214&663&1.0&0.454&0.452&0.815&0.818&0.85 \\
234&246.8241&19.5842&4061&55362&499&1.0&0.462&0.463&0.848&0.848&0.85 \\
235&200.1508&12.4425&5427&56001&700&1.0&0.528&0.529&0.818&0.821&0.85 \\
236&28.1217&9.5832&4530&55564&308&1.0&0.557&0.555&0.841&0.845&0.85 \\
237&11.9627&31.0756&6872&56540&443&1.0&0.643&0.643&0.844&0.847&0.85 \\
238&170.9226&13.9839&5370&56003&683&1.0&0.521&0.52&0.888&0.886&0.85 \\
239&207.1491&9.5015&5442&55978&711&1.0&0.578&0.577&0.812&0.812&0.85 \\
240&159.9814&8.136&5349&55929&379&1.0&0.617&0.619&0.873&0.875&0.85 \\
241&169.8367&65.4877&7110&56746&833&1.0&0.529&0.527&0.826&0.826&0.85 \\
242&182.5102&3.7887&4749&55633&381&1.0&0.596&0.597&0.885&0.885&0.85 \\
243&150.4387&37.4688&4637&55616&705&1.0&0.677&0.677&0.847&0.848&0.85 \\
244&248.9092&49.4879&8056&57186&769&1.0&0.671&0.672&0.89&0.891&0.85 \\
245&338.5431&32.4454&6509&56486&788&1.0&0.422&0.422&0.84&0.842&0.85 \\
246&195.6242&5.7652&4838&55686&467&1.0&0.492&0.493&0.879&0.883&0.85 \\
247&154.1187&9.2856&5336&55957&321&1.0&0.545&0.542&0.86&0.86&0.85 \\
248&160.0284&63.5812&7097&56667&865&1.0&0.528&0.528&0.836&0.836&0.85 \\
249&143.962&15.6466&5315&55978&623&1.0&0.605&0.608&0.86&0.86&0.85 \\
250&204.0295&33.6021&3985&55320&773&0.999&0.484&0.486&0.917&0.92&0.85 \\
\bottomrule 
\end{tabular} 

\newpage

\begin{tabular}{cccccccccccc}
\toprule
 & RA & DEC & PLATE & MJD & FIBER & $P_L$ & z & $z_{PG}$ & $z_{PE}$ & $mean_{eyeZE}$ & $P_T$ \\ 
 \hline
251&226.5216&40.6238&6054&56089&256&0.999&0.483&0.481&0.628&0.628&0.85 \\
252&171.3643&41.4413&4699&55684&447&0.999&0.487&0.485&0.914&0.912&0.85 \\
253&24.3379&16.7282&5130&55835&545&0.999&0.471&0.472&0.651&0.65&0.85 \\
254&154.1552&54.3047&6696&56398&487&0.999&0.487&0.488&0.859&0.862&0.85 \\
255&159.2973&10.5324&5346&55955&829&0.998&0.199&0.618&1.109&1.111&0.85 \\
256&38.4525&-7.1342&4388&55536&88&0.998&0.335&0.335&0.638&0.642&0.85 \\
257&180.8803&13.5498&5390&56002&607&0.997&0.52&0.517&0.993&0.994&0.85 \\
258&320.9945&5.4038&4081&55365&185&0.997&0.608&0.61&1.099&1.095&0.85 \\
259&327.7053&3.4418&4090&55500&539&0.996&0.571&0.567&0.687&0.685&0.85 \\
260&22.2645&3.9838&4310&55508&314&0.992&0.196&0.196&0.909&0.914&0.84 \\
261&353.3421&15.6571&6137&56270&267&0.992&0.378&0.379&0.687&0.69&0.84 \\
262&236.7188&40.2551&6051&56093&405&0.988&0.658&0.655&0.816&0.818&0.84 \\
263&4.8287&26.2703&6279&56243&741&1.0&0.262&0.262&0.604&1.013&0.8 \\
264&22.0923&22.3834&5115&55885&56&1.0&0.277&0.275&0.698&0.696&0.8 \\
265&32.8873&0.125&4235&55451&893&1.0&0.213&0.213&0.823&0.824&0.8 \\
266&155.9018&15.7562&5337&55987&126&1.0&0.368&0.368&0.762&0.762&0.8 \\
267&207.704&65.7459&6986&56717&439&1.0&0.376&0.376&0.706&0.705&0.8 \\
268&198.9484&46.7589&6624&56385&965&1.0&0.466&0.463&0.807&0.81&0.8 \\
269&205.3212&36.1224&3986&55329&887&1.0&0.555&0.555&0.68&0.679&0.8 \\
270&349.6495&26.6599&6587&56537&438&0.999&0.434&0.435&0.771&0.774&0.8 \\
271&329.2513&16.9616&5061&55806&704&0.999&0.28&0.278&0.419&0.417&0.8 \\
272&32.7654&0.838&3611&55128&717&0.999&0.291&0.29&0.716&0.715&0.8 \\
273&237.8022&44.8269&6033&56069&487&0.999&0.475&0.475&0.75&0.751&0.8 \\
274&221.0844&13.5378&5473&56033&642&0.999&0.262&0.262&0.737&0.734&0.8 \\
275&183.495&62.8751&6972&56426&825&0.999&0.374&0.37&0.651&0.65&0.8 \\
276&142.0039&24.1646&5774&56002&919&0.999&0.547&0.542&0.72&0.717&0.8 \\
277&209.3069&3.0451&4784&55677&284&0.999&0.346&0.347&0.633&0.632&0.8 \\
278&217.9071&40.9282&5171&56038&678&0.999&0.353&0.353&0.772&0.773&0.8 \\
279&219.354&17.344&5469&56037&660&0.998&0.264&0.263&0.662&0.66&0.8 \\
280&140.4015&32.5592&5808&56325&787&0.998&0.344&0.346&0.458&0.454&0.8 \\
281&21.0178&2.6268&4314&55855&763&0.998&0.511&0.511&0.698&0.695&0.8 \\
282&329.8912&24.1816&5952&56453&939&0.998&0.256&0.256&0.476&0.472&0.8 \\
283&131.9713&7.0192&5289&55893&216&0.997&0.406&0.406&0.761&0.763&0.8 \\
284&212.1806&-1.6042&4035&55383&557&0.997&0.469&0.468&0.788&0.794&0.8 \\
285&200.5318&26.3394&5997&56309&122&0.997&0.304&0.303&0.619&0.629&0.8 \\
286&205.7799&15.1811&5441&56017&574&0.996&0.276&0.274&0.718&0.635&0.8 \\
287&39.255&-5.8455&4386&55540&203&0.996&0.453&0.452&0.849&0.851&0.8 \\
288&192.5906&19.0848&5856&56090&868&0.996&0.5&0.5&0.82&0.821&0.8 \\
289&206.9027&-3.1424&4044&55359&255&0.996&0.448&0.448&0.585&0.585&0.8 \\
290&244.0409&32.2355&4956&55737&703&0.995&0.452&0.451&0.796&0.799&0.8 \\
291&235.7646&3.9591&4804&55679&397&0.995&0.485&0.487&0.763&0.764&0.8 \\
292&128.065&41.9915&3808&55513&884&0.995&0.336&0.337&0.468&0.468&0.8 \\
293&17.3649&22.7717&5126&55923&495&0.994&0.498&0.498&0.797&0.8&0.8 \\
294&237.7069&4.8213&4877&55707&322&0.994&0.362&0.363&0.664&0.671&0.8 \\
295&244.4578&50.5162&6316&56483&157&0.993&0.057&0.053&0.625&0.63&0.8 \\
296&204.0125&-1.3843&4046&55605&207&0.992&0.451&0.453&0.792&0.795&0.79 \\
297&177.1556&34.5531&4646&55622&69&0.992&0.481&0.482&0.816&0.814&0.79 \\
298&207.2167&7.1418&4865&55713&994&0.991&0.377&0.376&0.567&0.568&0.79 \\
299&145.6952&4.8124&4798&55672&609&0.991&0.369&0.37&0.676&0.676&0.79 \\
300&28.4645&29.1361&6269&56246&26&0.991&0.358&0.359&0.822&0.825&0.79 \\
\bottomrule 
\end{tabular} 

\newpage
 
\begin{tabular}{cccccccccccc}
\toprule
 & RA & DEC & PLATE & MJD & FIBER & $P_L$ & z & $z_{PG}$ & $z_{PE}$ & $mean_{eyeZE}$ & $P_T$ \\ 
 \hline
301&140.8969&44.7792&4696&56354&602&0.991&0.453&0.454&0.647&0.647&0.79 \\
302&207.8508&12.348&5443&56010&181&0.99&0.475&0.488&0.793&0.795&0.79 \\
303&16.1574&24.6193&5701&55949&938&0.99&0.262&0.261&0.483&0.48&0.79 \\
304&156.6061&49.5063&6694&56386&449&0.99&0.502&0.501&0.638&0.634&0.79 \\
305&189.6456&53.8166&6680&56420&276&0.99&0.389&0.39&0.735&0.737&0.79 \\
306&175.499&12.0599&5379&55986&804&0.989&0.32&0.319&1.022&1.02&0.79 \\
307&38.7375&-9.026&4389&55539&171&0.989&0.485&0.487&0.616&0.609&0.79 \\
308&163.4217&61.7198&7096&56683&949&0.989&0.521&0.522&0.804&0.805&0.79 \\
309&163.8089&63.6662&7104&56711&683&0.989&0.507&0.506&0.643&0.645&0.79 \\
310&332.7269&5.7945&4427&56107&478&0.988&0.177&0.177&0.776&0.776&0.79 \\
311&173.2826&36.7101&4615&55618&751&0.988&0.478&0.477&0.847&0.846&0.79 \\
312&169.3396&42.4221&4686&56013&402&0.987&0.324&0.324&0.781&0.78&0.79 \\
313&169.1691&54.9578&7093&56657&415&0.987&0.462&0.463&0.754&0.754&0.79 \\
314&120.3392&26.2055&4458&55536&341&0.987&0.388&0.442&1.039&0.513&0.79 \\
315&129.6789&19.4517&4484&55565&133&0.986&0.459&0.458&0.699&0.696&0.79 \\
316&337.6379&3.389&4291&55525&488&0.985&0.295&0.294&0.515&0.508&0.79 \\
317&177.8081&59.7521&7098&56661&382&0.984&0.26&0.26&0.788&0.781&0.79 \\
318&144.8292&62.9965&5720&56602&622&0.98&0.368&0.368&0.679&0.674&0.78 \\
319&205.5448&2.4691&4039&55648&533&0.98&0.49&0.491&0.604&0.606&0.78 \\
320&116.8439&21.8474&4473&55589&427&0.979&0.463&0.462&0.608&0.606&0.78 \\
321&157.8873&49.6349&6694&56386&290&0.978&0.364&0.365&0.693&0.685&0.78 \\
322&331.5924&-2.1985&4377&55828&729&0.971&0.502&0.504&0.809&0.816&0.78 \\
323&178.1606&11.3971&5384&55984&237&0.97&0.496&0.498&0.751&0.755&0.78 \\
324&179.9073&48.1052&6644&56384&832&0.966&0.206&0.207&0.66&0.684&0.77 \\
325&186.8933&3.9815&4753&55674&519&0.966&0.436&0.438&0.762&0.764&0.77 \\
326&200.5782&64.0063&6822&56711&571&0.964&0.482&0.484&0.785&0.785&0.77 \\
327&244.0409&32.2355&4960&55747&247&0.964&0.452&0.454&0.798&0.8&0.77 \\
328&137.8128&44.0708&4687&56338&436&0.961&0.426&0.429&0.771&0.771&0.77 \\
329&32.177&-3.5351&4393&55944&804&0.961&0.347&0.346&0.448&0.442&0.77 \\
330&206.3508&7.0963&4865&55713&836&1.0&0.229&0.229&0.621&0.625&0.77 \\
331&9.096&1.2598&4302&55531&47&1.0&0.546&0.544&0.812&0.806&0.77 \\
332&26.8826&8.4239&4532&55559&806&1.0&0.134&0.134&0.674&0.671&0.77 \\
333&135.1302&17.4353&5297&55913&653&0.999&0.426&0.426&0.729&0.73&0.77 \\
334&0.637&16.5119&6173&56238&368&0.996&0.173&0.172&0.3&0.295&0.76 \\
335&126.0904&53.3606&7376&56749&41&0.954&0.487&0.481&0.623&0.615&0.76 \\
336&229.1471&22.7604&3962&55660&988&0.954&0.381&0.383&1.05&1.047&0.76 \\
337&2.3299&3.2238&4297&55806&184&0.952&0.375&0.373&0.809&0.808&0.76 \\
338&157.7742&0.1498&3833&55290&514&0.993&0.259&0.258&0.424&0.419&0.76 \\
339&133.0062&2.9457&3813&55532&807&0.991&0.556&0.555&0.721&0.721&0.76 \\
340&122.4983&2.1159&4788&55889&628&0.986&0.402&0.404&0.811&0.802&0.76 \\
341&147.826&52.2992&5724&56364&14&0.986&0.521&0.524&0.687&0.686&0.76 \\
342&357.0739&11.6525&6150&56187&89&0.959&0.472&0.472&0.772&0.774&0.74 \\
343&143.3092&22.088&5789&56246&362&1.0&0.35&0.352&0.693&0.694&0.7 \\
344&343.6146&6.8472&5060&56181&169&1.0&0.469&0.467&0.83&0.832&0.7 \\
345&343.1472&30.7205&6507&56478&653&1.0&0.493&0.494&0.876&0.876&0.7 \\
346&341.3532&30.9137&6508&56535&199&1.0&0.483&0.484&0.907&0.908&0.7 \\
347&17.6062&-8.1183&7158&56956&375&1.0&0.475&0.476&0.822&0.829&0.7 \\
348&121.8303&48.0916&3687&55269&694&1.0&0.533&0.533&0.902&0.903&0.7 \\
349&120.4713&13.6811&4505&55603&553&1.0&0.651&0.649&0.856&0.857&0.7 \\
350&34.4954&-9.004&4395&55828&269&1.0&0.446&0.447&0.887&0.887&0.7 \\
\bottomrule 
\end{tabular} 

\newpage

\begin{tabular}{cccccccccccc}
\toprule
 & RA & DEC & PLATE & MJD & FIBER & $P_L$ & z & $z_{PG}$ & $z_{PE}$ & $mean_{eyeZE}$ & $P_T$ \\ 
 \hline
351&210.2792&41.0622&6630&56358&844&1.0&0.448&0.449&0.87&0.87&0.7 \\
352&15.0112&32.0889&6593&56270&403&1.0&0.487&0.487&0.873&0.874&0.7 \\
353&251.1224&27.3479&4189&55679&329&1.0&0.506&0.502&0.871&0.872&0.7 \\
354&203.2293&35.547&3986&55329&517&1.0&0.58&0.579&0.993&0.996&0.7 \\
355&144.2942&32.1162&5806&56310&529&1.0&0.577&0.575&1.019&1.019&0.7 \\
356&0.7187&7.0414&4535&55860&151&1.0&0.657&0.658&1.027&1.026&0.7 \\
357&166.1015&29.6335&6443&56367&205&1.0&0.459&0.458&0.824&0.826&0.7 \\
358&176.4516&21.7884&6432&56309&704&1.0&0.492&0.492&0.887&0.889&0.7 \\
359&171.9566&52.8471&6698&56637&183&1.0&0.504&0.527&0.752&0.766&0.7 \\
360&24.8965&31.2826&6599&56567&37&1.0&0.512&0.513&0.879&0.879&0.7 \\
361&198.4153&47.5507&6624&56385&909&1.0&0.536&0.537&0.872&0.872&0.7 \\
362&205.295&20.7845&5851&56075&383&1.0&0.683&0.681&0.857&0.858&0.7 \\
363&182.142&7.4393&5393&55946&461&1.0&0.46&0.461&0.808&0.812&0.7 \\
364&168.8302&4.3146&4770&55928&29&1.0&0.499&0.499&0.65&0.649&0.7 \\
365&175.5293&12.9521&5381&56009&447&1.0&0.563&0.562&0.913&0.911&0.7 \\
366&32.2812&-1.0957&4235&55451&207&1.0&0.426&0.424&0.762&0.758&0.7 \\
367&185.6856&59.482&6968&56443&425&1.0&0.595&0.59&0.923&0.922&0.7 \\
368&150.0294&3.0329&4737&55630&805&1.0&0.608&0.608&0.87&0.87&0.7 \\
369&121.0774&27.7734&4458&55536&833&0.999&0.496&0.495&0.934&0.933&0.7 \\
370&175.7924&45.4451&6645&56398&317&0.999&0.702&0.703&0.84&0.841&0.7 \\
371&236.7033&11.746&4886&55737&187&0.999&0.698&0.695&0.888&0.883&0.7 \\
372&333.159&2.0303&4321&55504&391&0.999&0.635&0.636&1.107&1.101&0.7 \\
373&234.2179&44.192&6041&56102&197&0.999&0.485&0.486&0.862&0.86&0.7 \\
374&233.2274&4.8999&4805&55715&605&0.999&0.426&0.422&0.804&0.804&0.7 \\
375&135.1777&37.8146&4608&55973&779&0.999&0.623&0.624&1.104&1.098&0.7 \\
376&162.3164&44.9276&4690&55653&877&0.998&0.585&0.584&0.907&0.909&0.7 \\
377&115.433&19.1082&4488&55571&847&0.998&0.353&0.351&0.854&0.859&0.7 \\
378&180.9264&53.1512&6682&56390&543&0.998&0.485&0.484&0.75&0.75&0.7 \\
379&150.0906&47.7961&6663&56338&640&0.998&0.248&0.249&0.702&0.694&0.7 \\
380&129.3629&61.4292&5707&56269&901&0.998&0.44&0.442&0.846&0.844&0.7 \\
381&336.4794&25.5091&5957&56210&37&0.998&0.489&0.489&0.745&0.746&0.7 \\
382&178.067&21.7956&6423&56313&215&0.997&0.617&0.619&0.805&0.806&0.7 \\
383&168.0662&58.2332&7102&56666&969&0.997&0.585&0.586&0.878&0.881&0.7 \\
384&143.208&37.5164&4576&55592&901&0.997&0.497&0.496&0.849&0.851&0.7 \\
385&191.2632&2.7318&4755&55660&213&0.997&0.495&0.496&0.831&0.835&0.7 \\
386&16.1325&29.194&6256&56323&917&0.997&0.462&0.46&0.667&0.667&0.7 \\
387&34.3508&-9.3814&4395&55828&297&0.996&0.6&0.601&0.945&0.943&0.7 \\
388&181.7261&36.5147&4610&55621&781&0.996&0.462&0.463&0.569&0.567&0.7 \\
389&191.6312&7.1626&4835&55688&705&0.996&0.495&0.496&0.921&0.923&0.7 \\
390&176.2923&0.7391&3841&55572&957&0.995&0.529&0.535&0.641&0.642&0.7 \\
391&22.1685&-7.7357&7161&56625&401&0.995&0.284&0.282&0.8&0.807&0.7 \\
392&151.19&41.5726&4565&55591&43&0.995&0.489&0.498&0.678&0.68&0.7 \\
393&127.4452&19.1464&4489&55545&627&0.994&0.455&0.453&0.831&0.835&0.7 \\
394&194.8462&2.3886&4005&55325&715&0.994&0.502&0.503&0.874&0.876&0.7 \\
395&189.0387&21.843&5985&56089&614&0.994&0.387&0.389&0.86&0.858&0.7 \\
396&169.2965&14.6982&5367&55986&407&0.994&0.4&0.397&0.883&0.881&0.7 \\
397&166.3501&44.9208&4685&55657&663&0.994&0.224&0.222&0.739&0.736&0.7 \\
398&162.0087&35.0297&4635&55615&111&0.994&0.599&0.603&0.721&0.721&0.7 \\
399&116.0158&33.825&4443&55539&942&0.993&0.602&0.604&1.103&1.097&0.7 \\
400&17.5428&23.4982&5699&55953&248&0.993&0.331&0.331&0.708&0.708&0.7 \\
\bottomrule 
\end{tabular} 

\newpage

\begin{tabular}{cccccccccccc}
\toprule
 & RA & DEC & PLATE & MJD & FIBER & $P_L$ & z & $z_{PG}$ & $z_{PE}$ & $mean_{eyeZE}$ & $P_T$ \\ 
 \hline
401&187.6474&5.381&4833&55679&107&0.992&0.47&0.471&0.743&0.744&0.69 \\
402&159.8156&39.5365&4633&55620&265&0.991&0.529&0.528&0.712&0.715&0.69 \\
403&133.6105&7.4505&4867&55924&865&0.989&0.461&0.457&0.767&0.768&0.69 \\
404&241.218&61.7941&6798&56485&834&0.989&0.365&0.367&0.706&0.692&0.69 \\
405&161.0795&3.6325&4773&55648&89&0.988&0.559&0.559&0.874&0.881&0.69 \\
406&318.4847&-1.2406&4191&55444&129&0.984&0.461&0.461&0.792&0.795&0.69 \\
407&235.4098&3.1014&4804&55679&443&0.983&0.341&0.341&0.794&0.497&0.69 \\
408&141.8848&52.4323&5726&56626&293&0.98&0.527&0.526&0.63&0.631&0.69 \\
409&216.2736&49.914&6724&56416&413&0.978&0.538&0.538&0.781&0.781&0.68 \\
410&210.0395&8.5377&5447&55958&193&0.978&0.652&0.655&1.112&1.108&0.68 \\
411&211.6662&5.3535&4783&55652&742&0.978&0.393&0.394&0.543&0.537&0.68 \\
412&259.601&37.6442&4990&55743&385&0.976&0.472&0.472&0.767&0.767&0.68 \\
413&179.4266&52.7289&6683&56388&861&0.976&0.482&0.48&1.096&1.109&0.68 \\
414&238.4449&4.3444&4806&55688&714&0.975&0.492&0.495&0.64&0.636&0.68 \\
415&161.4553&47.6463&6701&56367&411&0.975&0.474&0.473&0.866&0.862&0.68 \\
416&28.3279&18.3564&5117&55925&161&0.972&0.557&0.557&0.742&0.748&0.68 \\
417&248.8037&11.945&4075&55352&439&0.972&0.56&0.584&0.823&0.823&0.68 \\
418&184.0856&41.9727&6633&56369&387&0.972&0.563&0.562&1.037&1.034&0.68 \\
419&20.9638&27.9408&6262&56267&84&0.968&0.596&0.596&0.713&0.716&0.68 \\
420&170.8309&10.9069&5368&56001&131&0.967&0.43&0.43&0.916&0.923&0.68 \\
421&30.8703&30.1949&6271&56304&968&0.964&0.348&0.35&0.569&0.572&0.68 \\
422&254.7412&39.9168&6063&56077&379&0.964&0.334&0.334&0.747&0.749&0.68 \\
423&34.054&31.9979&6606&56304&887&0.962&0.389&0.387&0.848&0.847&0.67 \\
424&238.5759&53.5785&7562&56799&455&0.961&0.065&0.066&0.409&0.414&0.67 \\
425&16.0141&6.6393&4423&55889&546&0.96&0.175&0.176&0.351&0.331&0.67 \\
426&138.0252&41.5316&4641&55947&593&0.958&0.479&0.48&0.822&0.824&0.67 \\
427&183.2311&46.5743&6641&56383&959&0.957&0.519&0.519&0.93&0.935&0.67 \\
428&168.5867&53.1771&6699&56411&786&0.956&0.466&0.464&0.818&0.818&0.67 \\
429&237.944&16.4394&3927&55333&945&0.952&0.493&0.493&0.945&0.943&0.67 \\
\bottomrule 
\end{tabular}

\end{center}

\section{Low-resolution image grading from HSC vs. DECaLS}
\label{app:B}
In \S\ref{sec:confir_rate}, we have motivated the choice to base the estimation of the confirmation rate of the HQ catalog, presented in \S\ref{sec:HQ_catalog}, on the 56 candidates matching the HSC public image database, instead of the 279 matches we have found for the DECaLS image database, with the better image quality of the former sample. In this Appendix, we want to visually quantify this and possibly drive some conclusions about the impact that the choice of poorer quality imaging could have on our results. We start \Zhong{by clarifying} that our arguments cannot be extended to other analyses that might have made a different choice (e.g. T+21), because the grading of the imaging data has a large level of subjectivity, hence we cannot generalize the conclusions we can make on the basis of the criteria adopted in this paper.

\begin{figure*}
    \centering 
    \vspace{+0.5cm}
    \includegraphics[width=1\linewidth]{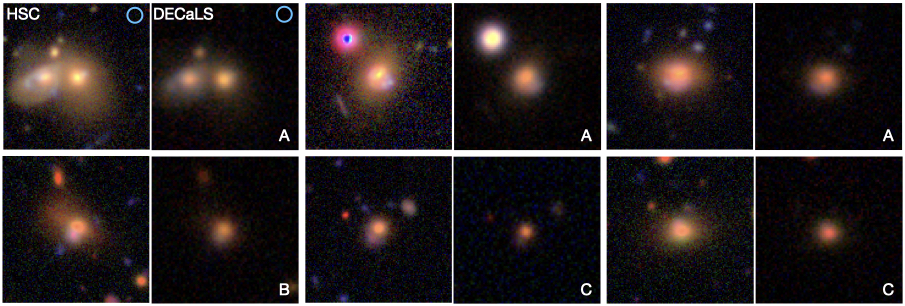}
     \caption{A sample of color images of the HSC lens candidates confirmed with the \Gs\ vs. the corresponding DECaLS images. For each strong lens, we show the $gri$ combined images from the HSC database left panel, against the $grz$ color image of the same pointing from DECaLS (right panel). Color combined images are used with the same set-up on the individual band. Part of the lower quality of DECaLS can be due to the $z$-band imaging which was the only redder band available. In each DECaLS \Zhong{panel,} we show the visual grading given by us. In the top left pair of figures, we also show the size of the fiber from eBOSS (top-right corner).}
\label{fig:HSC_vs_DECALS}
\end{figure*}

Given this necessary preamble, in Fig. \ref{fig:HSC_vs_DECALS}, we show the $gri$ color combined images of six of the eight HSC lens candidates confirmed with the \Gs\ (see \S\ref{sec:confir_rate}), which we consider rather solid having received high \Zhong{grades} both from HSC imaging and eBOSS spectroscopy from the \Gs. Compared \Zhong{to} face-to-face, we also show the $grz$ combined images obtained by the DECaLS database, for which we have given a visual grading according to the standard adopted in \ref{sec:confir_rate}. It looks fairly clear that HSC images have overall higher quality and that this \Zhong{impacts} the grading of the lensing features one is expected to judge from the images. The leverage of our grading has been even slightly lowered since we have started from \Zhong{the evidence} of some background \Zhong{sources} which might help convince our brain that the feature in the images can be a lens, but not too much because we had to consider the possibility of occasional overlap with standard galaxies along the line-of-sight. As a consequence of that, the grading of DECaLS images might be slightly higher than we typically give in \Zhong{the visual} inspection of imaging candidates (e.g. \citealt{2021ApJ...923...16L}). The net result is that for this sample of ``secure'' gravitational lenses \Zhong{if} we used the DECaLS images we would have possibly excluded the 50\%. If we take the test made over this small sample of galaxies at face value, we can predict that DECaLS imaging would lead us to an estimate of the success rate of the order of 20-25\%. To confirm this we have visually inspected the 274 matched to DECaLS and indeed find that maybe 1/5 of them had some convincing lensing features. T+21 used the DECaLS color images to find ``lensing \Zhong{evidence}'' for 447/1551 of their spectroscopical sample, corresponding to a 29\%, which is not far from what we have predicted above, meaning that their confirmation rate is also rather underestimated, and possibly closer to the upper limit found by us (57\%).

\end{document}